\documentclass[aos,preprint]{imsart}
\usepackage[margin=2.5cm]{geometry}
\RequirePackage[OT1]{fontenc}
\RequirePackage{amsmath,amssymb,amsthm}
\RequirePackage[numbers]{natbib}
\RequirePackage[colorlinks,citecolor=blue,urlcolor=blue]{hyperref}
\usepackage{graphicx}

\startlocaldefs
\numberwithin{equation}{section}
\theoremstyle{plain}

\endlocaldefs

\theoremstyle{plain}
\long\def\comment#1{}

\theoremstyle{definition}

\numberwithin{definition}{section}

\numberwithin{remark}{section}

\begin{document}

\begin{frontmatter}
\title{Insights from Optimal Pandemic Shielding in a Multi-Group SEIR Framework\thanksref{T1}
}
\runtitle{Optimal Policy for Fighting a Pandemic}
\thankstext{T1}{This version: \today. We are grateful for helpful comments by Axel B\"orsch-Supan, Sven Klaassen and participants of the MEA Seminar and the Franco-German Fiscal Policy Seminar.}

\begin{aug}
\author{\fnms{Philipp} \snm{Bach}\thanksref{m1}\ead[label=e1]{} \thanksref{T2}}
\and
\author{\fnms{Victor} \snm{Chernozhukov}\thanksref{m2}\ead[label=e2]{}}
\and
\author{\fnms{Martin} \snm{Spindler}\thanksref{m1}\ead[label=e3]{}}

\thankstext{T2}{Corresponding author. Code is available upon request.}

\runauthor{Bach, Chernozhukov, Spindler}

\affiliation{University of Hamburg\thanksmark{m1}, MIT\thanksmark{m2} and University of Hamburg\thanksmark{m1}}

\address{Philipp Bach\\
University of Hamburg\\
Hamburg Business School\\
Moorweidenstr. 18\\
20148 Hamburg\\
Germany\\
E-mail: philipp.bach@uni-hamburg.de}

\address{Victor Chernozhukov\\
Massachusetts Institute of Technology\\
Economics Department\\
USA\\
E-mail: vchern@mit.edu}

\address{Martin Spindler\\
University of Hamburg\\
Hamburg Business School\\
Moorweidenstr. 18\\
20148 Hamburg\\
Germany\\
E-mail: martin.spindler@uni-hamburg.de}
\end{aug}

\begin{abstract}
\textbf{Abstract.} The COVID-19 pandemic constitutes one of the largest threats in recent decades to the health and economic welfare of populations globally. In this paper, we analyze different types of policy measures designed to fight the spread of the virus and minimize economic losses. Our analysis builds on a multi-group SEIR model, which extends the multi-group SIR model introduced by Acemoglu et al.~(2020). We adjust the underlying social interaction patterns and consider an extended set of policy measures. The model is calibrated for Germany. Despite the trade-off between COVID-19 prevention and economic activity that is inherent to shielding policies, our results show that efficiency gains can be achieved by targeting such policies towards different age groups. Alternative policies such as physical distancing can be employed to reduce the degree of targeting and the intensity and duration of shielding. Our results show that a comprehensive approach that combines multiple policy measures simultaneously can effectively mitigate population mortality and economic harm.
\end{abstract}
%


\end{frontmatter}


\begin{center}
\begin{flushleft}
\textit{We will never get tired of saying that the best way out of this pandemic is to take a comprehensive approach. [...] 
Not testing alone. 
Not physical distancing alone. 
Not contact tracing alone. 
Not masks alone. 
Do it all.}
\end{flushleft}
\begin{flushright}
Tedros Adhamom Ghebreyesus, WHO Director-General, July 1, 2020
\end{flushright}
\end{center}

\section{Introduction}

The COVID-19 pandemic constitutes one of the largest threats in recent decades to the health and economic welfare of populations globally. A key challenge for policy makers everywhere is to prevent SARS-CoV-2 infections while avoiding economic losses of a magnitude that would result, in the long run, in an unacceptable level of negative effects on population health and well-being. Policy makers in most countries have reacted to the pandemic by imposing strict lockdown policies. In some countries and regions, strict lockdowns have remained in effect for many months or have been reimposed after initially being relaxed. Although such policies have slowed the spread of the virus by reducing social interactions, the more severe lockdowns have been accompanied by a large decline in economic activity. While protecting health and saving lives must, of course, take the highest priority, an optimal policy has to weigh both health and economic losses $–$ that is, keeping mortality as low as possible, on the one hand, and mitigating an economic downturn on the other. In doing so, the goal is to identify a so-called efficient frontier – in other words, possible combinations of measures that that achieve a certain, ideally very low level of population mortality with minimal economic loss or vice versa. Once an efficient frontier for a set of different lockdown strategies has been constructed, a comparison of these strategies allows policy makers to achieve efficiency gains. The point on the efficient frontier that is considered desirable is a decision that must be made by policy makers and, ideally, society as a whole.

In a recent contribution, \cite{Acemoglu2020} extend the classical SIR model, which is well-known from the epidemiological literature, by explicitly incorporating the trade-off that policy makers must consider in times of the pandemic. The authors derive the efficient frontier for different policies and show that efficiency gains can be achieved by targeting lockdown policies at different age groups, each of which is, in turn, characterized by different productivity and mortality risks.\footnote{\cite{Acemoglu2020} employ the term \textquotedblleft lockdown\textquotedblright{} to denote policies that limit social interactions, such as leisure activities or face-to-face interactions at work. In the following, we will refer to these policies as \textquotedblleft shielding\textquotedblright{} measures to underscore the underlying concept of protecting people with higher mortality risks due to higher age or comorbidities.} In a setting calibrated to the U.S. population and economy, they show that protecting the most vulnerable group (i.e., those aged 65 and older) with stricter shielding rules (i.e., targeted shielding) is associated with fewer losses than a blanket shielding policy (also referred to as a uniform shielding, i.e., a policy that applies equally to all groups). \cite{Acemoglu2020} briefly mention and discuss a potential extension of the multi-group SIR model to the SEIR case. Here, we continue their analysis and analyze a variety of policy measures within the SEIR model. We explicitly state the key equations of this model and calibrate it to social interaction patterns as estimated in \cite{klepac2020}.

In this paper, we consider a model that is calibrated to Germany $–$ that is, we adjust it to the country’s demographic and economic characteristics, as well as its system of health care provision. Germany and the U.S. differ in many regards, such as  the demographic structure of the population,   age-specific employment and income patterns, and the capacities of the health system. We present the results of the model and discuss various policy measures, such as group distancing, test strategies, contact tracing, and combinations of these. We also discuss in detail how a targeted policy, protecting vulnerable groups like old people, might be implemented in practice and discuss some policy examples.

Mortality from COVID-19 is particularly high among older people, \cite{Ferguson2020}, whose productivity is relatively low. Hence, a targeted shielding policy that limits face-to-face contacts with persons aged 65 or older might lead to lower mortality in this population group and less damage to the economy. Additionally, a set of potentially voluntary policies that reduce transmission rates and social contacts could, in principle, be considered as an alternative to age-targeted shielding. Indeed, in our analysis, we find that testing, contact tracing, group distancing and improved conditions for working from home help to reduce the economic costs of the pandemic and the intensity and duration of age-targeted shielding. Moreover, if these measures are combined in a comprehensive  approach as described in the initial quote by Tedros Adhamom Ghebreyesus, population mortality and economic outcomes improve substantially. Throughout our analysis, the efficiency gains associated with age-targeting remain relatively stable and sizable, and we recommend exploiting these gains by improving conditions for individuals at high risk, for example by providing services such as special shopping or consultation hours for older people, as well as testing capacities   for those who have contact with high-risk groups to decrease the probability of infections.

The rest of this paper is structured as follows: In Section \ref{SEIRmodel} we briefly introduce the multi-group SEIR model. In Section \ref{calibration} we describe our specification of the parameters for the SEIR model for Germany. Section \ref{results} presents the results and describes the optimal policies comprising measures such as group distancing, testing, contact tracing and improved medical treatment. Finally, a conclusion summarizes the results and makes a range of policy recommendations.

Because there is still so much that we do not know about SARS-CoV-2, including the transmission rate, mortality rates and aspects related to immunity, all of the results reported throughout the paper must be interpreted with caution. As in the study by \cite{Acemoglu2020}, we do not focus on presenting absolute quantitative results, such as GDP forecasts, but rather qualitative insights into potential policy measures that are considered in variation-of-parameters analyses.

\subsubsection*{Literature review} 
    
The classical SIR and SEIR models are used widely in epidemiology and described in many standard textbooks. Driven by the COVID-19 crisis, various extensions of the standard epidemiological models have been developed and modified to consider economic factors. For example, \cite{Tertilt2020} include individual choices about the amount of time spent on activities outside the house, such as work or consumption, to the standard SIR epidemiological model. These activities are associated with externalities, i.e., higher risk of transmission to and from others. The model also incorporates heterogeneity in terms of age and different policy measures, such as testing or quarantines. \cite{berger2020} provide an extended SEIR model focusing on testing and quarantine measures and thereby explicitly address the imperfect information that arises due to the fact that cases can be symptomatic or asymptomatic. A recent study by \cite{Grimm2020} extends a classical SEIR model by introducing a high and low risk group that differ, for example, in hospitalization and mortality rates. Their study focuses on the evolution of infected, recovered and deceased, i.e., the epidemiological aspects of the SEIR model in a parametrization calibrated to Germany. While a blanket shielding policy (i.e., for the entire population) is, of course, the optimal way to protect everyone from infection, the associated economic losses might become substantial. The multi-group SEIR model incorporates economic costs that arise due to sick leave, productivity losses when individuals work from home and discounted lifetime income losses from deaths due to COVID-19. Moreover, important indirect health consequences are associated with strict shielding measures, such as missed appointments for other conditions, less exercise, mental health issues, increased   alcohol consumption, social isolation and increased levels of domestic abuse. While these indirect, non-pecuniary costs are not incorporated in our study, it  might be useful to model them in future work. 

We build on the work of \cite{Acemoglu2020}, who study targeted shielding policies in a multi-group SIR model, and thereby address the trade-off between mortality and economic losses. They consider two possible targeting strategies: finding separate, optimal shielding policies for the young, middle-aged and senior groups  (the so-called “fully targeted” policy) or imposing two separate shielding policies, one for the senior group and the other for the young and the middle-aged (so-called “semi-targeted” shielding). In their baseline results, semi-targeted policies are associated with substantial efficiency gains that cannot be improved substantially by fully targeted policies.

While \cite{Acemoglu2020} analyze the optimal policy for the U.S., we extend their framework and calibrate it to Germany. Our baseline model is a SEIR model that incorporates contact patterns as estimated by \cite{klepac2020}, who evaluate data from the BBC pandemic project in 2017 and 2018. Moreover, we consider a broader set of policy measures, such as testing and contact tracing, as well as various forms of group distancing.

\section{Multi-Group SEIR Model} \label{SEIRmodel}

\begin{figure}
\centering
\includegraphics[scale=0.4]{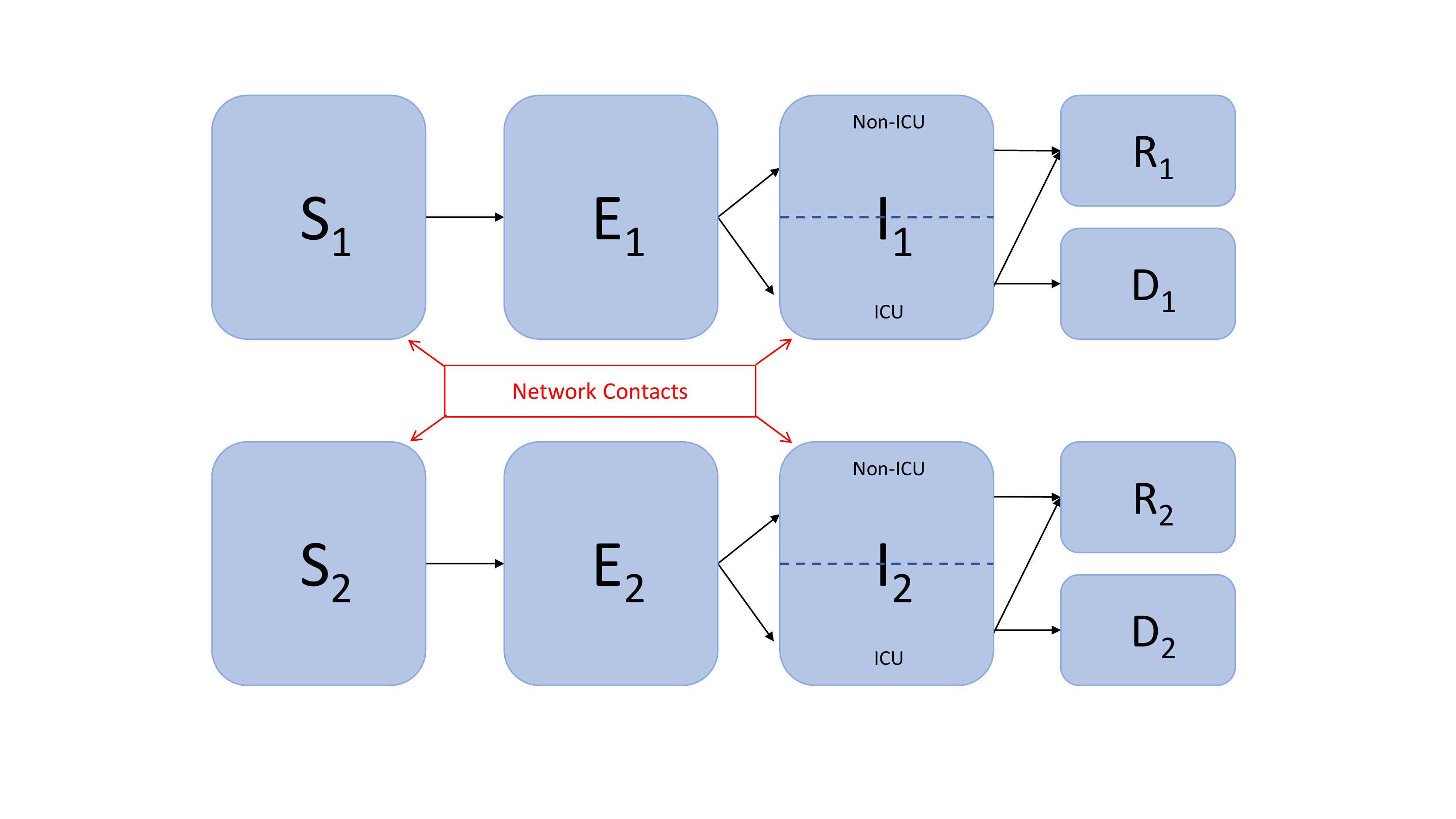}
\caption{Compartments in a two-group SEIR model. The red arrows illustrate the potential channels of infections through physical contacts.}
\label{compartments}
\end{figure}

In this section, we briefly describe a SEIR model based on \cite{Acemoglu2020}, who focus in their analysis on the SIR model and state that their conclusions also hold for the SEIR version. For an in-depth discussion with additional information on the theoretical set up of the original SIR model, we refer to \cite{Acemoglu2020}. One of the major features of the framework is that it allows the population to be partitioned into subgroups that are heterogeneous in terms of their productivity and mortality rates. In particular, we consider the following three subgroups: young (20-49 years), middle-aged (50-64 years) and senior citizens (65+ years). Accordingly, there are age-group specific compartments for susceptible ($S_j$),  infectious ($I_j$),  recovered ($R_j$) and deceased ($D_j$) persons, with $j=y,m,s$ referring to the young, middle-aged and senior groups. The epidemiological SEIR model extends the SIR model by the compartment of exposed individuals $-$ that is, those who have been infected by the virus but whose infection is not yet sufficiently severe that they have symptoms or are infectious. Hence, the model considered in the following incorporates a compartment $E_j$ for each age group in addition to compartments  $S_j$, $I_j$, and $R_j$ at each point in time $t \in [0, \infty)$. 
\[
S_j(t) + E_j(t) + I_j(t)+ R_j(t) + D_j(t) = N_j.
\]
$N_j$ is the number of initial members in each group, $j=y,m,s$. The compartment structure of a two-group SEIR model is illustrated in Figure \ref{compartments} with the red arrows indicating the paths of transmissions through contacts of infectious and susceptible.

Without any policy intervention that enforces shielding of the population or isolation of those who are infected, the (gross) number of new infections in the segment of exposed ($E_j$) and infectious ($I_j$) is governed by the following equations
\begin{align}
\text{New exposed in group $j$} &= M_j(S, E, I, R; \alpha) \cdot \beta \cdot S_j \cdot \sum_k \rho_{jk}I_k \label{newE} \\ \nonumber \\ 
\text{New infected in group $j$} &= \gamma_j^I \cdot E_j, \label{newI}
\end{align}
where $\{\rho_{jk}\}$ are parameters for the contact rate between group $j$ and $k$ and $M_j(\cdot)$ refers to a matching technology, with $M_j(\cdot)=1$ if $\alpha=2$ which is our baseline case. The parameter $\beta$ denotes the transmission rate from contacts between individuals in $I_j$ and $S_j$ and $\gamma^E_j$ is the exit rate from the latent state to the infectious state.
	
\subsection{Model Assumptions}  \label{assump}
	
In this section we describe and discuss the model assumptions.
\subsubsection*{Infection, ICU, Fatality and Recovery}
In the SEIR model described above, a transmission of SARS-CoV-2 arises through contact of susceptible individuals with infectious individuals. After an average latent period $\frac{1}{\gamma_j^E}$, they become infectious themselves. Individuals in compartment $I_j$ may require ICU care. We assume for simplicity that a need for ICU is apparent immediately after entering state $I_j$. ICU patients either recover with Poisson rate $\delta_j^r$ or die at Poisson rate $\delta_j^d$. Non-ICU patients will always recover at Poisson rate $\gamma^I_j$. The death rate can vary with total ICU needs relative to capacity. We assume that
	\[ \gamma_j = \delta^d_j(t)  + \delta^r_j(t). \]
	This means that the proportions of ICU and non-ICU patients among the infected do not change over time in group $j$.  $H_j(t)$ denotes the number of individuals needing ICU care at time $t$ in group $j$, so that $H_j(t)= \iota_j I_j(t)$. $H(t)=\sum_j H_j(t)$ is the total need for ICU. The probability of death is a non-decreasing function of the number of patients, such that the probability of death will rise if the capacity is exceeded:
	\[ \delta_j^d(t)=\psi_j (H(t)), \]
	for a given non-decreasing function $\psi_j$.

\subsubsection*{Testing, Contact Tracing and Isolation}
   Detection and isolation of infected individuals is not perfect. In the SIR model, \cite{Acemoglu2020} denote the probability that an individual in compartment $I_j$ is not detected and put in isolation by $\eta_j$. In their analysis, comparative statics are performed to illustrate the consequences of variation of $\eta_j$, for example due to intensified testing. Incorporating the group of exposed ($E_j$) in a SEIR model allows tests to be performed for those who have had contact with an infected person. 
 This setting could be considered a simplified form of contact tracing, for instance enabled by a smartphone application that records physical contacts. Hence, quarantining those who have been in contact with infected individuals might enable policy makers to exclude these infected but not yet infectious individuals from social interactions.  Accordingly, we denote the probability that a person in compartment $E_j$ or $I_j$ is not detected and isolated by $\eta_j^E$ and $\eta_j^I$, respectively, and thereby avoid including additional state variables. In this manner, we can model the fact that only those infected who have not been detected and isolated in stage $E_j$ or $I_j$ contribute to the spread of the disease via their contacts. 

\subsubsection*{Shielding and Physical Distancing}
	Shielding policies describe all measures that reduce the rate of transmission of infections in social and business life and physical distancing. The productivity of members of $j$ is $w_j$ without shielding and $\xi_j w_j$ with shielding, with $\xi_j \in [0,1]$. $L_j(t)=1$ refers to a full shielding policy and $L_j(t)=0$ to a situation without any restrictions to social interactions. $L_j(t) \in (0,1)$ would be partial shielding, for example by shielding a (potentially randomly and independently drawn) fraction of the population. It is assumed that shielding cannot be perfectly enforced and that, with shielding, the effective reduction in social interaction is only $1-\theta_jL_j(t)$ with $\theta_j<1$.	

\subsubsection*{Contact Rates}
We implement a version of the SEIR model that incorporates social interaction patterns to capture the major findings in \cite{klepac2020} $–$ that is, high rates of interaction within and by the group of young and decreasing intensity of interactions with age. The study evaluates large-scale data on the frequency and intensity of social interactions that were collected in the BBC Pandemic project in the UK in 2017 and 2018 and make it possible to derive age-specific contact rates. To model the group interaction within and between groups, let denote $\rho^{0}_{jk}$ the elements of the contact matrix
\begin{align*}
\rho^{0} = \left( \begin{array}{rrr}
1.0 & 0.5 & 0.4 \\
0.5 & 0.6 & 0.4 \\
0.4 & 0.4 & 0.5 \\
\end{array}\right),
\end{align*}
with the first row and column referring to the young group, the middle row and column referring to the middle-aged group and the third row and column referring to the senior citizen group.\footnote{An example: The entry of a contact matrix $\rho_{23}$ represents the contact rate that applies to interactions of members of the middle-aged and the senior age group, i.e., $\rho_{ms}$. Due to symmetry of the matrix, it holds that $\rho_{23}=\rho_{32}$.}
The contact estimates of \cite{klepac2020} refer to a pre-pandemic setting and, hence, constitute the benchmark scenario for comparison to social distancing policies. To incorporate voluntary reductions of physical contacts, we base our baseline results in Section \ref{results} on a rescaled contact matrix that presumes a 25\% reduction in physical contacts. 
\begin{align*}
\rho = 0.75 \cdot \rho^{0} = \left( \begin{array}{rrr}
0.750 & 0.375 & 0.300 \\
0.375 & 0.450 & 0.300 \\
0.300 & 0.300 & 0.375 \\
\end{array}\right),
\end{align*}
Incorporating more realistic contact patterns in the SEIR model with multiple groups is important for evaluating policy measures that are targeted at different age groups. For example, lower rates of contact between the vulnerable group (i.e., senior citizens) and younger people might allow for less intense shielding patterns. 

\subsubsection*{Physical Distancing, Face Masks and Additional Hygiene Measures}
Various mandatory or voluntary policies can be employed to reduce the transmission rate of SARS-CoV-2. These measures range from a general reduction in face-to-face or physical contacts (for example, by imposing strict physical distancing measures that apply equally to all age groups) or specific interventions that aim to protect especially those who are most vulnerable. The latter include, for example, a reduction in face-to-face contacts with senior citizens – for instance by placing restrictions  on visits to nursing homes or prescribing mandatory (reusable or disposable) face masks during for contacts with senior citizens. For example, \cite{chu2020} undertook a systematic review and meta-analysis of studies that examined the effectiveness of face masks and physical distancing for COVID-19 and related diseases (e.g., MERS and SARS). Accordingly lower transmission rates are associated with greater physical distance and the use of N95 face masks and comparable respirators rather than disposable surgical masks. There are a huge number of potential policy measures that aim to reduce the transmission of SARS-CoV-2, all of which can be employed in combination. We list a few examples of such measures in Section \ref{recomm}. Something that all of these measures have in common is that they effectively change or rescale the elements in the contact matrix $\rho$. In our analysis, we focus mainly on two variants of group distancing, namely (i) so-called uniform group distancing, which effectively reduces the contact rates in $\rho$ for all groups (corresponding to a multiplication of the matrix (corresponding to a multiplication of the matrix $\rho$ with a scalar $\nu$), and (ii) group distancing policies with a focus on the vulnerable that refer only to interactions with the group of seniors and the elements $\rho_{sj}$ with $j=y,m,s$, and $\rho_{js}$, respectively. Moreover, it is possible to simulate settings in which the level of interactions within the senior group might be left unchanged, thus reducing the impact on daily interactions with others at the same age.

\subsubsection*{Improved Conditions for Working from Home}
Working from home can be an effective way to reduce the costs of the pandemic and of shielding policies. To host a scenario with improved conditions for working from home, we (i) implement a parameter constellation with respect to the contact rates within and between the young and middle-aged group and between these groups and the senior group and (ii) decrease the productivity loss associated with working from home, $\xi_j$.We believe that this captures some aspects of working from home in that those who are most likely to be employed can reduce their social interactions with lower economic losses. Changes in terms of (i) are imposed by scaling the entries of the contact matrix $\rho_{yy}, \rho_{mm}, \rho_{ym}$ by a factor $\pi_1$ and a scaling the contact rates $\rho_{ys}$ and $\rho_{ms}$ by $\pi_2$ with $\pi_1<\pi_2$.

\subsubsection*{Vaccine and Cure}
 \cite{Acemoglu2020} assume that a vaccine and an effective drug for all infected individuals becomes available at some date $T$ and that full immunity is achieved and maintained after an infection.\footnote{In line with \cite{Acemoglu2020} we focus on the case with deterministic arrival of a vaccine.} In our analysis, we will evaluate changes in $T$ resulting from a faster development of a vaccine - for example after one year or six months. 

Currently, there are various treatments for COVID-19 that have been approved or are being evaluated in clinical trials. We assess the implications of a medical treatment with respect to the optimal shielding policy. Put simply, a new treatment could have any of the following three effects: (i) reduce the length of hospitalization, (ii) reduce the probability of dying from COVID-19, (iii) reduce the probability that an infection with SARS-CoV-2 becomes severe. We will focus on the availability of a treatment that leads to a reduction in mortality from COVID-19 for the group of senior citizens because most deaths and severe cases have been observed in this age group (e.g., as reported for Germany in \cite{lageRKI}). 

\subsection{Dynamics in the MG-SEIR Model}

If vaccine and cure are unavailable, the number of individuals in the exposed compartment for group $j$ evolves according to the differential equations for all $t \in (0,T)$
\[ 
\dot{E}_j =  M_j(S,E,I,R,L; \alpha) \beta (1- \theta_j L_j) S_j
 \sum_k \rho_{jk} \eta^E_k \eta^I_k (1 - \theta_k L_k) I_k - \gamma^{E}_j E_j,\]
for nonnegative $\beta$ and contact coefficients $\rho_{jk}$ and where
\[
M_j(S,E,I,R,L; \alpha) \equiv \left( \sum_k \rho_{jk} \left[ (S_k + \eta_k^E E_k + \eta_k^E \eta^I_k I_k + (1-\kappa_k)R_k)(1-\theta_jL_k) + \kappa_k R_k \right] \right)^{\alpha-2}.
\]
In the quadratic case $M_j(S,E,I,R,L)=1$. The parameter $\kappa_j$ refers to the share of recovered individuals that can return to work and social life while being exempted from shielding policies due to immunity.\footnote{We acknowledge that there is not yet a consensus on whether individuals become immune to SARS-CoV-2 after an infection and whether such immunity, if achieved, is maintained for a substantial period. The empirical evidence on both points is mixed. We follow the baseline setting in  \cite{Acemoglu2020} with $\kappa_j=1$ for all $j$ and repeat the robustness checks with setting $\kappa_j=0$ for all $j$. The main conclusion remains valid and results are omitted for the sake of brevity.} Setting $\eta_j^E=1$ for all $j$ refers to a setting where it is not possible to test and isolate exposed individuals. However, a value $\eta_j^E<1$ means that the effective number of individuals who contribute to further spread of the disease can be reduced by contact tracing and isolating those who have been exposed.

The rest of the laws of motion for $t \in (0,T)$ are
\begin{eqnarray}
\dot{S}_j &=& - \dot{E}_j - \gamma^{E}_j E_j,\\ 
\dot{I}_j &=& \gamma_j^{E} E_k - \gamma_j^{I}I_j, \\
\dot{D}_j &=& \delta_j^d H_j,\\
\dot{R}_j &=& \delta^r_j H_j + \gamma^{I}_j(I_j-H_j),
\end{eqnarray}
where $H_j=\iota_j I_j$ denotes the number of ICU patients in group $j$. 
After discovery of a vaccine and cure at $T$, every individual alive is in the recovered category.

\subsection{Efficient Frontier}
The government can control the degree of shielding $L_j(t)$ for each group $j$ at any point in time $t \in [0,T)$. In particular we will compare uniform policies (i.e., blanket policies with $L_j(t)=L(t)$) and group-specific (i.e., targeted) policies. The goal of the social planner is to minimize the overall costs of the pandemic, which consist of two parts:
\begin{enumerate}
\item Lives Lost $=\sum_j D_j(T).$
\item Economic Losses $=\int_0^T \sum_j \Psi_j(t)dt.$
\end{enumerate}

The economic losses for group $j$ are given by
\begin{eqnarray}
\Psi_j(t) &=& (1-\xi_j) w_j S_j(t) L_j(t) +  \label{econloss} \\
	&+& (1-\xi_j)w_jE_j(t)(1-\eta^E_k (1-L_j(t))) + \nonumber\\
  &+& (1-\xi_j)w_jI_j(t)(1-\eta^E_k \eta^I_k(1-L_j(t))) +  \nonumber \\
	&+& (1-\xi_j)w_j(1-\kappa_j)R_j(t) L_j(t) + \nonumber \\ 
	&+& w_j \Delta_j \iota_j \delta_j^d(t) I_j(t), \nonumber
\end{eqnarray}
where the second term refers to the income loss of exposed individuals under shielding. The third term in the economic cost function is now adjusted to the case with the testing and isolation of exposed individuals, as well. $\Delta_j$ captures the present discounted value of a group $j$ member’s remaining employment time until retirement, which is lost due to death.
The objective function is a weighted sum of both losses with weight factor $\chi$ and the task is to choose a shielding policy which minimizes
\[
\int_0^T \sum_j \Psi_j(t) dt + \chi \sum_j D_j(T). 
\]
Varying the values for $\chi$ makes it possible to identify the efficient frontier - in other words, to find the policy that minimizes the objective function for a given $\chi$. Hence, the policy recommendations that can be obtained from an analysis of the efficient frontiers do not depend heavily on a specific choice of $\chi$ but rather reflect the difficult trade-off that policy makers face in the pandemic \citep{Acemoglu2020}.

\section{Specification and Calibration} \label{calibration}
Before we discuss optimal shielding policies in the multi-group SEIR model, we will first comment on how we set and calibrated the parameters for Germany. We will present adaptations of country-specific parameters that would also apply to a calibration of the initial multi-group SIR model in \cite{Acemoglu2020}. These parameters refer to demographic and economic conditions, as well as to characteristics of health care provision in Germany. Second, we will discuss the adaptations of the SIR model parameters to a SEIR version based on information from the Robert Koch Institute (RKI) as of July 2020. Finally, we will comment on the ation of the basic reproduction number $R_0$.

\subsection{Country-specific Parameters}

\subsubsection*{Calibration of Socio-Demographic and Economic Parameters}

\begin{table}
\centering
\begin{tabular}{l r r}
\hline
 Parameter & US & GER \\ \hline
 \multicolumn{3}{c}{\textit{1. Socio-demographic and economic}} \\ 
& & \\ 
 Population shares & \{0.53, 0.26, 0.21\} & \{0.46, 0.28, 0.26\} \\
 \{$N_y$, $N_m$, $N_o$\} & & \\
 Per capita income & \{1.00, 1.00, 0.26\} & \{1.00, 1.00, 0.085\} \\
 \{$\omega_y$, $\omega_m$, $\omega_o$\} & & \\
 Remaining years in empl.  & \{32.50, 10.00, 2.50\} & \{32.43, 10.44, 2.50\}\\
\{$\Delta_y$, $\Delta_m$, $\Delta_o$\}& &   \\ \hline  
 \multicolumn{3}{c}{\textit{2. Health-care related}} \\ & & \\ 
 Mortality penalty & 1.00 & \{0.20, 0.40, $\underline{0.60}$, 0.80\} \\
 $\lambda$ & & \\
 ICU constraint & \{0.016, 0.020\} & \{0.020, 0.030, 0.040\} \\
$\bar{H}(t)$ & & \\ \hline
\end{tabular}
\caption{\centering Parameters for the United States and Germany. The underlined mortality parameter indicates the choice in the baseline setting. The remaining values for $\lambda$ and $\bar{H}(t)$ are used in robustness checks.}
\label{params}
\end{table}

Germany has a demographic com- position that is substantially different from that of the U.S. In particular, the share of the group aged 65 and older is larger and that of the young group is smaller than in the U.S. For example, the median age in the U.S. is around 38 \citep{census2019} years whereas it is around 45 years in Germany  \citep{Destatis2020}. Using data from German micro census from 2018 as provided by the German Federal Statistical Office \cite{mikrozensus2018} and \cite{mikrozensus2020}, we calculated the remaining lifetime earnings as displayed in Table \ref{params} assuming retirement at age 67. 


An interesting difference that we observed in the comparison of Germany and the U.S. is the distinct employment patterns in the group aged 65 and above. Whereas approximately 20\% of individuals in this group are still employed in the U.S., the corresponding share for Germany amounts only to around 7\%, leading to the re-weighted per-capita earnings in Table \ref{params}. In both countries, the median earnings are relatively similar for those who are employed in the middle-aged group and the senior groups.

The demographic distribution of the population in Germany implies that the share of persons who have a higher risk of dying from COVID-19 is relatively large. Thus, uniform shielding policies that aim to keep mortality in the entire population at a low level are expected to be more costly in terms of economic damage. At the same time, the group of senior citizens accounts for a relatively low share of GDP, implying that targeted policies are more favorable. Shielding targeted only towards the elderly therefore makes it possible to reduce overall mortality while allowing the younger and economically more productive groups to continue working. 

\subsubsection*{Calibration of Health and Medical Variables}

Calibrating the model in terms of parameters that are related to health care provision is challenging - for example due to limited comparability of hospital capacities and their dynamic expansion in reaction to the pandemic \citep{OECD2020}. We performed various variations to parameters of the original SIR model of \cite{Acemoglu2020} and its SEIR version which are provided, in part, in the Appendix and chose one of these parameter configurations as a baseline setting in our analysis as described in the following. 

\begin{table}
\begin{tabular}{l r r}
\hline & US & GER \\ \hline 
Acute Care beds/10,000 pop. & 24.00 & 60.00 \\
(OECD, 2020) & & \\
ICU beds/10,000 pop. & 2.58 & 3.39 \\
(OECD, 2020) & & \\
ICU beds/10,000 pop. & 3.13 & 3.89 \\
(AHA, 2020, DIVI, 2020) & & \\
\hline
\end{tabular}
\caption{Hospital beds and ICU beds. Source: \cite{OECD2020}, \cite{AHA2020}, \cite{DIVI2020}. \cite{AHA2020} refers to adult ICU beds and population only.}
\label{hosp}
\end{table}

Health care provision in Germany is considerably different from that in the U.S. In a recent report, the Organisation for Economic Cooperation and Development (OECD)  \citep{OECD2020} compares health care provision across different countries. We list the numbers of hospital and ICU beds for the U.S. and Germany in Table \ref{hosp}. Due to the dynamic expansion of hospital capacities during the COVID-19 pandemic in both countries, we add more recent, constantly updated data from the \href{https://www.aha.org/statistics/fast-facts-us-hospitals}{American Hospital Association} (AHA)\citep{AHA2020} and the German Interdisciplinary Association for Intensive and Emergency Medicine (DIVI) \citep{DIVI2020}. Compared to the OECD data, the number of ICU beds reported by AHA and DIVI has increased by around $21\%$ in the U.S. and around $15\%$ in Germany. The number of ICU beds is frequently reported to be one of the crucial measures of whether countries are able to keep mortality from COVID-19 low. According to the report by \cite{OECD2020}, Germany is the country with the highest ICU capacities among all OECD members. Germany not only has more ICU beds per capita than the U.S.; other measures, such as the number of hospital beds or coverage with public health insurance \citep{OECD2020}, suggest that the health care system in Germany has comparably greater capacities (per capita) than that of the U.S.

To take account of these differences, we adjust the parameter $\lambda$, which enters the relationship of the daily mortality rate, $\delta^{d}_j$, and hospital capacities at time $t$, $H(t)$, to a default value $\lambda = 0.6$, which is smaller than $\lambda =1$ as chosen in the analysis of \cite{Acemoglu2020}.
\begin{align}
\label{mort}
\delta^{d}_j = \underline{\delta}_j^d \cdot \left[ 1 + \lambda \cdot H(t) \right],
\end{align}
where $\bar{\delta}_j^d$ is the baseline mortality rate for group $j$ with $\delta_y=0.001$, $\delta_m = 0.01$ and $\delta_s=0.06$.\footnote{These mortality rates are based on \cite{Ferguson2020}. We repeat the robustness check performed in \cite{Acemoglu2020} with $\delta_s=0.12$ and confirm that the main conclusions remain unchanged. We omit the resulting plot for the sake of brevity because we provide two robustness checks with regard to a \textit{lower} mortality rate in Section \ref{optimalpolicies}.} We refer to Figure \ref{lambdavar} in the Appendix for illustrations of the variation of health-provision-related parameters. An alternative to specifying the parameter $\lambda$ would be to impose a hard ICU constraint by enforcing $H(t)<\bar{H}(t)$ as was done in the original study of \cite{Acemoglu2020}. This would reflect more generous capacities than in the U.S. 

Allowing for a less sensitive relationship between mortality and ICU needs (i.e., lowering the value of $\lambda$ in \ref{mort}) reflects the policy maker being able to achieve lower mortality rates at a given (possibly high) number of infections. Similarly, a higher bound on available ICU beds (i.e., increasing $\bar{H}(t)$) implies that the policy maker faces a trade-off between mortality and economic damage under relaxed capacity constraints.  We performed several variations with respect to the health-related parameters and refer to some examples illustrated in Figure  \ref{lambdavar} in the Appendix. These changes can all be summarized generally as restrictions to the set of possible options that are available to policy makers in situations in which infection rates are high.

\subsection{Parameters for the SEIR model}

The multi-Group SEIR model is able to adapt to some    characteristics specific to SARS-CoV-2. For example, one characteristic of the virus is that infections are frequently caused by some exposure to infectious persons through personal contact. To model the period spent in states $E$ (i.e., carrying the virus but not exhibiting symptoms and neither being infectious) and $I$ (i.e., potentially exhibiting symptoms and being infectious), we base the rates $\gamma_j^E$ and $\gamma^I_j$ on the conclusions of the Robert Koch Institute as provided in the RKI COVID-19 report \cite{RKI2020}. In our analysis, we assume that the latent period is 6 days ($\gamma^E_j = \frac{1}{6}$) and the infectious period is 9 days ($\gamma^I_j = \frac{1}{9}$).

\subsection{Calibrating $R_0$}

The parameter $\beta$ has been calibrated to match a basic reproduction number $R_0=2.4$ under the parameter constellation as described above. Setting $R_0=2.4$ corresponds to the lower bound on $R_0$ as reported by the RKI \cite{RKI2020} as of July 2020. The calibration is performed in a setting without any policy intervention based on the contact matrix $\rho_0$, i.e., no shielding, i.e., $L_j=0$, no testing and isolation, i.e., $\eta^I_j=1$, and no contact tracing, i.e., $\eta^E_j=1$, is imposed for any $j$ in an almost entirely susceptible population.

\section{Results and Optimal Policies} \label{results}

In this section, we present our results and discuss the optimal policies that can be derived from our model. We will first refer to the efficient frontier according to the German parametrization and then illustrate the effectiveness of various policies. Lastly, we will comment on the implementation of these policies in practice.

\subsection{Efficient Frontier} \label{frontierres}

\begin{figure}
\includegraphics[scale=1]{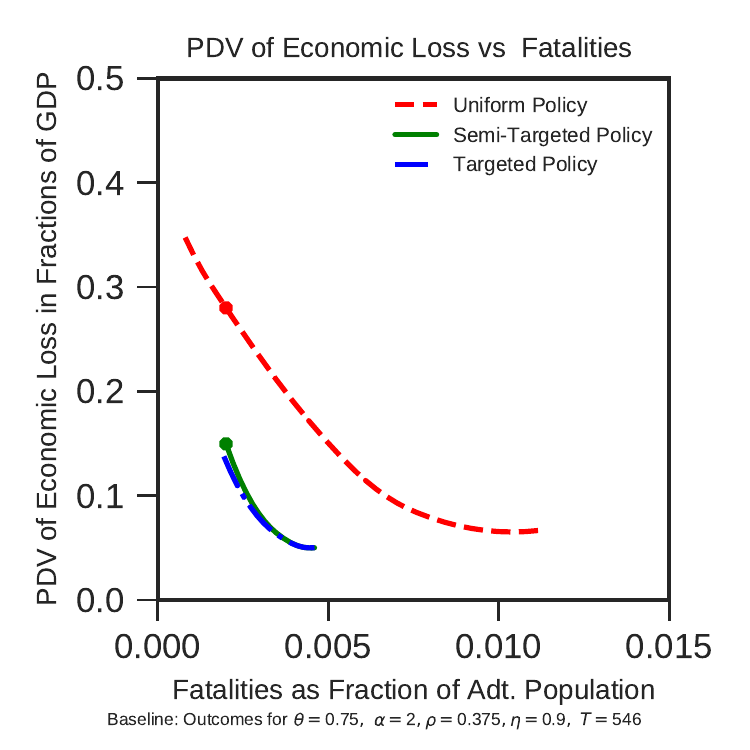} 
\caption{Efficient frontier in SEIR model with adjusted contact matrix $\rho$ and demographic and health care parameters adjusted to Germany.} 
\label{frontecon}
\end{figure}

Adapting the model to the socio-demographic, economic and health-care-related parameters for Germany leads to the baseline policy frontier shown in Figure  \ref{frontecon}. In line with the results reported by \cite{Acemoglu2020}, the economic cost of shielding can be reduced substantially at a given mortality level by employing targeted shielding policies.\footnote{Analogously to \cite{Acemoglu2020}, we will refer to all policies that involve a different degree of shielding $L_j$ for at least one group $j$ generally as \textit{targeted} policies. We follow the distinction of \textit{fully targeted} policies with shielding intensities $L_j$ that are determined separately for all three groups and \textit{semi-targeted} policies that distinguish only one level for the senior group and one level that applies to the young and middle-aged group.} Due to the non-uniform contact rates, fully targeted policies provide improvements over semi-targeted policies, similar to the findings of \cite{Acemoglu2020}. However, in many cases, these improvements are moderate to small. Assuming that the costs of implementing fully targeted policies are non-negligible and are likely to outweigh their gains, we will focus mostly on the comparison of semi-targeted and uniform policies in the following.

\subsection{Optimal Policy} \label{optimalpolicies}

	

%
%
%

%
%

In the following, we shed light on the effectiveness of targeting shielding towards different age groups and the combination of this with additional measures, such as testing activities. Similar to \cite{Acemoglu2020}, we analyze the impact of various policy measures by varying parameters and making comparisons to a baseline setting.
In this benchmark, the parameters are chosen in line with  the German socio-demographic and economic calibration discussed in Section \ref{calibration}. 
When presenting the results, we will frequently refer to a safety-focused scenario entailing policies that do not allow population mortality to exceed 0.2\%. In figures with policy frontiers, the results that correspond to this setting are indicated by a dot on the respective policy frontier line.

The policy measures considered in the following refer to improving testing for those in compartment $I_j$ (referred to as testing) and those in compartment $E$  (referred to as contact tracing), two variants of group distancing, improved conditions for working from home, and a combination of these. In addition, we analyze the implications of a medical treatment that makes it possible to lower the mortality rate for those in ICU treatment, as well as of a vaccine arriving early. We list additional results and robustness checks in the Appendix.

\begin{figure}
\begin{center}
\textit{(i)}
\end{center}
\includegraphics[scale=1]{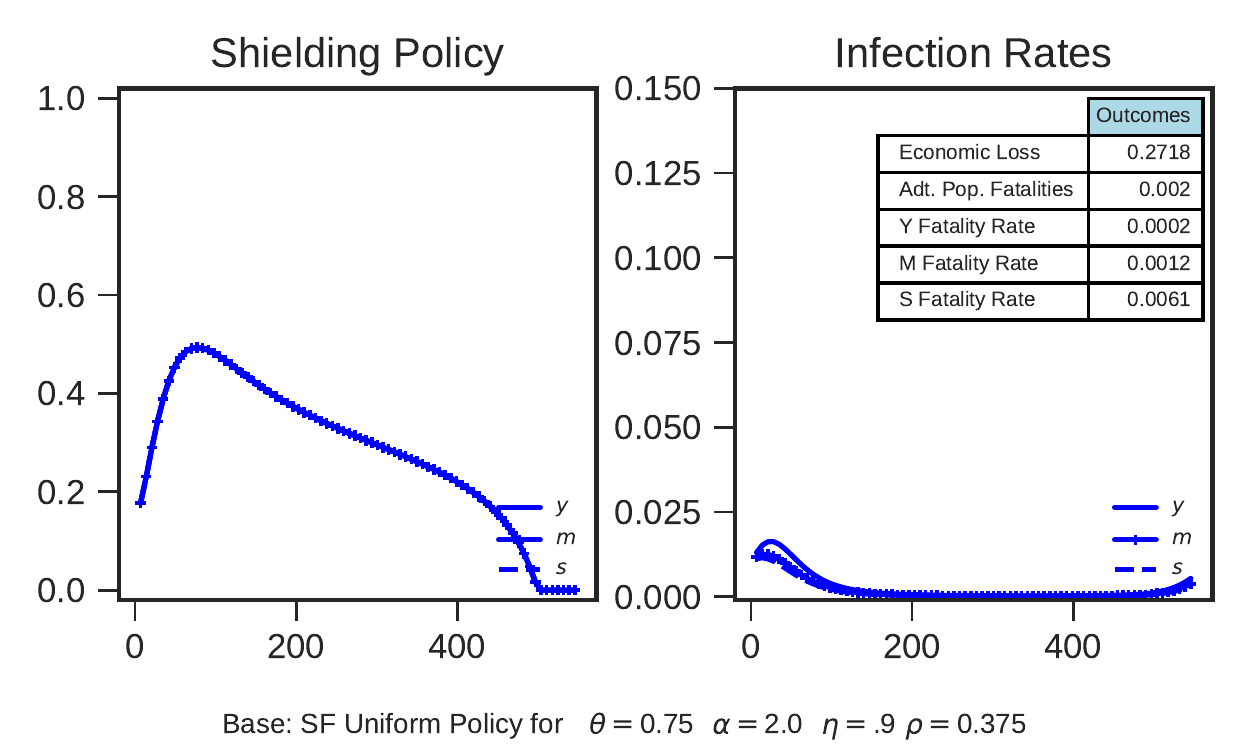}\\
\begin{center}
\textit{(ii)}
\end{center}
\includegraphics[scale=1]{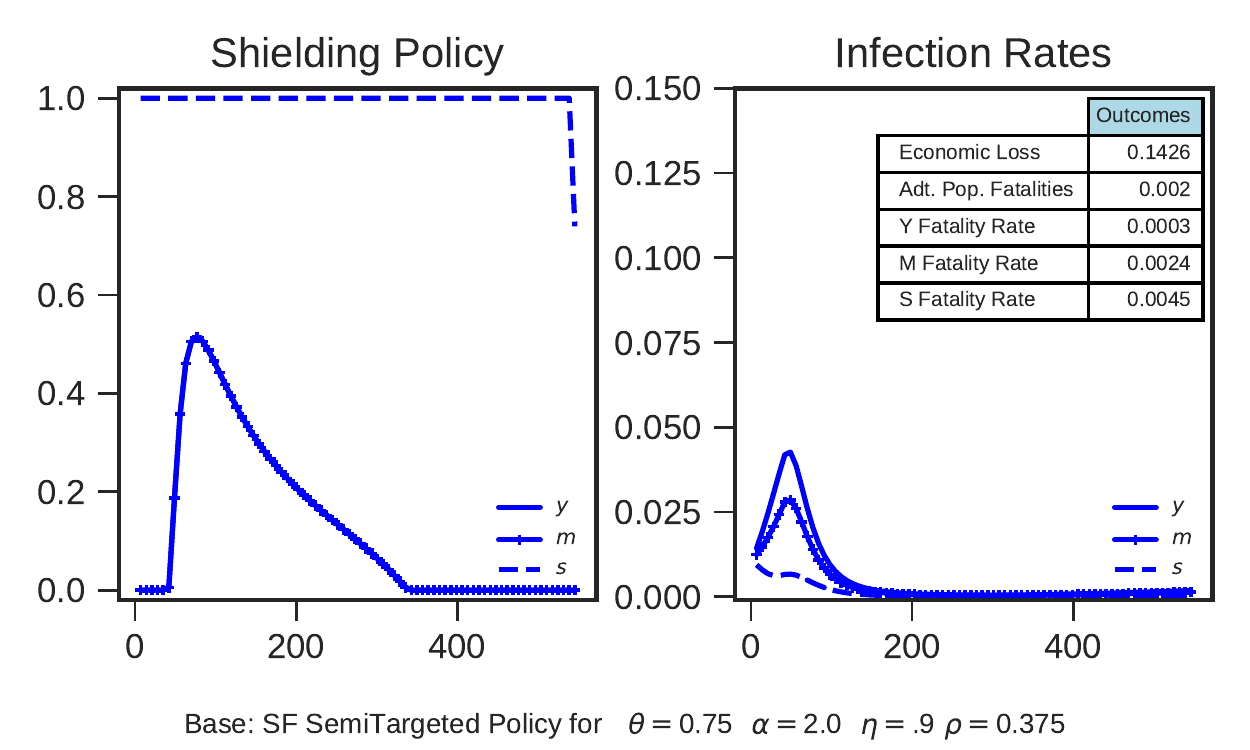}
\caption{Panel ($i$): Optimal uniform shielding policy with safety focus, baseline setting. Panel ($ii$): Optimal semi-targeted shielding policy with safety focus, baseline setting.}
\label{baselinepolicy} 
\end{figure}

\begin{figure}
\begin{center}
\textit{(i)}
\end{center}
\includegraphics[scale=1]{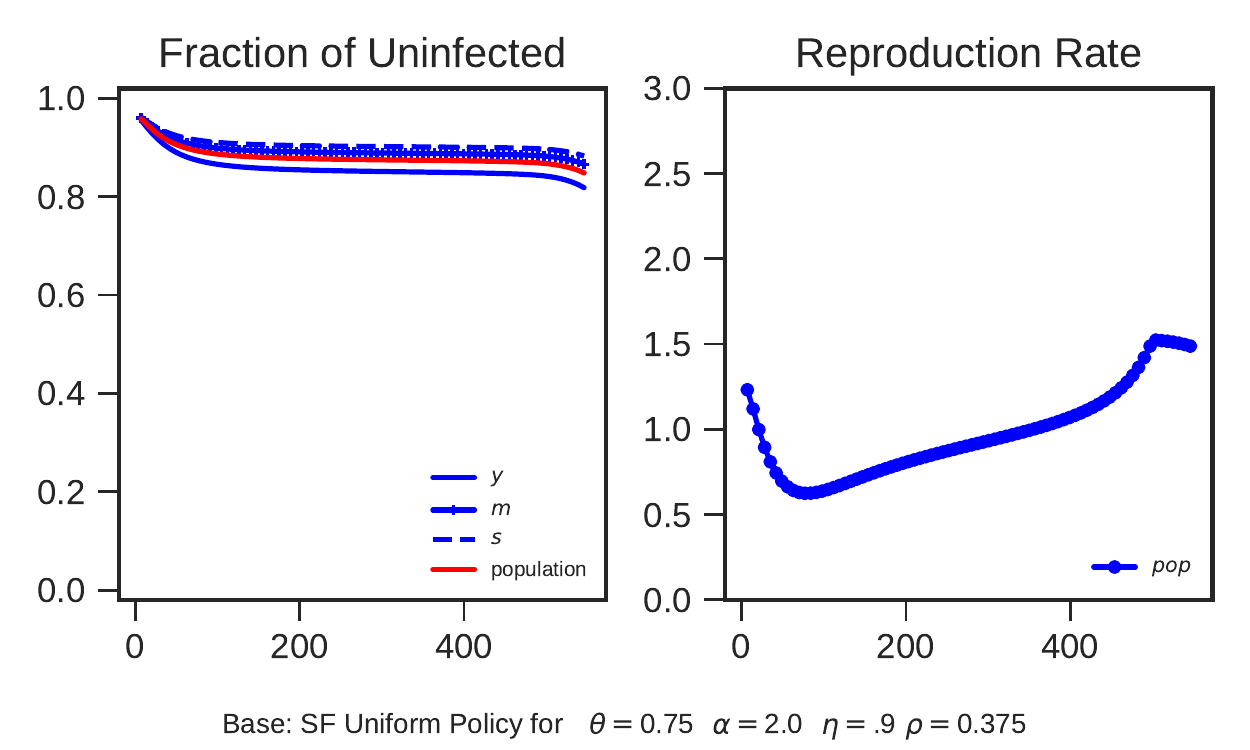}\\
\begin{center}
\textit{(ii)}
\end{center}
\includegraphics[scale=1]{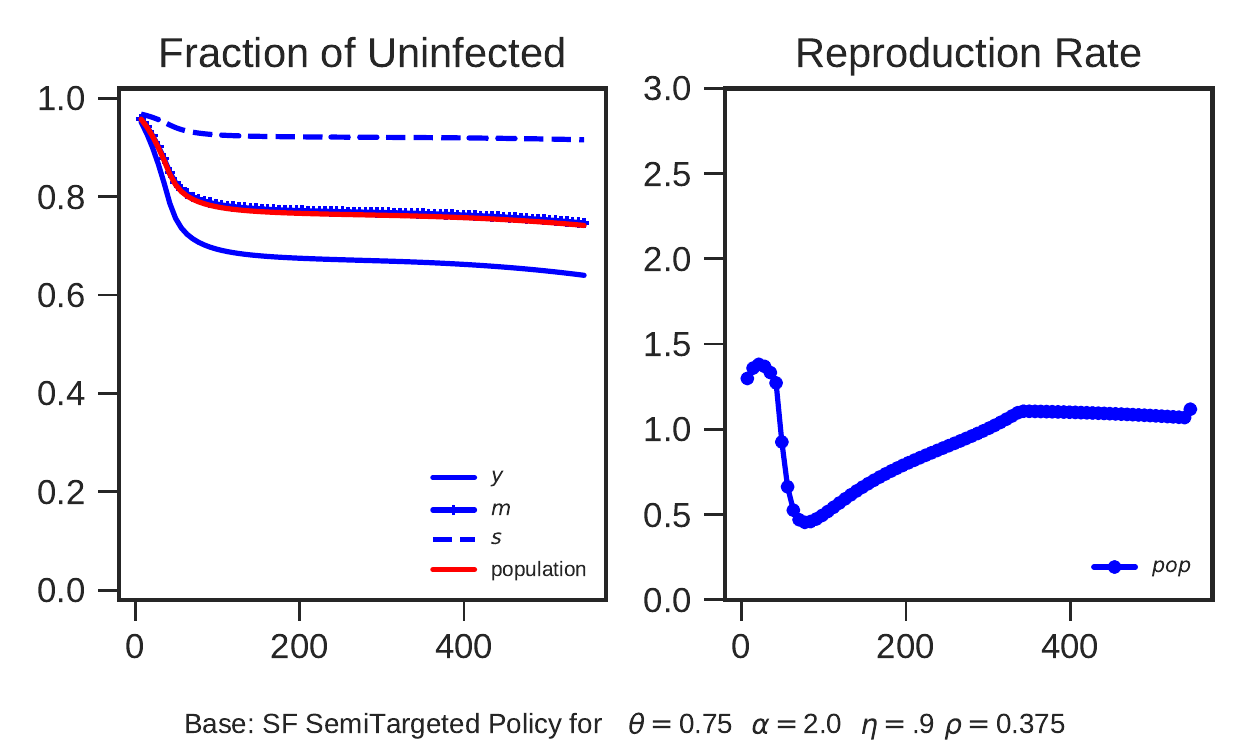}
\caption{Share of uninfected (left) and reproduction rate $R(t)$ (right) in the baseline setting with safety focus. Panel ($i$): Optimal uniform shielding policy. Panel ($ii$): Optimal semi-targeted shielding policy.}
\label{baselineepi} 
\end{figure}

Figure \ref{baselinepolicy} illustrates the optimal policy in the baseline setting with uniform and semi-targeted shielding. The results with regard to the economic loss at the fixed mortality level of 0.2\% illustrate the gains that can be achieved by targeted shielding. In the baseline case with semi-targeted policies, a high shielding intensity is imposed on the elderly until a vaccine arrives, whereas the intensity for the other groups is lowered gradually after an initial peak. Figure \ref{baselineepi} illustrates the evolution of the share of uninfected in each age group and the reproduction rate over time. Semi-targeted shielding policies as illustrated in Panel (ii) are associated with different infection rates across the age groups. Hence, the share of infected in the vulnerable group is relatively low whereas infections are more prevalent in the group of young. However, if uniform policies are considered (Panel ($ii$)) the variation in the share of uninfected across the age groups is much smaller.

\begin{figure}
\begin{center}
\textit{(i)}
\end{center}
\includegraphics[scale=0.5]{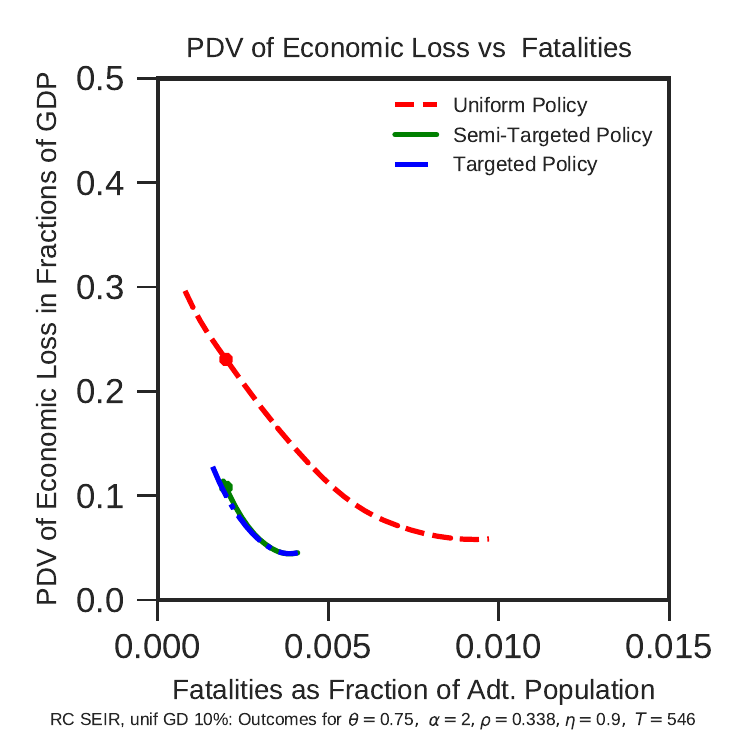}
\includegraphics[scale=0.5]{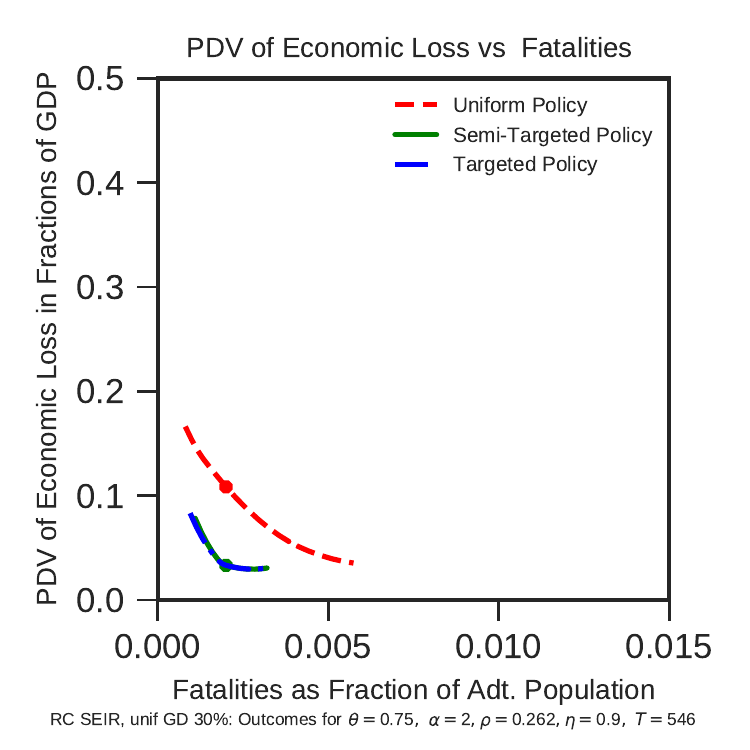}
\includegraphics[scale=0.5]{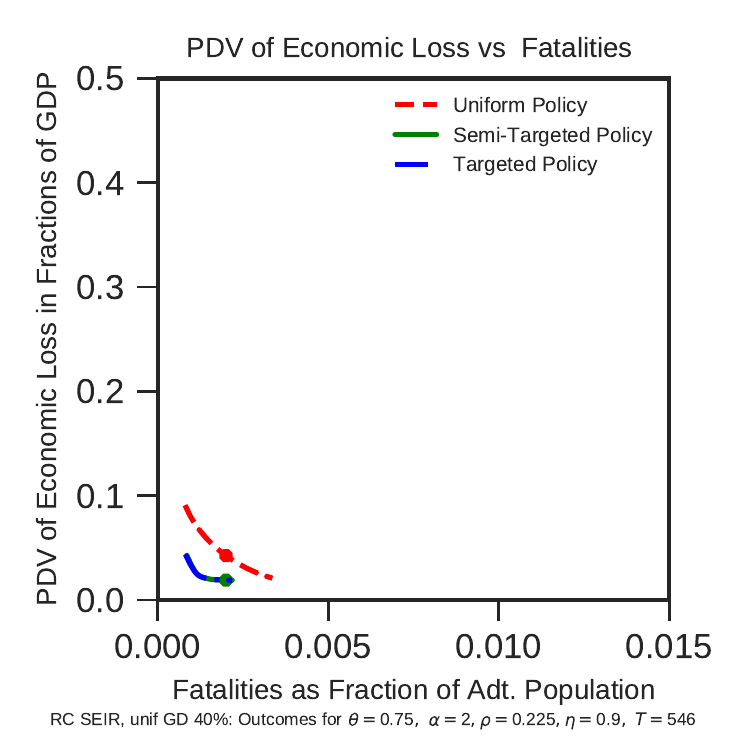}\\
\begin{center}
\textit{(ii)}
\end{center}
\includegraphics[scale=0.5]{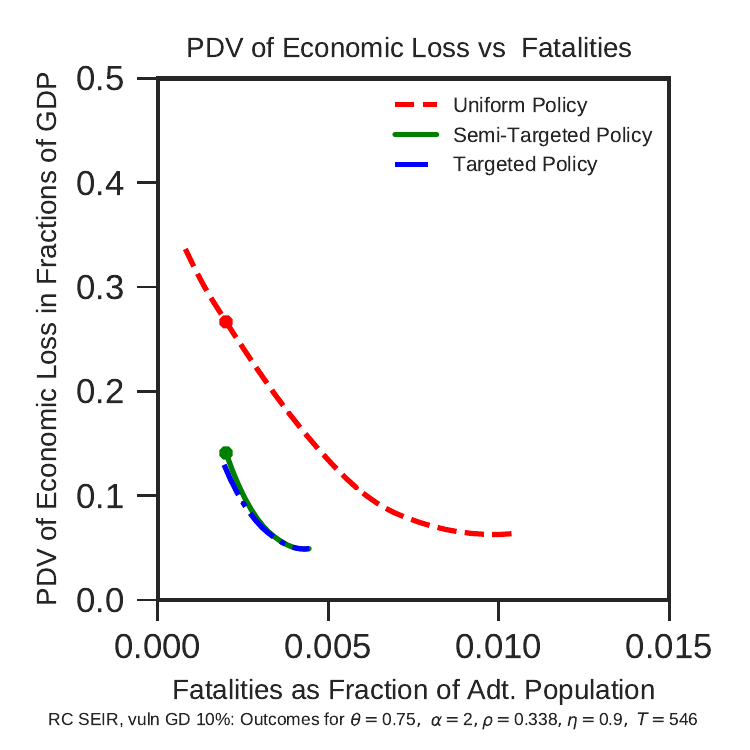}
\includegraphics[scale=0.5]{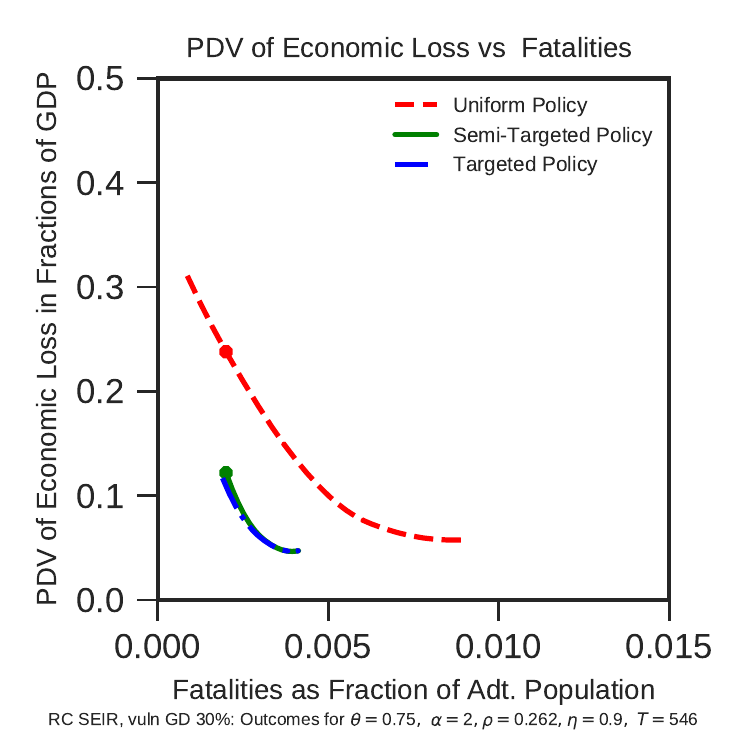}
\includegraphics[scale=0.5]{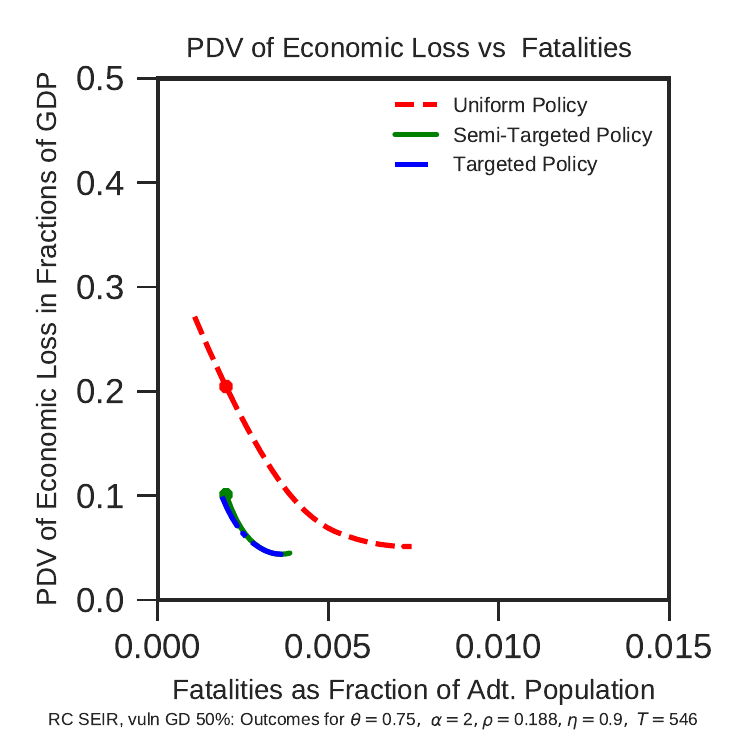}

\caption{Policy frontiers with group distancing or reduced transmission rates between and within groups. Panel ($i$): Uniform reduction in all contact rates in the contact matrix $\rho$ of 10\% (left), 30\% (center) and 40\% (right).  Panel ($ii$): Group distancing focusing on the most vulnerable group (i.e., the group of those aged 65+) with a reduction in the between-group contact rates $\rho_{ys}, \rho_{ms}$ of 10\% (left), 30\% (center) and 50\% (right).}
\label{GDfront}
\end{figure}

\begin{figure}
\begin{center}
\textit{(i)}
\end{center}
\includegraphics[scale=1]{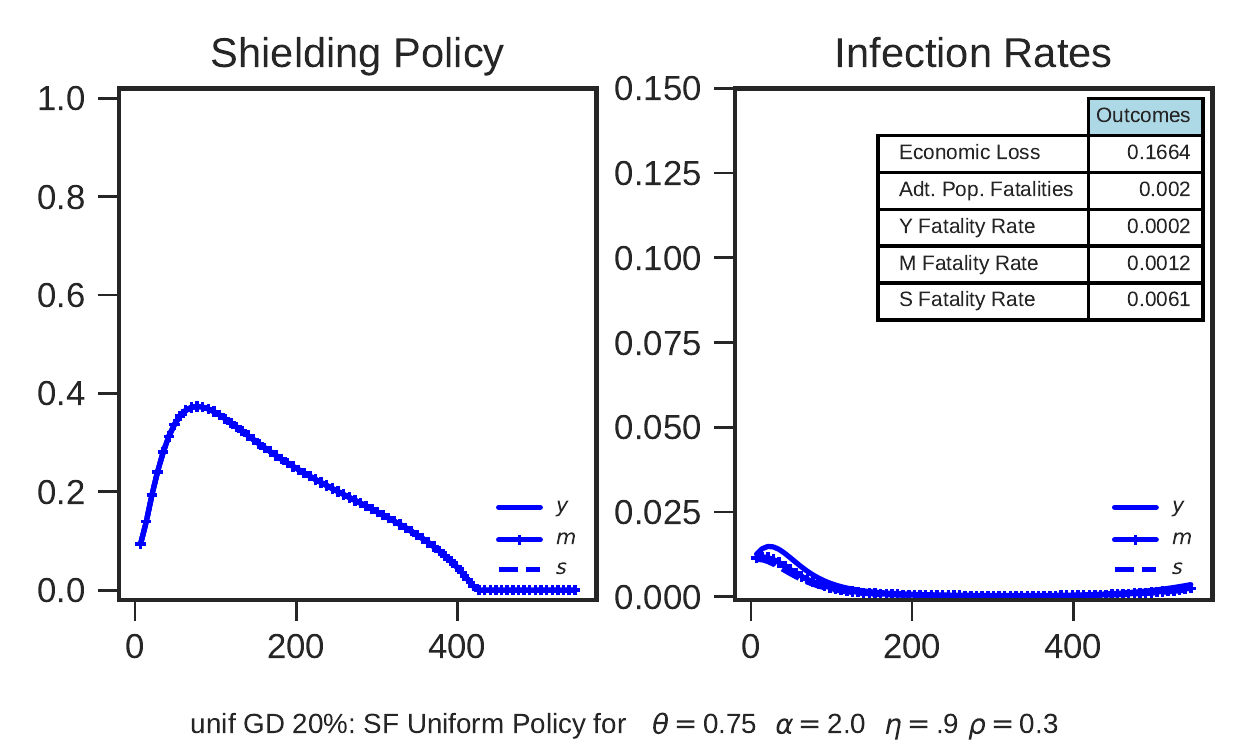}\\
\begin{center}
\textit{(ii)}
\end{center}
\includegraphics[scale=1]{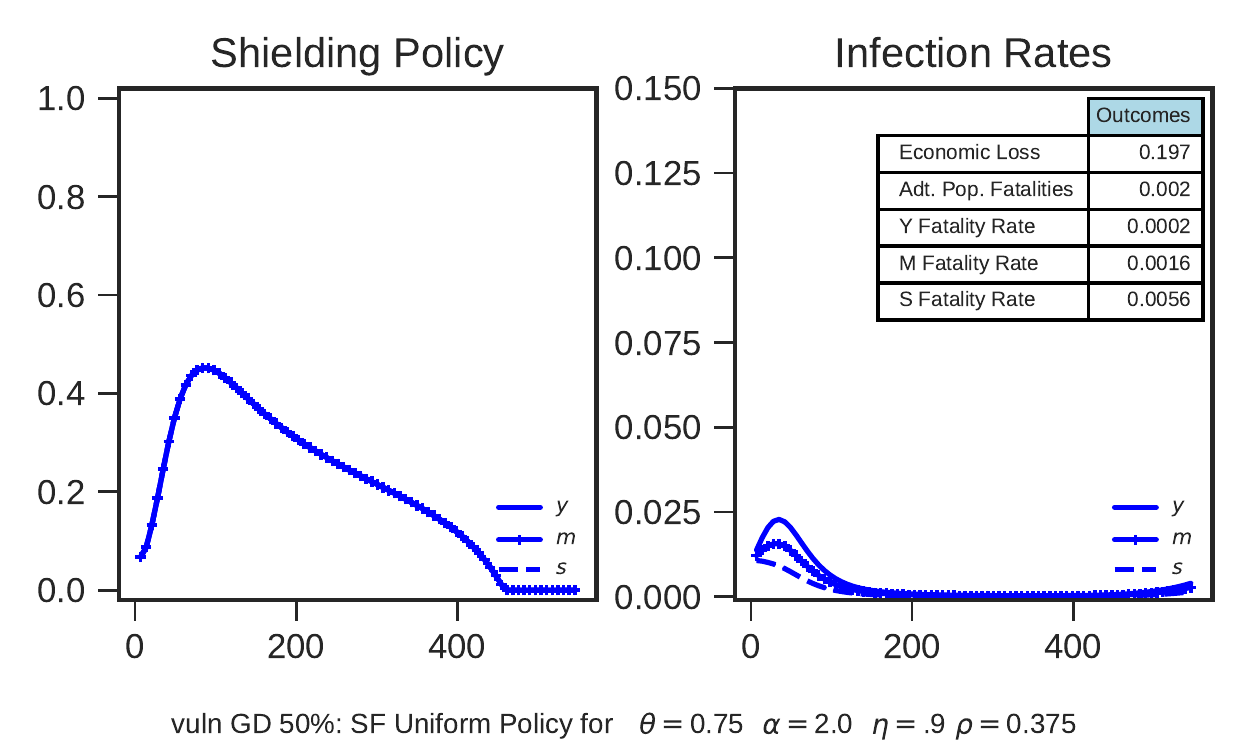}
\caption{Optimal uniform shielding policy with safety focus and group distancing. Panel ($i$): Uniform group distancing with 20\% reduction in social interactions between and within all groups. Panel ($ii$): Group distancing with focus on interactions with vulnerable groups and reduction in contact rates $\rho_{ys}, \rho_{ms}$ by 50\%.}
\label{GDunif} 
\end{figure}

\begin{figure}
\begin{center}
\textit{(i)}
\end{center}
\includegraphics[scale=1]{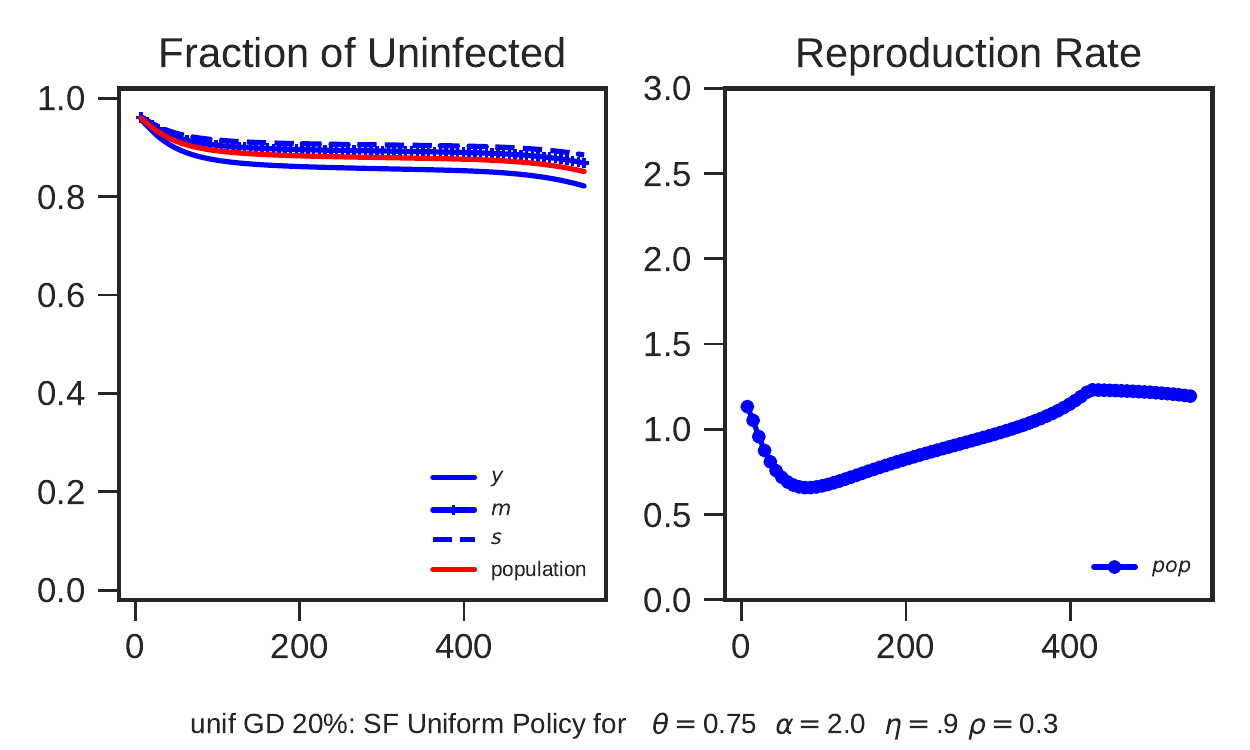}\\
\begin{center}
\textit{(ii)}
\end{center}
\includegraphics[scale=1]{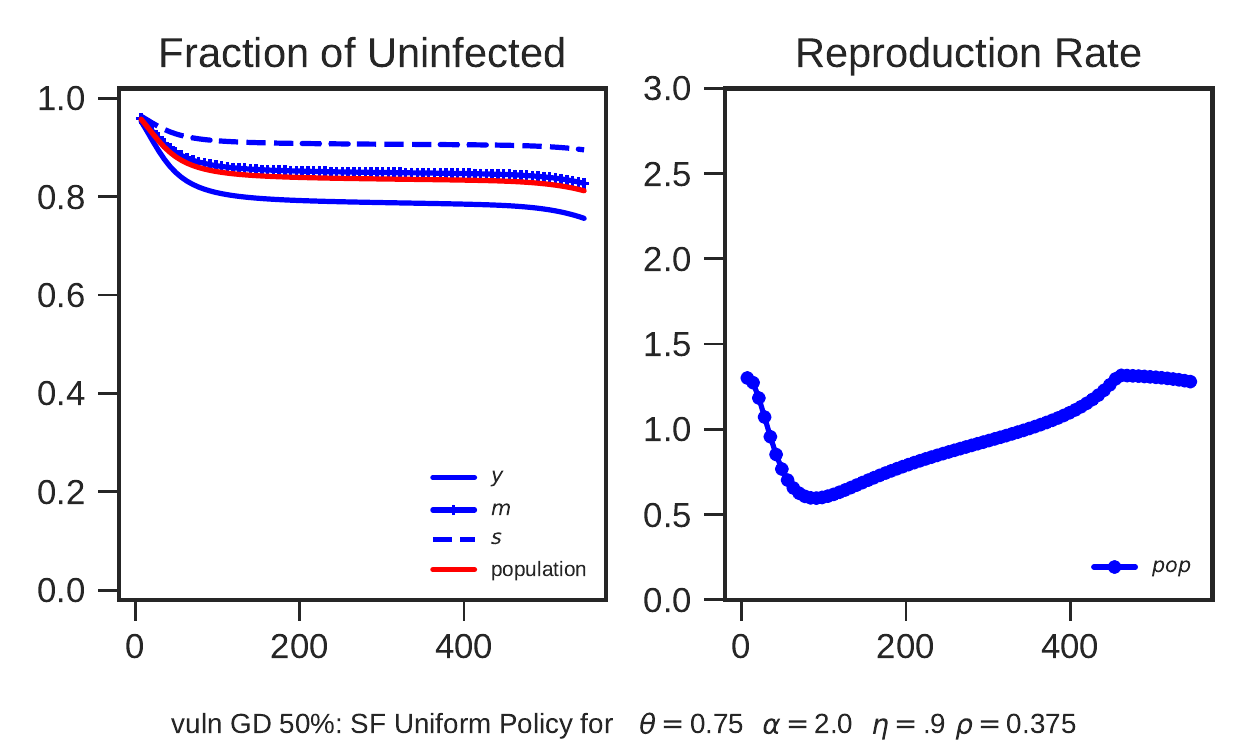}
\caption{Share of uninfected (left) and reproduction rate $R(t)$ (right) in the setting with safety focus and group distancing. Uniform shielding policies. Panel ($i$): Uniform group distancing with 20\% reduction in social interactions between and within all groups. Panel ($ii$): Group distancing with focus on interactions with vulnerable groups and reduction in contact rates $\rho_{ys}, \rho_{ms}$ by 50\%.}
\label{GDunifepi} 
\end{figure}

\begin{figure}
\begin{center}
\textit{(i)}
\end{center}
\includegraphics[scale=1]{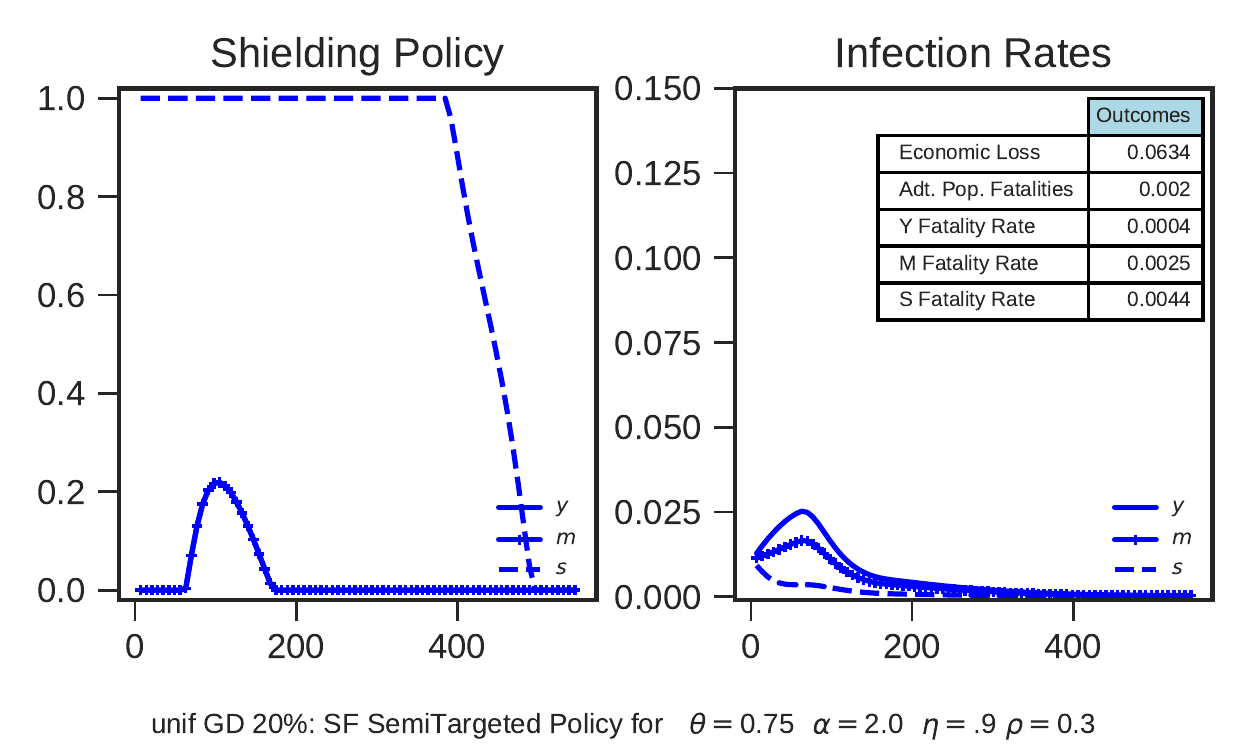}\\
\begin{center}
\textit{(ii)}
\end{center}
\includegraphics[scale=1]{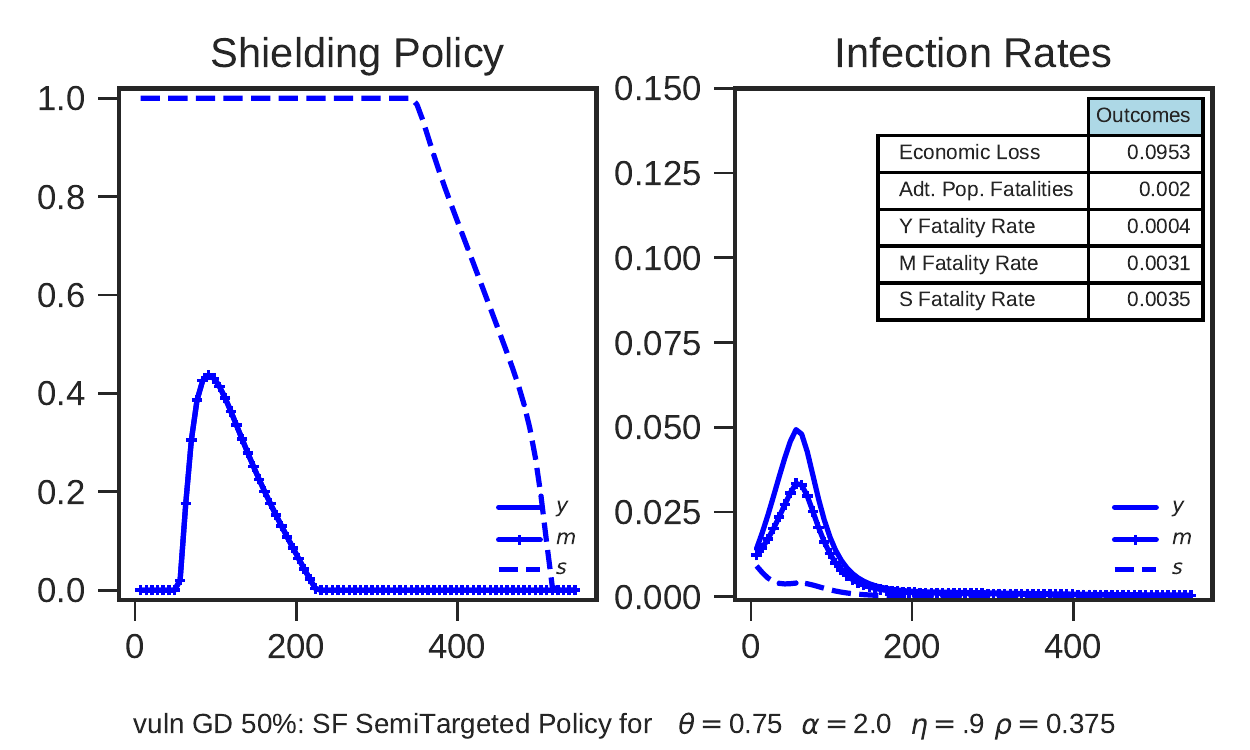}
\caption{Optimal semi-targeted shielding policy with safety focus and group distancing. Panel ($i$): Uniform group distancing with 20\% reduction in social interactions between and within all groups. Panel ($ii$): Group distancing with focus on interactions with vulnerable groups and reduction in contact rates $\rho_{ys}, \rho_{ms}$ by 50\%.} 
\label{GDsemi}
\end{figure}

\begin{figure}
\begin{center}
\textit{(i)}
\end{center}
\includegraphics[scale=1]{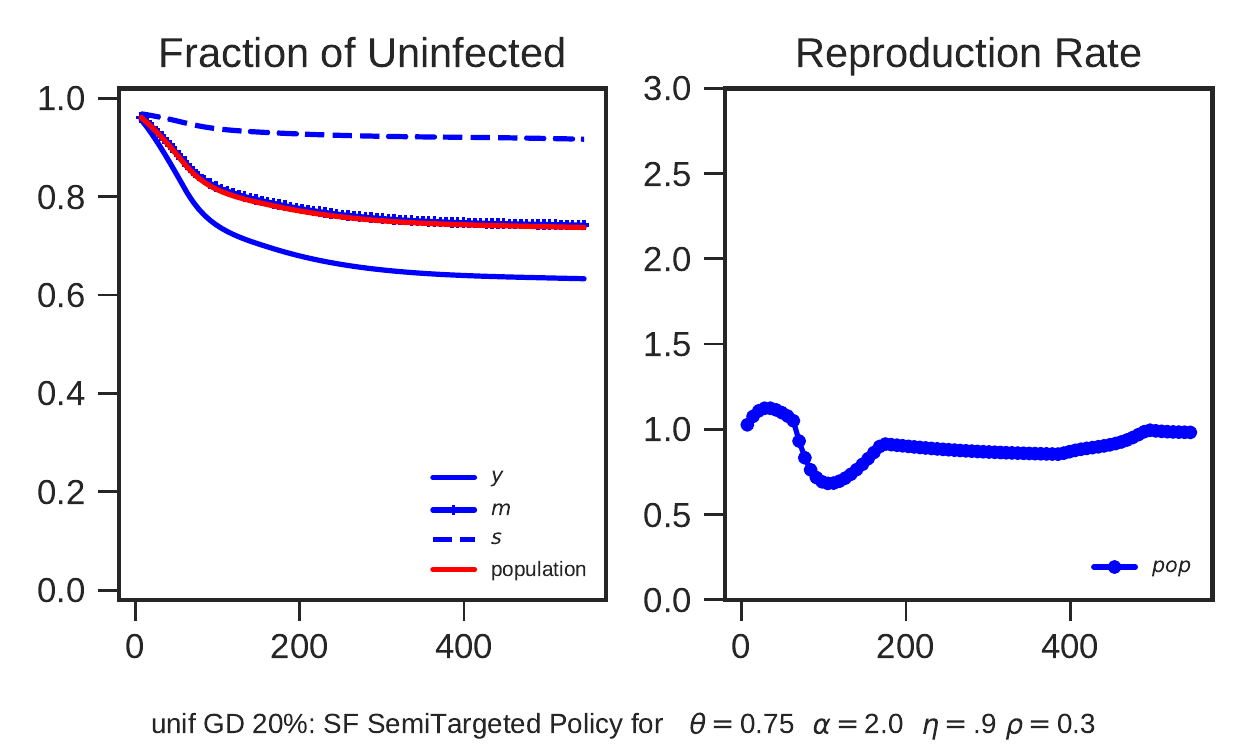}\\
\begin{center}
\textit{(ii)}
\end{center}
\includegraphics[scale=1]{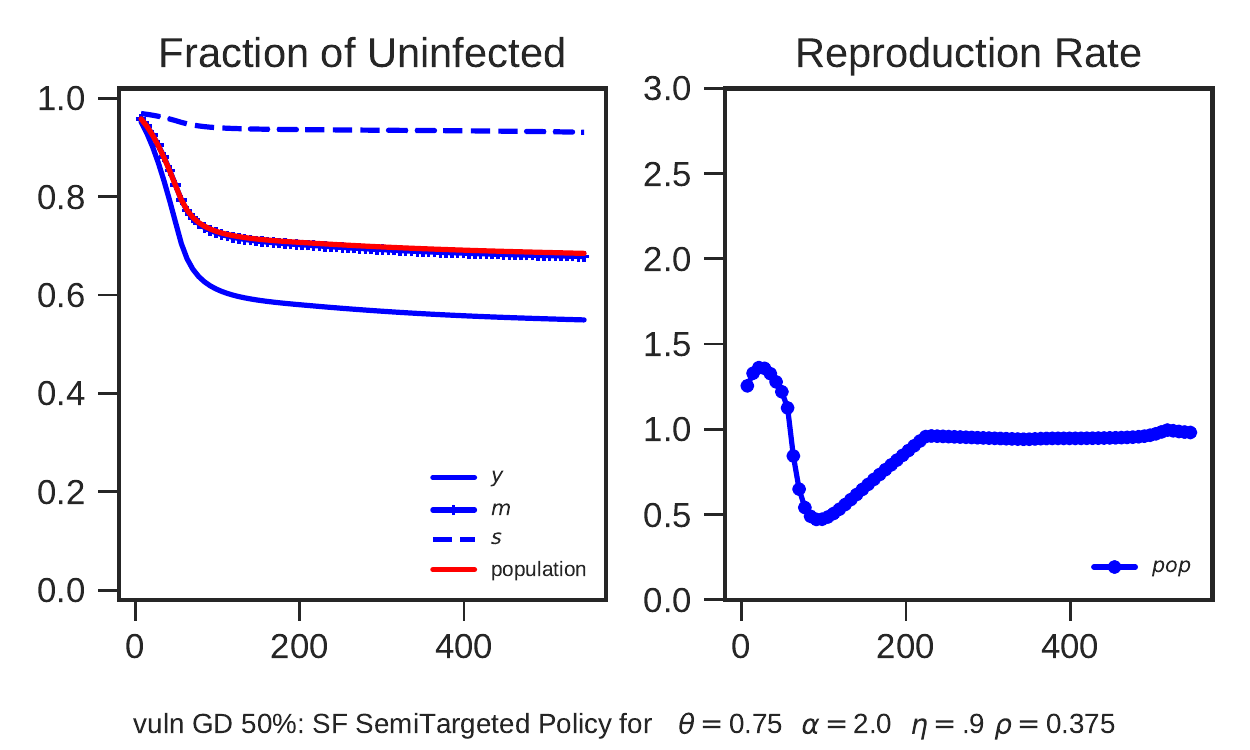}
\caption{Share of uninfected (left) and reproduction rate $R(t)$ (right) in the setting with safety focus and group distancing. Semi-targeted shielding policies.  Panel ($i$): Uniform group distancing with 20\% reduction in social interactions between and within all groups. Panel ($ii$): Group distancing with focus on interactions with vulnerable groups and reduction in contact rates $\rho_{ys}, \rho_{ms}$ by 50\%.} 
\label{GDsemiepi}
\end{figure}

\begin{figure}
\centering
\includegraphics[scale=0.75]{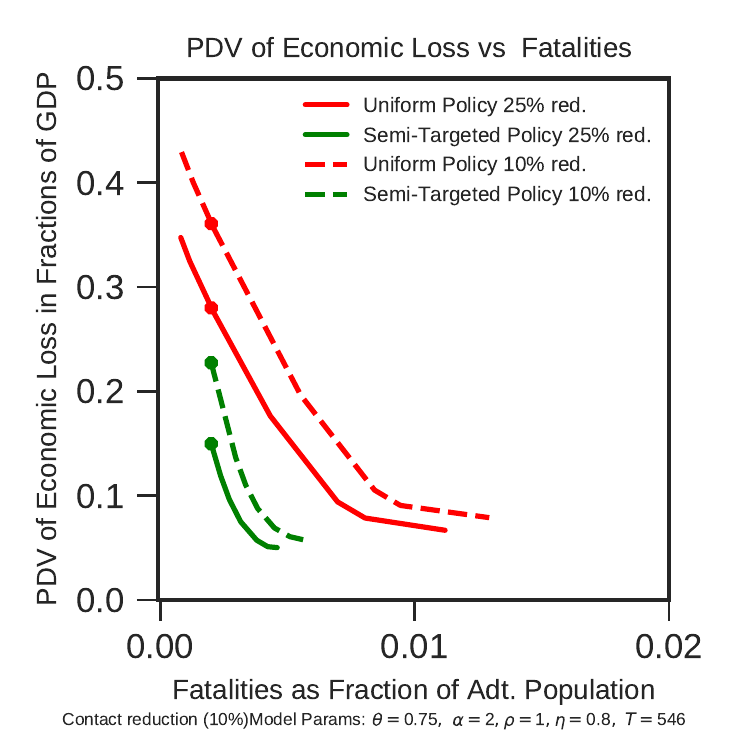} 
\includegraphics[scale=0.75]{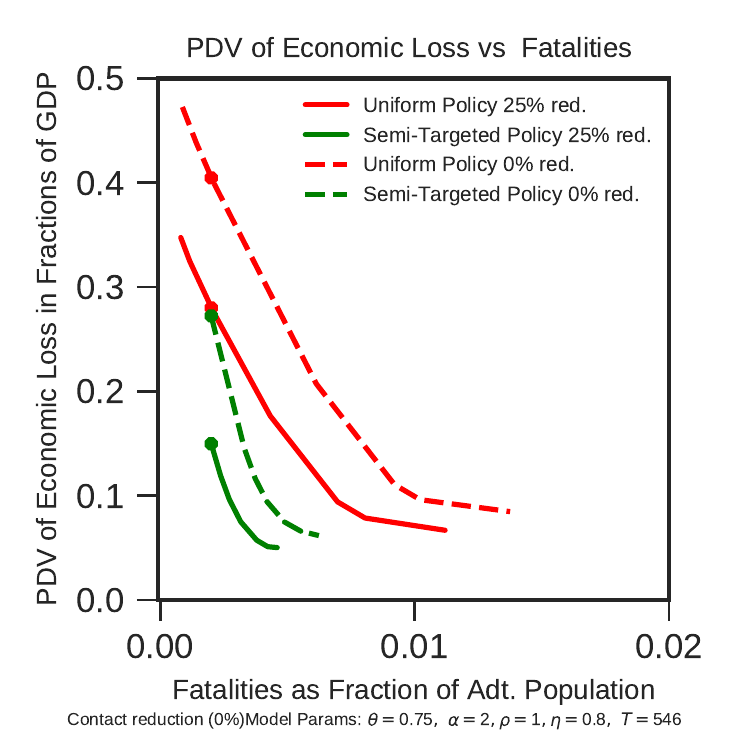}
\caption{Efficient frontier with alternative contact rate adjustment. Reduction of contact rates in $\rho^{0}$ by $10\%$ (left), i.e., $\rho = 0.9 \cdot \rho^{0}$, and without any reduction of contact rates in $\rho^{0}$, i.e., $\rho = \rho^{0}$ (right). Solid lines refer to baseline scenario with $\rho = 0.75 \cdot \rho^{0}$.} 
\label{backtonormal}
\end{figure}

\subsubsection*{The Effect of Physical Distancing, Face Masks and Additional Hygiene Measures}

In general, the policy maker could reduce the intensity and duration of the shielding policy if the transmission rates in personal contacts could be decreased, for example by a voluntary limitation of contacts or reducing the transmission probability by wearing face masks, as described for instance in \cite{chu2020}. Among other recommendations, we provide a list of possible mandatory or voluntary group distancing measures in Section \ref{recomm}. 

We consider two variants of group distancing by manipulating the entries of the contact matrix $\rho$. The first scenario of mandatory or voluntary group distancing we consider refers to a uniform reduction in contact rates in $\rho$. This setting could be considered similar to a general call to the population to reduce all personal interactions, irrespective of the age or vulnerability of the persons considered. The second scenario we consider refers to a change in the contact rates with respect to the senior group – in other words $\rho_{ys}$ and $\rho_{ms}$ and leaving all other entries in $\rho$ unchanged. This scenario corresponds to \textquotedblleft breaking the infection chain\textquotedblright{} with regard to the vulnerable group. In this scenario, the within-group and between-group contacts for the young and middle-age groups are left unchanged. Moreover, in the setting considered, we also leave the contacts within the group of senior citizens unchanged and thus attempt to model a scenario with an impact on the daily contacts of the elderly that is as low as possible. Figure \ref{GDfront} illustrates the policy frontier corresponding to changes in the contact matrix $\rho$ according to uniform group distancing policy (panel ($i$)) and group distancing focusing on the vulnerable (panel ($ii$)). As expected, scaling all entries in $\rho$ simultaneously is substantially more effective in reducing transmissions than is targeted group distancing. However, the social and psychological costs of a uniform group distancing policy are probably high and panel (ii) in Figure 
\ref{GDfront} shows that a reduced, but targeted approach might also help mitigate the health and economic costs of the pandemic.

The results in Figures \ref{GDunif} and \ref{GDsemi} illustrate that group distancing can reduce the intensity and duration of uniform shielding policies while mitigating the economic damage. Substantial efficiency gains can be achieved by targeting  shielding towards the separate groups. Comparing (i) a uniform physical group distancing policy (corresponding to a 20\% reduction in contact rates across all groups, shown at the top of Figures \ref{GDunif} and \ref{GDsemi}) and (ii) targeted group distancing towards the vulnerable (corresponding to a 50\% reduction in contact rates between young and middle group and the senior citizens, shown at the bottom of Figures \ref{GDunif} and \ref{GDsemi}) illustrates that an intense reduction in contacts and/or transmission rates between the vulnerable group and the other age groups can be an effective tool for mitigating the health and economic consequences of the COVID-19 pandemic if combined with targeted shielding. Figures \ref{GDunifepi} and \ref{GDsemiepi} present the evolution of the share of susceptibles and the reproduction rate over time and shed light on the epidemiological consequences of uniform group distancing (Panel (i) in Figure \ref{GDunifepi} and Figure \ref{GDsemiepi}) and a group distancing policy with a focus on the vulnerable (Panel (ii) in Figure \ref{GDunifepi} and Figure \ref{GDsemiepi}). Hence, a more targeted form of group distancing is associated with a greater difference in terms of age-group specific infection rates. Similar to the baseline scenario, the variation in terms of age-specific infection rates is higher if semi-targeted policies are considered. 

As a robustness check, we employ two settings with reduced physical distancing and illustrate the corresponding frontiers in Figure \ref{backtonormal}. Doing so, we intend to illustrate the consequences if individuals are less compliant to physical distancing guidelines, for example, because they underestimate the risk of transmissions.

\subsubsection*{The Effect of Testing and Contact Tracing}

Improved testing and isolation with respect to infectious individuals refers to a reduction in the probability $\eta_j^I$ from the baseline value of $\eta_j^I=0.9$ $-$ that is, the probability that an infectious individual is not detected and isolated to avoid subsequent infections.\footnote{In our analysis, we will only focus on changes in $\eta^I_j$ and $\eta^E_j$ that apply equally to all groups, e.g., consider cases with $\eta^I_y=\eta^I_m=\eta^I_s=0.9$.} A second testing strategy could refer to those who have had contact with the infectious individuals $-$ that is, decreasing the probability that someone who was exposed to an infectious person is not detected and isolated, $\eta^E$, with default value $\eta_j^E=1$. 

\begin{figure}
\begin{center}
\textit{(i)}
\end{center}
\includegraphics[scale=0.5]{figures/GER_SEIR_EconHealth_FigA14_frontier_R024.pdf} \includegraphics[scale=0.5]{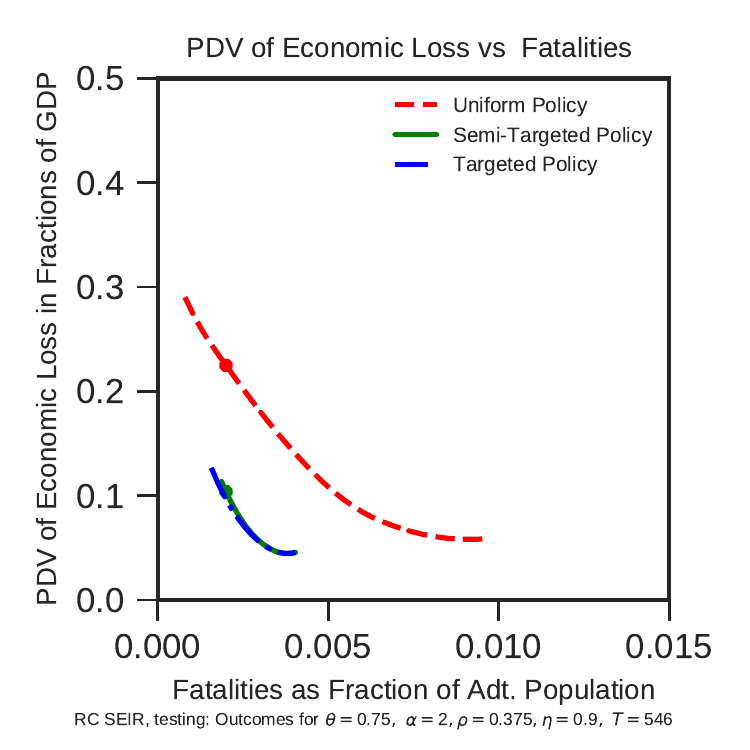}
\includegraphics[scale=0.5]{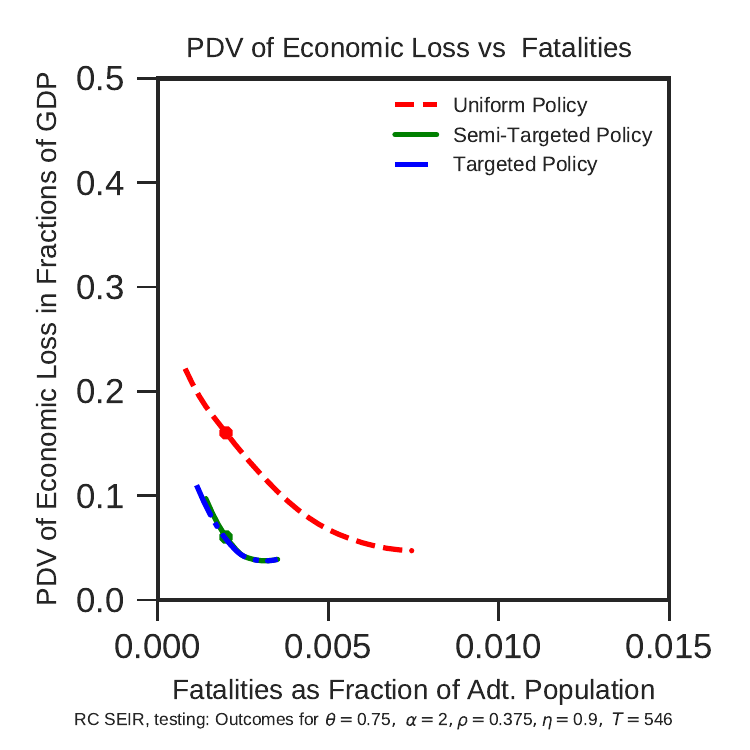}
\\
\begin{center}
\textit{(ii)}
\end{center}
\includegraphics[scale=0.5]{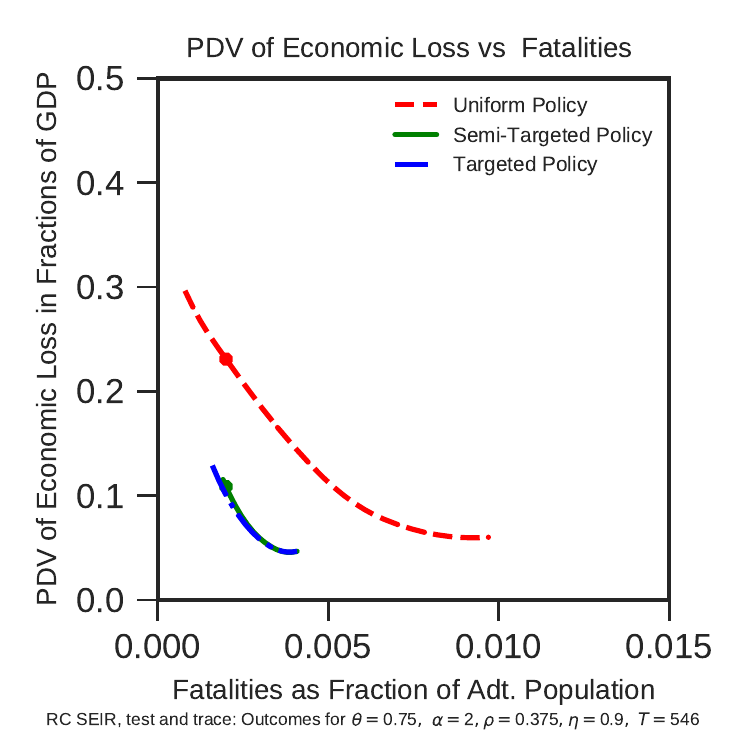}
\includegraphics[scale=0.5]{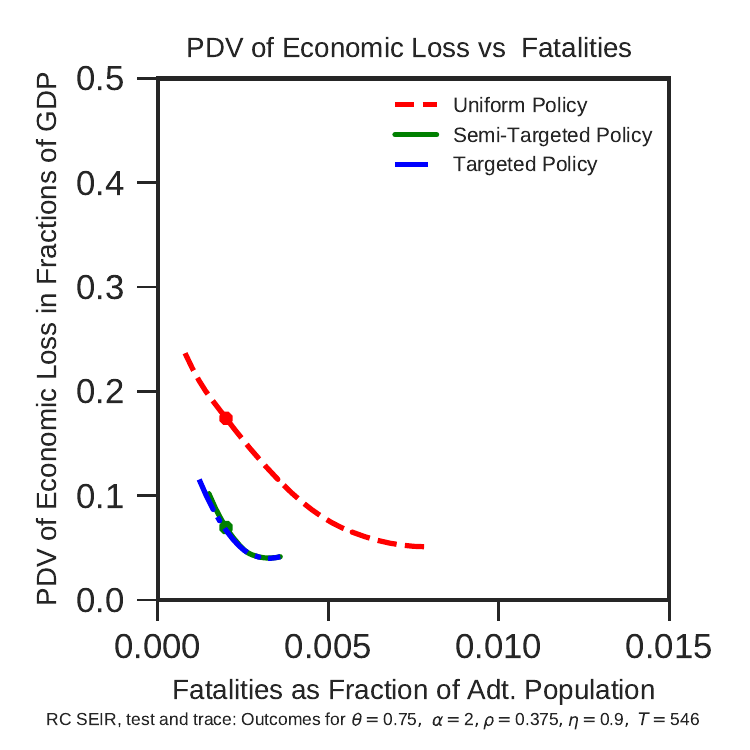}
\includegraphics[scale=0.5]{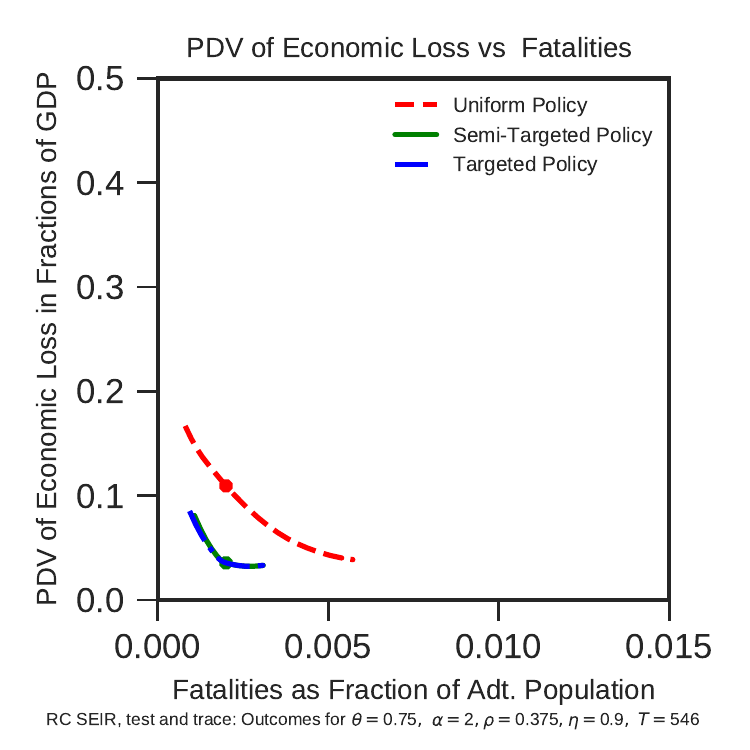} \\
\begin{center}
\textit{(iii)}
\end{center}
\includegraphics[scale=0.5]{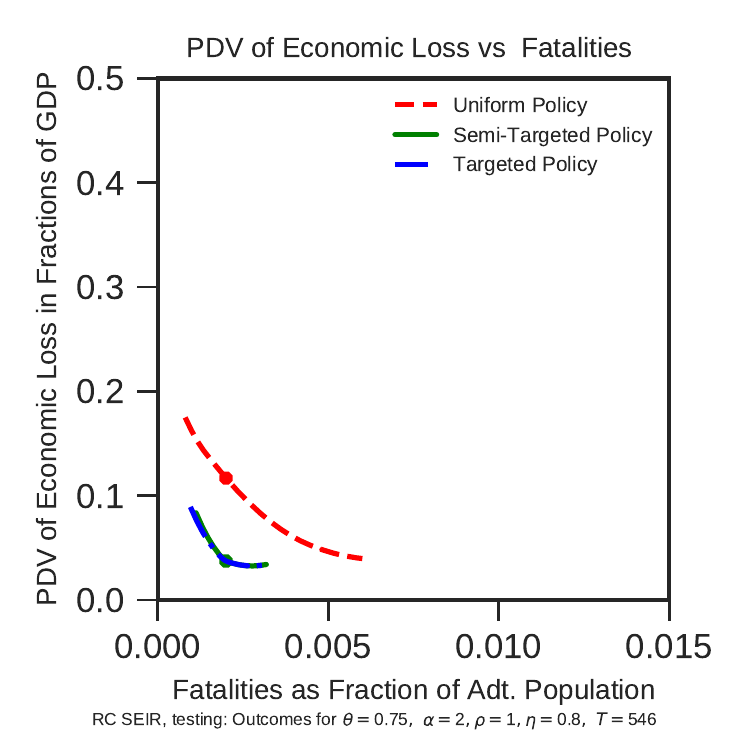}
\includegraphics[scale=0.5]{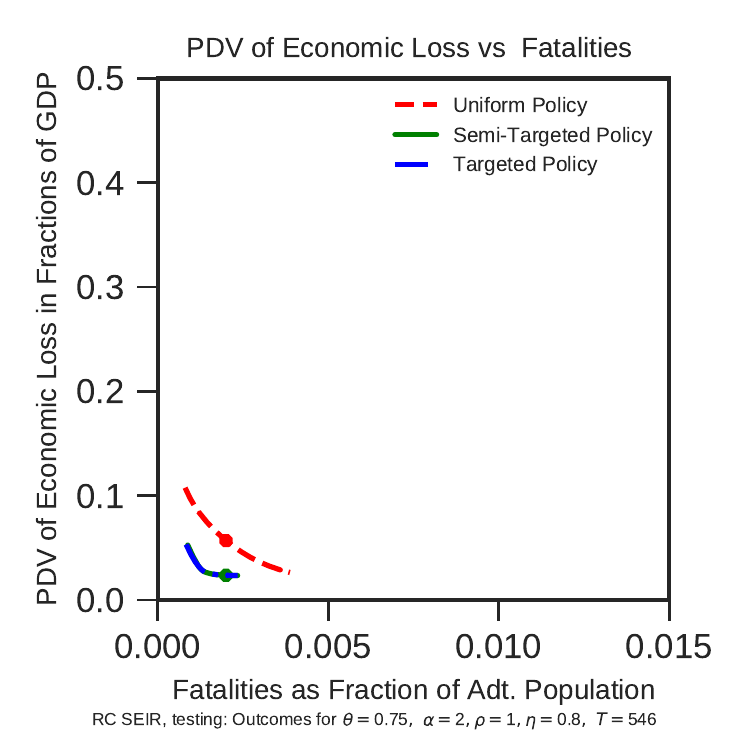}
\includegraphics[scale=0.5]{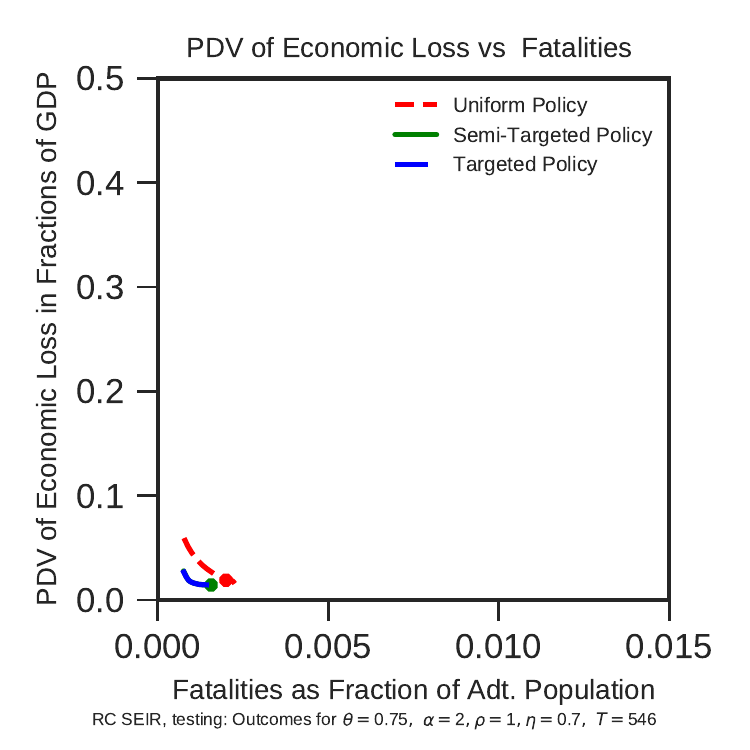}
\caption{Policy frontiers with improved testing and isolation. Panel ($i$): Tests for infectious persons, parameters in order from left to right $(\eta_j^I, \eta_j^E)=(0.9, 1), (0.8, 1), (0.7, 1)$. Panel ($ii$): Improved test and trace policy for exposed individuals, parameters $(\eta_j^I, \eta_j^E)=(0.9, 0.9), (0.9, 0.8), (0.9, 0.7)$. Panel ($iii$): Combination of testing infectious and test and trace policy with parameters $(\eta_j^I, \eta_j^E)=(0.8, 0.8), (0.7, 0.8), (0.7, 0.7)$.}
\label{test}
\end{figure}

Figure \ref{test} illustrates the beneficial implications of improved testing with regard to persons in state $I_j$ (panel ($i$)), improved testing and tracing for persons in $E_j$ (panel ($ii$)) and a combination of these measures (panel ($iii$)). A reduction of $\eta^I$ allows the policy frontier to be shifted closer to the origin, and therefore for efficiency gains to be realized compared to the baseline setting with $\eta_j^I=0.9$. The corresponding frontier is shown in the first plot (on the left) of panel ($i$). Similar conclusions can be drawn for the contact tracing policy as illustrated in panel ($ii$) of Figure \ref{test}. However, simultaneously improving tests both for the infectious and tests for the exposed leads to a substantial improvement in the menu of potential alternatives for policy makers. For example, the safety-focused scenario indicated by the dot on the frontiers involves substantially lower economic costs if the probabilities for undetected infections are reduced to $\eta_j^I=\eta_j^E=0.8$.

\begin{figure}
\begin{center}
\textit{(i)}
\end{center}
\includegraphics[scale=0.75]{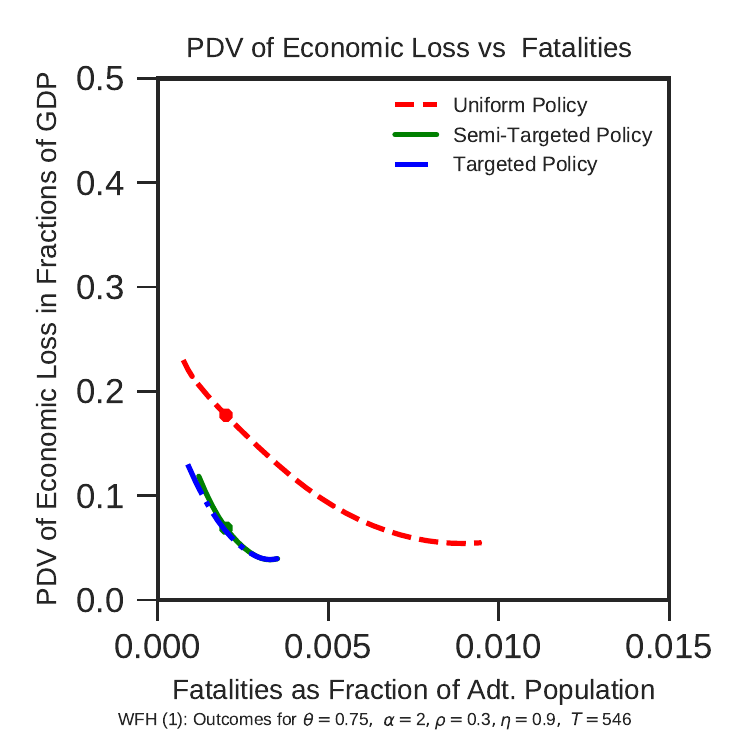} \includegraphics[scale=0.75]{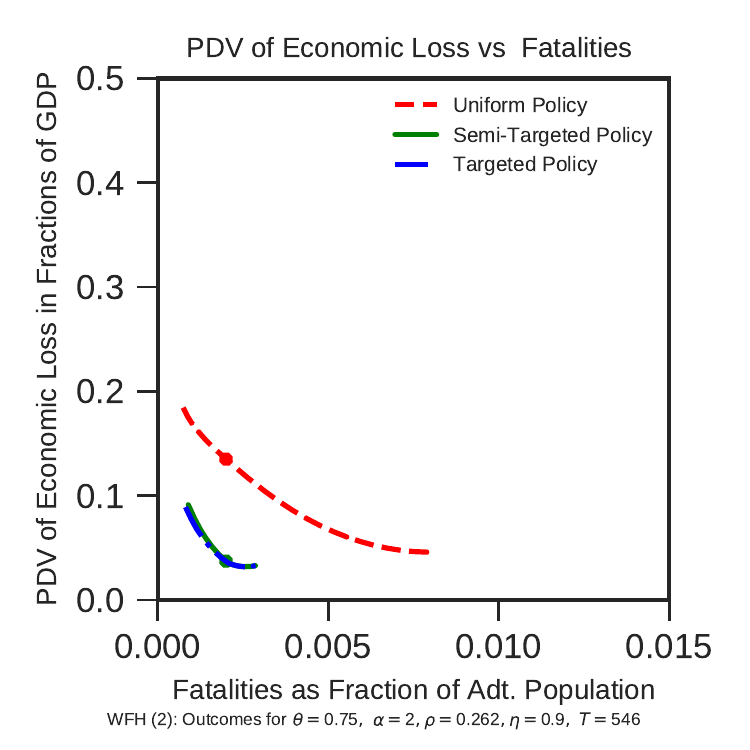}
\\
\begin{center}
\textit{(ii)}
\end{center}
\includegraphics[scale=0.88]{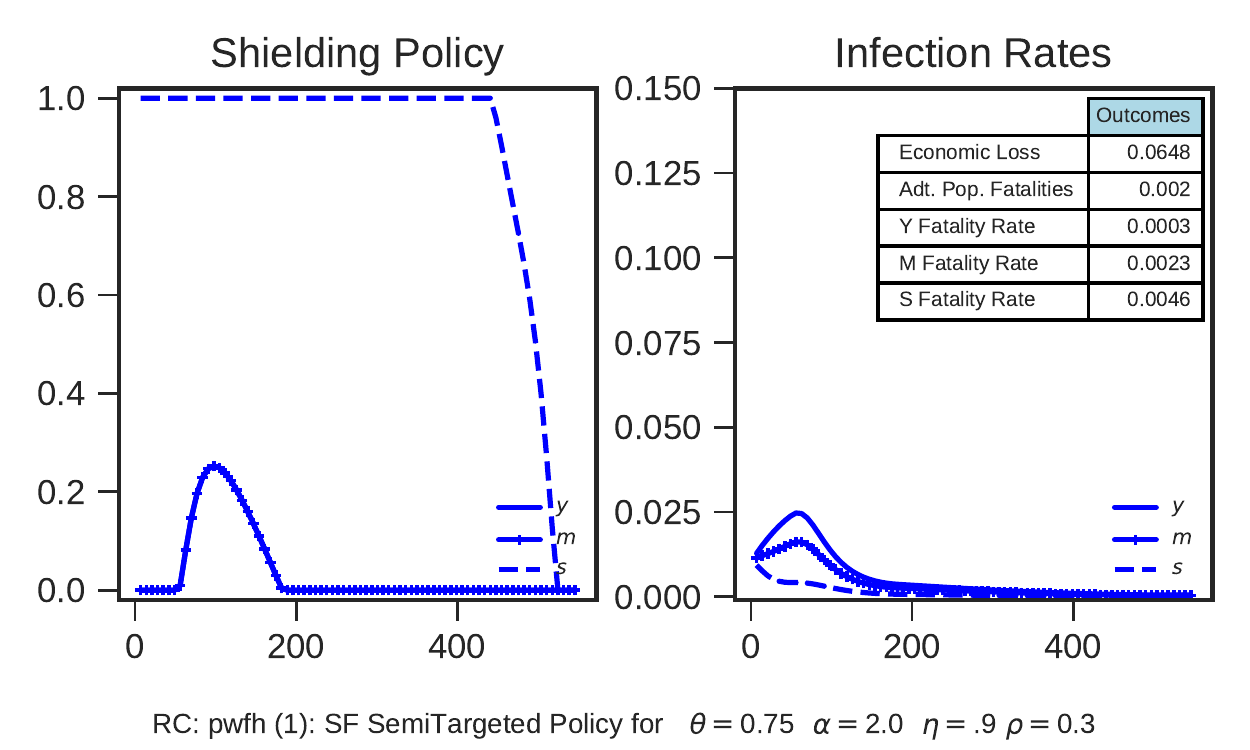}
\caption{Panel ($i$): Frontiers with two variants of improved conditions for working from home, with $\pi_1=20\%, \pi_2=5\%, \xi=0.4$ (left) and $\pi_1=30\%, \pi_2=10\%, \xi=0.4$ (right). Panel ($ii$): Optimal semi-targeted policy with safety focus with scaling $\pi_1=20\%, \pi_2=5\%, \xi=0.4$.}
\label{WFH} 
\end{figure}

\begin{figure}
\begin{center}
\textit{(i)}
\end{center}
\includegraphics[scale=0.75]{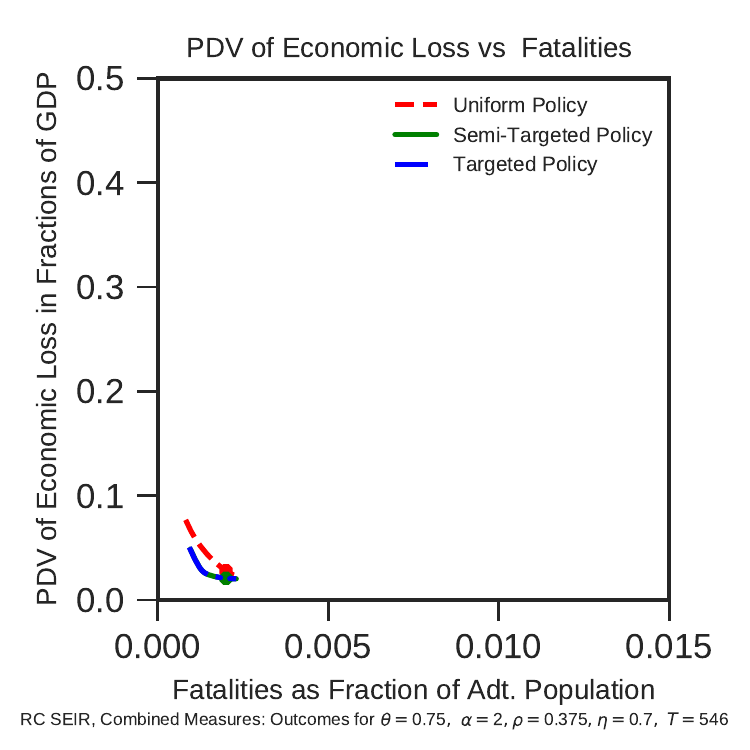}\\
\begin{center}
\textit{(ii)}
\end{center}
 \includegraphics[scale=0.8]{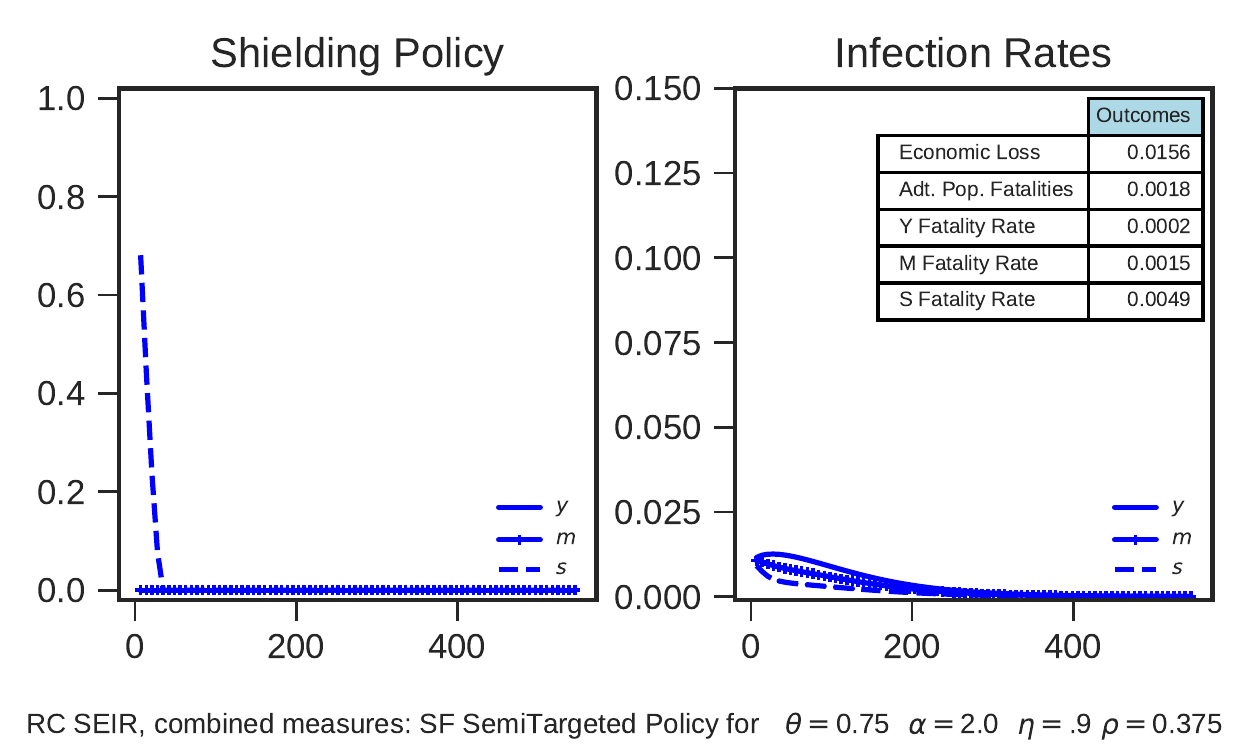}
\\
\begin{center}
\textit{(iii)}
\end{center}
 \includegraphics[scale=0.8]{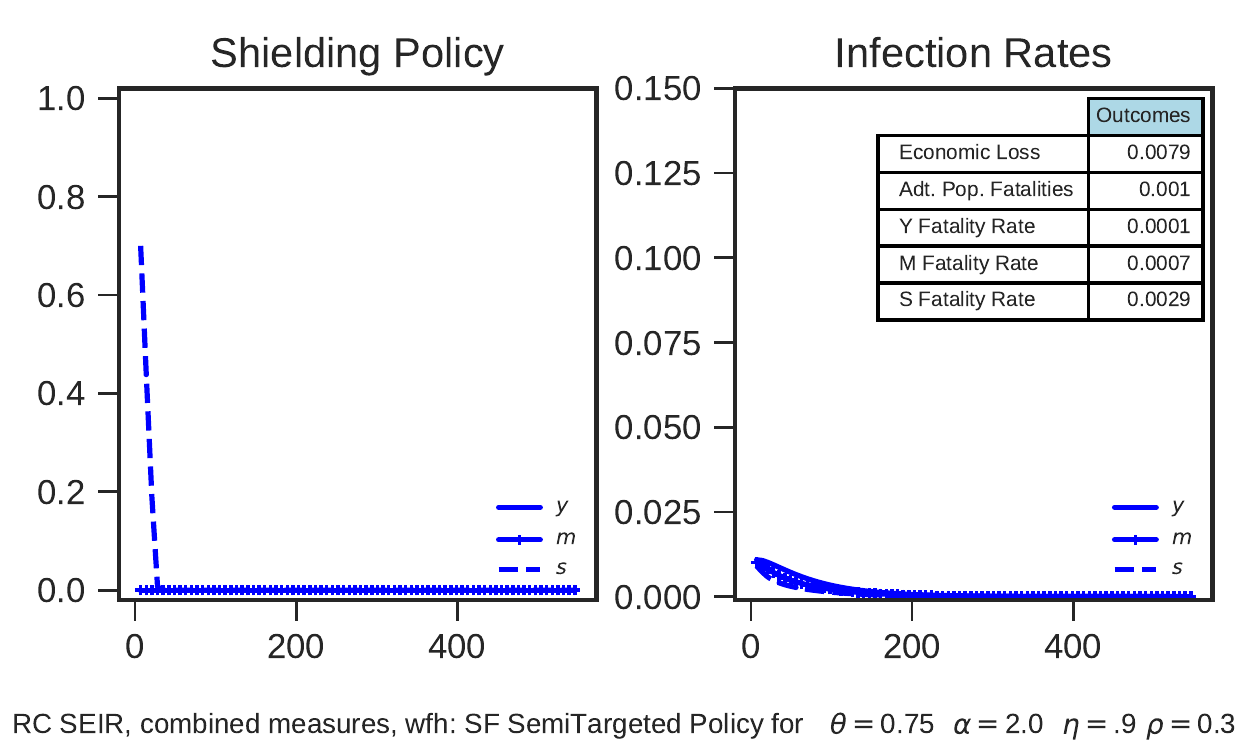}
\caption{Optimal safety focused policy with combination of policy measures (comprehensive approach). Panel ($i$) and ($ii$): Improved testing and isolation for infected ($\eta_I=0.7$) and exposed ($\eta_E=0.8$), reduced contact rates for interactions with the senior group ($\rho_{ys}=\rho_{ms}=0.2$). Panel ($iii$): Improved testing and isolation for infected ($\eta_I=0.7$) and exposed ($\eta_E=0.8$), reduced contact rates for interactions with the senior group ($\rho_{ys}=\rho_{ms}=0.2$), and improved conditions for working from home ($\pi_1=20\%, \pi_2=5\%, \xi=0.4$.)}
\label{comb} 
\end{figure}

\begin{figure}
\begin{center}
\textit{(i)}
\end{center}
 \includegraphics[scale=1]{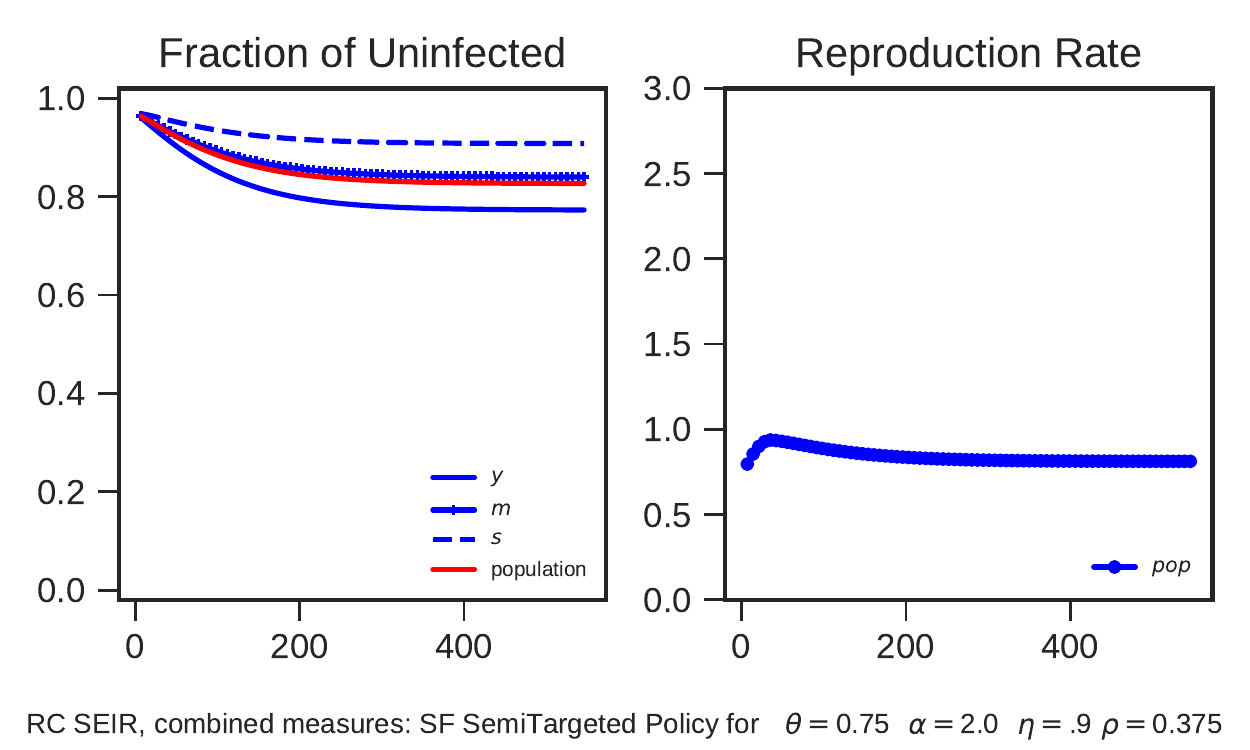}
\\
\begin{center}
\textit{(ii)}
\end{center}
 \includegraphics[scale=1]{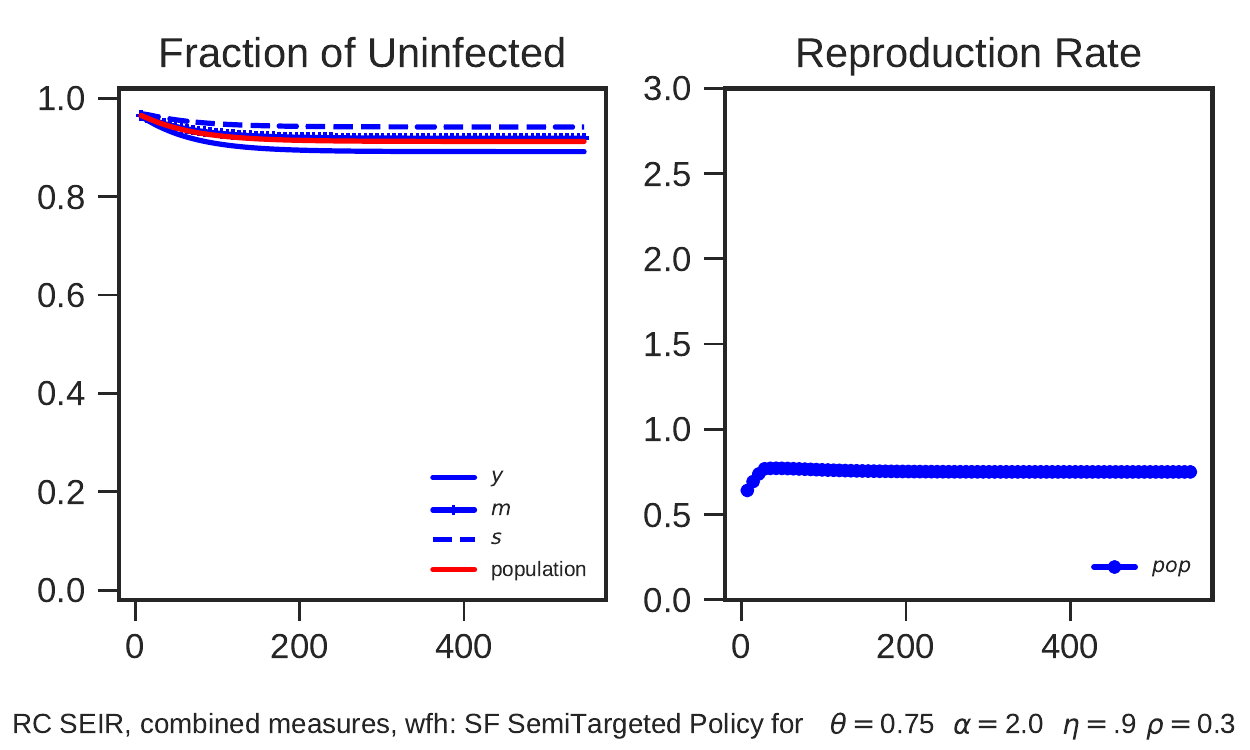}
\caption{Share of uninfected (left) and reproduction rate $R(t)$ (right) in the setting with combination of policy measures (comprehensive approach). Semi-targeted shielding policies. Panel ($i$): Improved testing and isolation for infected ($\eta_I=0.7$) and exposed ($\eta_E=0.8$), reduced contact rates for interactions with the senior group ($\rho_{ys}=\rho_{ms}=0.2$). Panel ($ii$): Improved testing and isolation for infected ($\eta_I=0.7$) and exposed ($\eta_E=0.8$), reduced contact rates for interactions with the senior group ($\rho_{ys}=\rho_{ms}=0.2$), and improved conditions for working from home ($\pi_1=20\%, \pi_2=5\%, \xi=0.4$.)}
\label{combepi} 
\end{figure}

\subsubsection*{The Effect of Improved Conditions for Working From Home}

In addition to voluntary or mandatory group distancing and test and trace policies, governments could provide incentives to promote working from home. To implement improved working from home conditions, we consider a setting with fewer physical interactions and increased productivity at home. Figure \ref{WFH} illustrates two scenarios with (i) $\pi_1=20\%$ and $\pi_2=5\%$ and (ii) $\pi_1=30\%$ and $\pi_2=10\%$. In both settings, the efficiency loss is reduced by 10 percentage points $-$ that is, in the baseline setting, the productivity loss under shielding was set to 70\% and is now changed to 60\%. Panel ($i$) of Figure \ref{WFH} shows the policy frontiers that correspond to setting (i) (left) and setting (ii) (right). Panel ($ii$) illustrates the optimal semi-targeted shielding policy with a safety focus. We can see that improved conditions for working from home make it possible to reduce substantially the economic costs associated with the pandemic and with shielding. Moreover, better conditions for working from home make it possible to reduce the duration and intensity of shielding measures as compared to the baseline setting.

\subsubsection*{A Comprehensive Approach}
The positive effect of improved testing and group distancing can be amplified if these measures are combined with other measures to form a comprehensive approach. As illustrated in panel ($i$) of Figure \ref{comb}, combining improved testing and contact tracing with group distancing focusing on interactions of the other groups with the group of senior citizens allows the policy frontier to be shifted closer to the origin. According to the efficient frontiers in Figure \ref{comb} (panel($i$)) and the optimal policies in panel ($ii$), policy makers can almost refrain entirely from imposing shielding rules. Optimally, only a short shielding is imposed on the senior group at the early stage of the pandemic. The optimal policies associated with uniform shieldings as presented in the Appendix, Figure \ref{unifcomb} do not involve any shielding at comparable economic damage and a slightly higher mortality in the population. 
  
Panel ($ii$) of  Figure \ref{comb} illustrates that a comprehensive approach that also includes improved conditions for working from home, with $\pi_1 =20\%$ and $\pi_2=5\%$, allows for even lower economic damages at a very short shielding phase. Moreover, the result on the infection rates and the reproduction rate in panel($ii$) of Figure \ref{combepi} show that the combined approach with improved conditions for working from home help to reduce the share of infected in all age groups. Furthermore, it can be observed that the comprehensive approach allows the reproduction rate being kept below the critical threshold of 1. 

 


\begin{figure}
\includegraphics[scale=0.75]{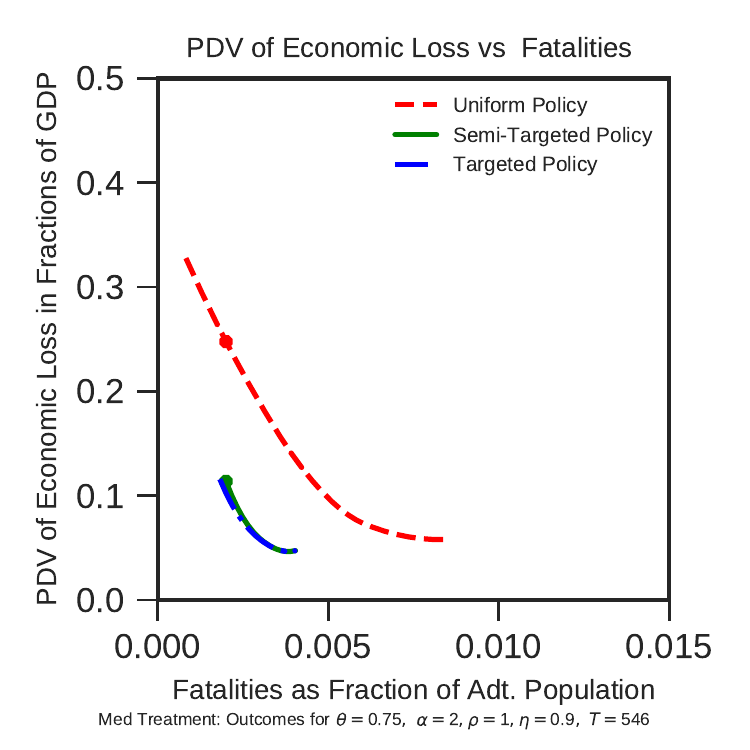} 
\includegraphics[scale=0.75]{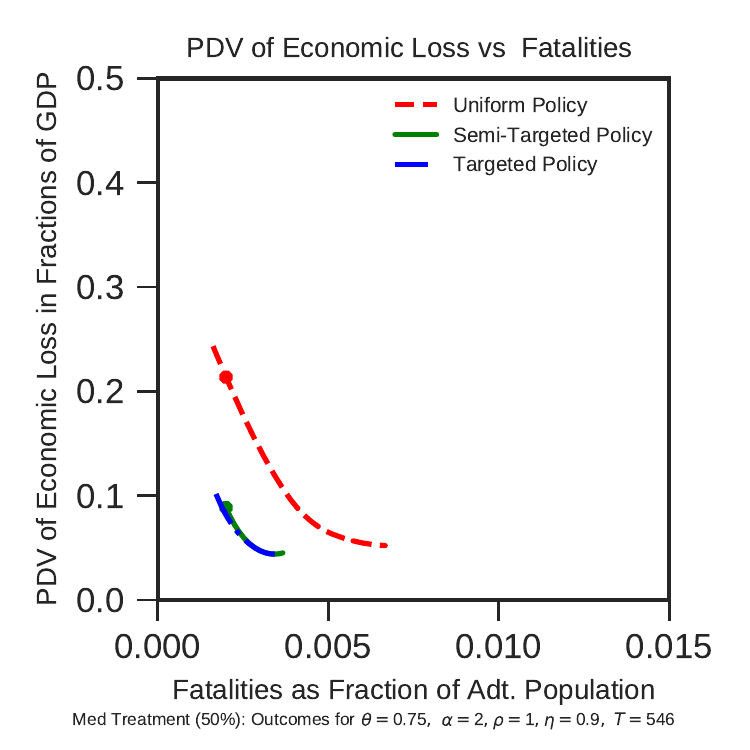}
\caption{Efficient frontier in SEIR model with improved medical treatment corresponding to 30\% (left) and 50\% (right) lower mortality for the senior group.} 
\label{frontiertreat}
\end{figure}

\subsubsection*{The Effect of Improved Medication and Treatment}

Finally, we assess the effect associated with an improved treatment for COVID-19, which corresponds with a 30\% and 50\% lower baseline mortality rate for the group of senior citizens, $\underline{\delta}_s^d$.
For example, a recent study by \cite{horby2020} shows that treatment with dexamethasone can reduce the mortality of severe hospitalized cases by up to one third.
We acknowledge that the different effects of an approved drug for treating COVID-19 patients, as described in (i) to (iii) in Section \ref{assump} (Vaccine and Cure), might lead to different results in terms of optimal policies. \cite{Acemoglu2020} provide robustness checks by increasing  the mortality rate for the senior group and also varying this group's per-capita income. By doing so, they conclude that the efficiency gains of targeted policies arise due to the high vulnerability rather than the low productivity of that group. Hence, effective medical treatments might soften the distinction between the vulnerable group and the groups with lower mortality risk.

Comparing the efficient frontier with improved treatment in Figure \ref{frontiertreat} with that in the baseline setting in Figure \ref{frontecon} illustrates that the economic costs at a given mortality level can be reduced substantially. At the same time, the distance between the frontiers of targeted and uniform shielding policies becomes smaller, which is in line with the observation in \cite{Acemoglu2020} that the efficiency gains of targeting accrue due to high vulnerability.\footnote{We perform various robustness checks (results omitted) with respect to the income parameters $\omega_j$ and mortality rates $\underline{\delta}^d_s$ and confirm the conclusions in \cite{Acemoglu2020}.} However, even with a substantially improved medical treatment that leads to a 50\% lower mortality among the senior group, targeted shielding is still associated with considerable efficiency gains compared to uniform approaches.

\subsubsection*{The Effect of a Vaccine Arriving Early}
\begin{figure}
\centering
\begin{center}
\textit{(i)}
\end{center}
\includegraphics[scale=1]{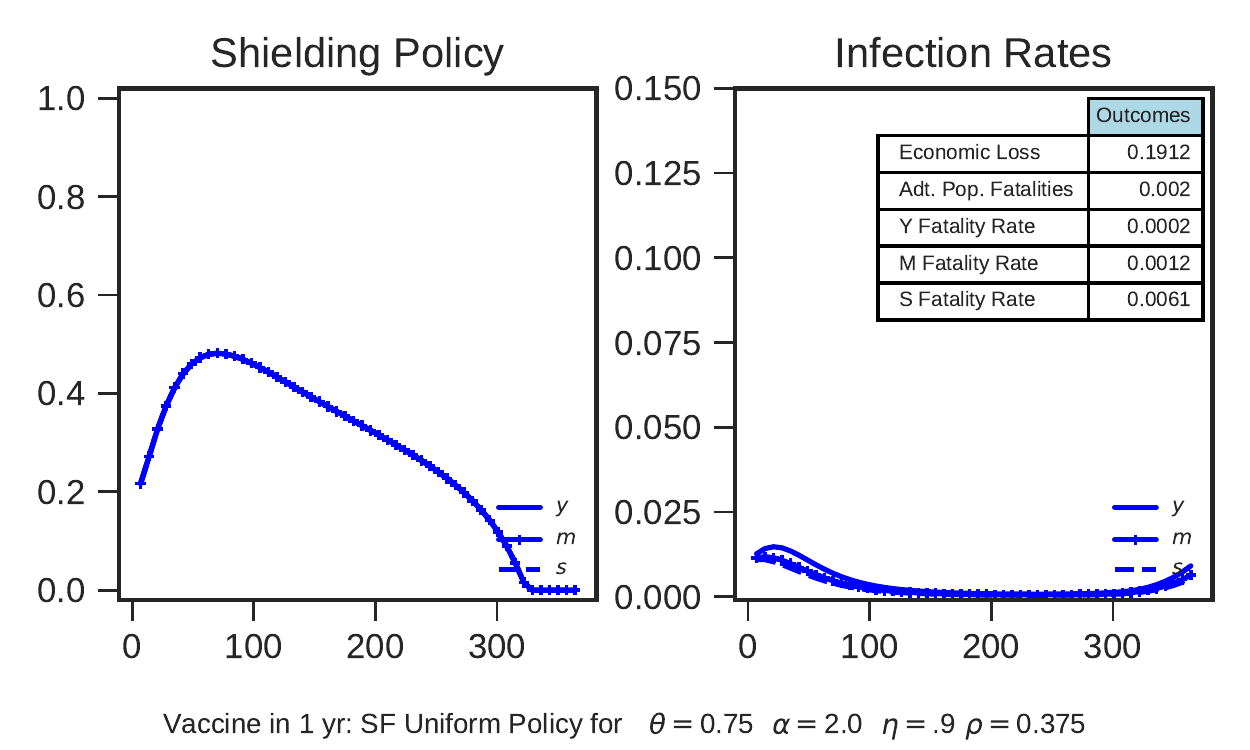} \\
\begin{center}
\textit{(ii)}
\end{center}
\includegraphics[scale=1]{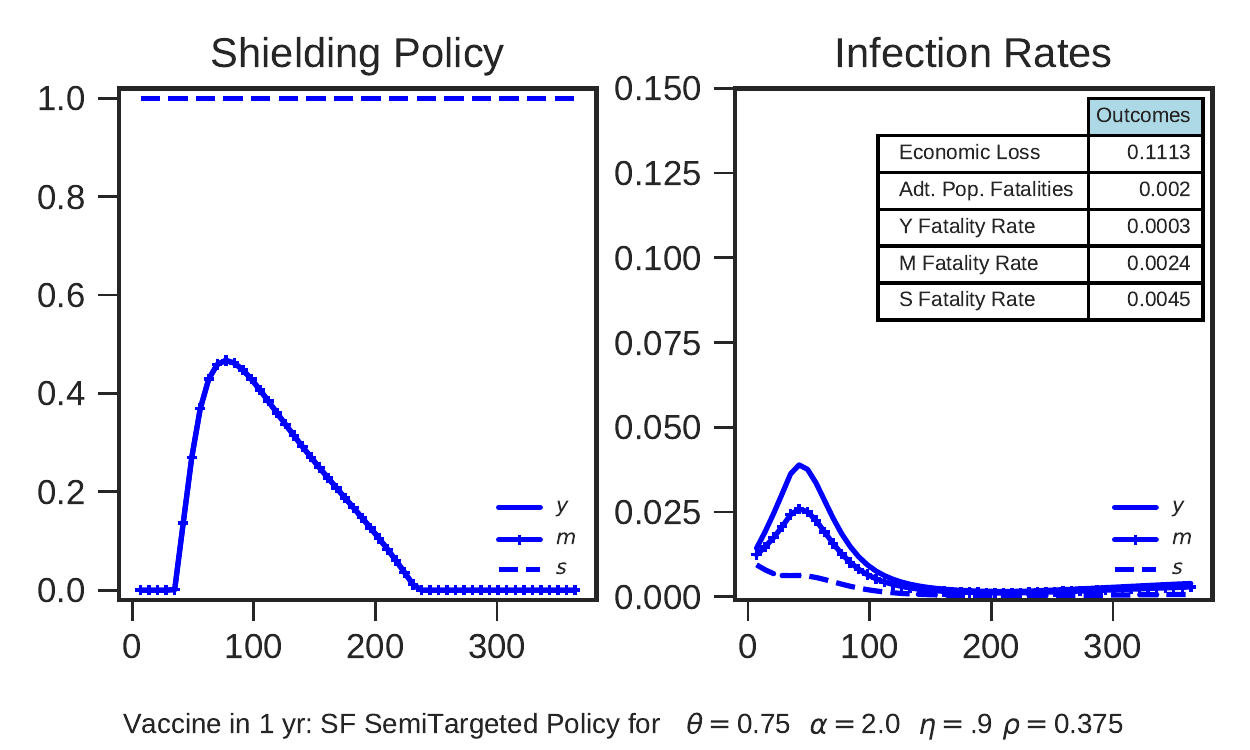}
\caption{Optimal shielding policy with a vaccine arrival after 1 year. Panel ($i$): Optimal uniform policy. Panel ($ii$): Optimal semi-targeted policy.}
\label{vaccin1}
\end{figure}

\begin{figure}
\centering
\begin{center}
\textit{(i)}
\end{center}
\includegraphics[scale=1]{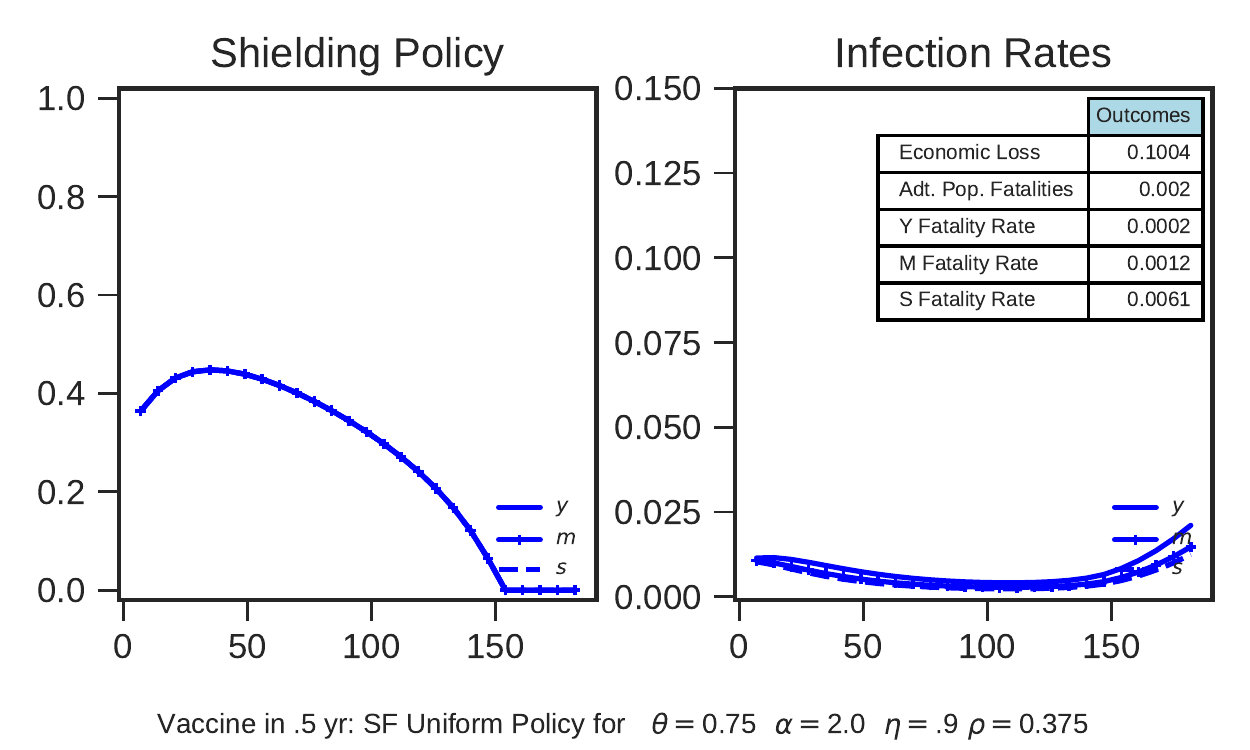} \\
\begin{center}
\textit{(ii)}
\end{center}
\includegraphics[scale=1]{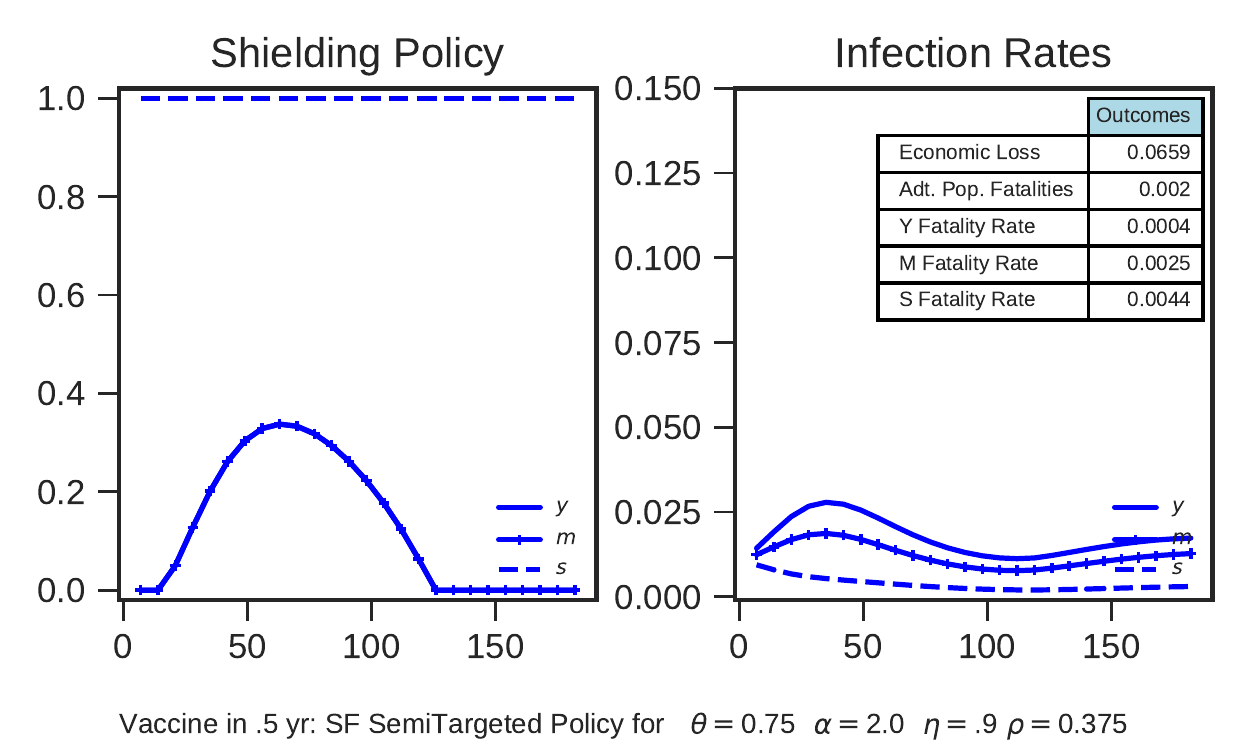}
\caption{Optimal shielding policy with a vaccine arriving after six months. Panel ($i$): Optimal uniform policy. Panel ($ii$): Optimal semi-targeted policy.}
\label{vaccin05}
\end{figure}

Since the early phase of the pandemic, governments around the world have encouraged research activities to develop a vaccine for SARS-CoV-2. In the initial study by \cite{Acemoglu2020} and the settings considered so far, we make a deterministic assumption that a vaccine arrives in 1.5 years. Figures \ref{vaccin1} and \ref{vaccin05} illustrate the optimal uniform and semi-targeted shielding policies if an effective vaccine is available after one year and after six months, respectively. The results highlight the economic importance of an effective vaccine being available early because this would substantially reduce the loss in GDP, which in the baseline scenario decreases by approximately $26\%$ under uniform shielding policies and $13\%$ under semi-targeted policies if a vaccine becomes available after 1.5 years. If, in contrast, a vaccine becomes available in one year, the loss under uniform shielding reduces to $18\%$ and $9\%$ under semi-targeted policies. In the scenario that a vaccine becomes available after six months, these numbers are $8\%$ and $5\%$, respectively. As a consequence of a shorter period $T$, the shielding policies are maintained over a shorter time span, wherease their intensity does not change substantially.

\clearpage

\subsection{Implementation of Optimal Policies} \label{recomm}

The results of the model indicate that it is favorable to protect vulnerable groups, in particular persons at higher ages, while relieving other groups of the shielding measures. Of course such a policy must be implemented with a sense of proportion and be supported by accompanying measures to reduce the burden as much as possible. We propose that implementing such a policy should take into account the following measures and points:


\begin{enumerate}
\item Prioritize the use of masks (N-95/FFP2; surgical masks) and personal protection equipment among seniors and other individuals with comorbidities. These groups and contact persons (like nurses) should be equipped with masks that are of high quality.
	
\item Limit and reduce potential transmissions by decreasing the number of contacts with persons who are at higher risk, in particular with those who are likely to be exposed to other infectious individuals. The risk of transmission could be reduced by generally requiring people to cover their nose and mouth with simple disposable or reusable face masks when interacting with the elderly or other vulnerable groups, or by reducing the risk of individuals being infectious at the time of the contact, for example through intensive testing and requiring quarantine or reduced contacts during a defined period before visits to nursing homes.

\item Set up special shopping and medical consultation hours for the elderly and vulnerable to allow them to do shopping for daily essentials and attend important medical appointments. This has been practiced in the U.S., UK and other countries. Also encouraging the use and potentially the expansion of various home-delivery services could be valuable in this regard.

\item Ensure older people who are still participating in the labor market and other high- risk individuals are able to work from home easily, for example by providing them with the appropriate equipment, infrastructure and training.

\item Provide additional benefits and compensation, such as job guarantees and prioritized paid leave, for employees who behave in a socially responsible way, caring for the elderly or other vulnerable individuals. For example, health insurers could offer monetary benefits to individuals who commit themselves to reducing social interactions with others in order to care for individuals at high risk

\item Provide mental health and social support via teleconferencing and other safe means of interaction, particular through online consultation hours and tele-medicine, as well as video-conferencing systems in nursing homes so that residents can stay in contact with their families.

\item Implement a stay-at-home policy for older people on a voluntary basis. High compliance with this policy might be achieved through an incentive scheme. Given the right incentives a senior citizen should follow a stay-at-home policy in his or her own interest.

\item Provide frequent, easy-to-understand and non-contradictory information and communication and assistance to members of vulnerable groups who live in their own home; also, create incentives for  members of the young and middle age groups to protect the vulnerable members of society.

\item Frequently update shielding policies according to new scientific evidence on the transmission of SARS-CoV-2  in order to increase the efficiency of such measures, reduce economic costs and achieve higher compliance with group distancing recommendations. 

\end{enumerate}

\section{Conclusion} 

In this paper, we adopt the extended SIR model of \cite{Acemoglu2020} to Germany. Germany differs from the U.S. both in its socio-demographics and its system of health care coverage and provision. The model allows for a comparison of the impact of different policies both on survival rates and economic losses, thus providing policy makers with information to derive optimal policies. We evaluate several scenarios in a quantitative manner and find that semi-targeted shielding makes it possible to achieve efficiency gains, which might be used to fund measures that improve the conditions of vulnerable groups, such as senior citizens and people with comorbidities. Most importantly, we find that the intensity and duration of shielding policies can be reduced by employing additional measures, such as group distancing, testing and contact tracing. Indeed, a comprehensive approach that combines these measures and implements them simultaneously can keep both economic losses and population mortality at a low level $-$ even with uniform shielding measures. Lastly, we highlight the importance of finding effective medical treatments and of timely vaccine development.

There are several extensions of our analysis that could be considered in future research. First, the estimates on contact rates  that are used in the baseline setting   are  based on a study from the UK    and might be re-adjusted to country-specific contact patterns. Additional work could be performed to provide comparable data for other countries, including Germany and the U.S. Second, the SEIR model incorporates contact tracing by including a parameter on the probability that a person who was exposed to an infectious individual is tested and isolated. This approach allowed us to maintain a relatively concise model structure. A more complex structure might involve separate compartments for exposed and infectious individuals who are either in quarantine or not in quarantine, allowing the social interactions between these two groups to be modeled. 
The model in \cite{Grimm2020} is an example of such an evolved compartment structure. Moreover, the infectiousness of individuals could be modelled in a more granular way. Several studies, such as \cite{Grimm2020}, distinguish between symptomatic, asymptomatic and severe cases and allow for transmissions of SARS-CoV-2 by asymptomatic cases. Alternatively, the SEIR model in \cite{berger2020} allows for infectiousness of the exposed individuals. Lastly, as soon as more information is available on whether people develop long-term immunity to SARS-CoV-2 after infection, this might be used to adapt the SEIR model to a SEIRS structure.

\newpage
\footnotesize
\pagebreak
\bibliographystyle{imsart-number}
\bibliography{mybib}

\begin{thebibliography}{17}

\bibitem{Acemoglu2020}
\begin{btechreport}[author]
\bauthor{\bsnm{Acemoglu},~\bfnm{Daron}\binits{D.}},
  \bauthor{\bsnm{Chernozhukov},~\bfnm{Victor}\binits{V.}},
  \bauthor{\bsnm{Werning},~\bfnm{Iv�n}\binits{I.}} \AND
  \bauthor{\bsnm{Whinston},~\bfnm{Michael~D}\binits{M.~D.}}
(\byear{2020}).
\btitle{Optimal Targeted Lockdowns in a Multi-Group SIR Model}.
\btype{Working Paper} No. \bnumber{27102},
\bpublisher{National Bureau of Economic Research}.
\bdoi{10.3386/w27102}
\end{btechreport}
\endbibitem

\bibitem{AHA2020}
\begin{bmisc}[author]
\bauthor{\bsnm{{American Hospital Association (AHA)}}}
(\byear{2020}).
\btitle{Fast Facts on U.S. Hospitals, 2020 (accessed July 3, 2020)}.
\bnote{https://www.aha.org/statistics/fast-facts-us-hospitals}.
\end{bmisc}
\endbibitem

\bibitem{berger2020}
\begin{btechreport}[author]
\bauthor{\bsnm{Berger},~\bfnm{David~W}\binits{D.~W.}},
  \bauthor{\bsnm{Herkenhoff},~\bfnm{Kyle~F}\binits{K.~F.}} \AND
  \bauthor{\bsnm{Mongey},~\bfnm{Simon}\binits{S.}}
(\byear{2020}).
\btitle{An SEIR Infectious Disease Model with Testing and Conditional
  Quarantine}
\btype{Technical Report},
\bpublisher{National Bureau of Economic Research}.
\end{btechreport}
\endbibitem

\bibitem{Tertilt2020}
\begin{btechreport}[author]
\bauthor{\bsnm{Brotherhood},~\bfnm{Luiz}\binits{L.}},
  \bauthor{\bsnm{Kircher},~\bfnm{Philipp}\binits{P.}},
  \bauthor{\bsnm{Santos},~\bfnm{Cezar}\binits{C.}} \AND
  \bauthor{\bsnm{Tertilt},~\bfnm{Mich\`{e}le}\binits{M.}}
(\byear{2020}).
\btitle{{An Economic Model of the Covid-19 Epidemic: The Importance of Testing
  and Age-Specific Policies}}
\btype{{CRC TR 224 Discussion Paper Series}} No.
  \bnumber{{crctr224\_2020\_175}},
\bpublisher{University of Bonn and University of Mannheim, Germany}.
\end{btechreport}
\endbibitem

\bibitem{mikrozensus2018}
\begin{bmisc}[author]
\bauthor{\bsnm{{Statistisches Bundesamt}}}
(\byear{2019}).
\btitle{Bev{\"o}lkerung und Erwerbst\"atigkeit, Erwerbsbeteiligung der
  Bev\"olkerung, Ergebnisse des Mikrozensus zum Arbeitsmarkt}.
\end{bmisc}
\endbibitem

\bibitem{mikrozensus2020}
\begin{bmisc}[author]
\bauthor{\bsnm{{Statistisches Bundesamt}}}
(\byear{2020}).
\btitle{Bev\"olkerung und Erwerbst\"atigkeit, Bev\"olkerungsfortschreibung auf
  Grundlage des Zensus 2011}.
\end{bmisc}
\endbibitem

\bibitem{census2019}
\begin{bmisc}[author]
\bauthor{\bsnm{{United States Census Bureau}}}
(\byear{2019}).
\btitle{Population Estimates Show Aging Across Race Groups Differs (accessed
  July 10, 2020)}.
\bnote{https://www.census.gov/newsroom/press-releases/2019/estimates-characteristics.html}.
\end{bmisc}
\endbibitem

\bibitem{chu2020}
\begin{barticle}[author]
\bauthor{\bsnm{Chu},~\bfnm{Derek~K}\binits{D.~K.}},
  \bauthor{\bsnm{Akl},~\bfnm{Elie~A}\binits{E.~A.}},
  \bauthor{\bsnm{Duda},~\bfnm{Stephanie}\binits{S.}},
  \bauthor{\bsnm{Solo},~\bfnm{Karla}\binits{K.}},
  \bauthor{\bsnm{Yaacoub},~\bfnm{Sally}\binits{S.}},
  \bauthor{\bsnm{Sch{\"u}nemann},~\bfnm{Holger~J}\binits{H.~J.}},
  \bauthor{\bsnm{El-harakeh},~\bfnm{Amena}\binits{A.}},
  \bauthor{\bsnm{Bognanni},~\bfnm{Antonio}\binits{A.}},
  \bauthor{\bsnm{Lotfi},~\bfnm{Tamara}\binits{T.}},
  \bauthor{\bsnm{Loeb},~\bfnm{Mark}\binits{M.}} \betal{et~al.}
(\byear{2020}).
\btitle{Physical Distancing, Face Masks, and Eye Protection to Prevent
  Person-to-Person Transmission of SARS-CoV-2 and COVID-19: A Systematic Review
  and Meta-Analysis}.
\bjournal{The Lancet}.
\end{barticle}
\endbibitem

\bibitem{Ferguson2020}
\begin{barticle}[author]
\bauthor{\bsnm{Ferguson},~\bfnm{Neil}\binits{N.}},
  \bauthor{\bsnm{Laydon},~\bfnm{Daniel}\binits{D.}},
  \bauthor{\bsnm{Nedjati~Gilani},~\bfnm{Gemma}\binits{G.}},
  \bauthor{\bsnm{Imai},~\bfnm{Natsuko}\binits{N.}},
  \bauthor{\bsnm{Ainslie},~\bfnm{Kylie}\binits{K.}},
  \bauthor{\bsnm{Baguelin},~\bfnm{Marc}\binits{M.}},
  \bauthor{\bsnm{Bhatia},~\bfnm{Sangeeta}\binits{S.}},
  \bauthor{\bsnm{Boonyasiri},~\bfnm{Adhiratha}\binits{A.}},
  \bauthor{\bsnm{Cucunuba~Perez},~\bfnm{ZULMA}\binits{Z.}},
  \bauthor{\bsnm{Cuomo-Dannenburg},~\bfnm{Gina}\binits{G.}} \betal{et~al.}
(\byear{2020}).
\btitle{Report 9: Impact of non-pharmaceutical interventions (NPIs) to reduce
  COVID19 mortality and healthcare demand}.
\end{barticle}
\endbibitem

\bibitem{OECD2020}
\begin{bmisc}[author]
\bauthor{\bsnm{{Organisation for Economic Cooperation and Development}}}
(\byear{2020}).
\btitle{Beyond Containment: Health systems responses to Covid-19 in the OECD}.
\end{bmisc}
\endbibitem

\bibitem{DIVI2020}
\begin{bmisc}[author]
\bauthor{\bsnm{{Deutsche Interdisziplin\"are Vereinigung f\"ur Intensiv- und
  Notfallmedizin (DIVI)}}}
(\byear{2020}).
\btitle{Daily Report DIVI Intensivregister (July 3, 2020)}.
\end{bmisc}
\endbibitem

\bibitem{Destatis2020}
\begin{bmisc}[author]
\bauthor{\bsnm{{Bundeszentrale f{\"u}r politische Bildung und Statistisches
  Bundesamt (Destatis)}}}
(\byear{2018}).
\btitle{{ Ein Sozialbericht f{\"u}r die Bundesrepublik Deutschland, Datenreport
  2018 Bd. 2018}}.
\end{bmisc}
\endbibitem

\bibitem{Grimm2020}
\begin{barticle}[author]
\bauthor{\bsnm{Grimm},~\bfnm{Veronika}\binits{V.}},
  \bauthor{\bsnm{Mengel},~\bfnm{Friederike}\binits{F.}} \AND
  \bauthor{\bsnm{Schmidt},~\bfnm{Martin}\binits{M.}}
(\byear{2020}).
\btitle{Extensions of the SEIR Model for the Analysis of Tailored Social
  Distancing and Tracing Approaches to Cope with COVID-19}.
\bjournal{medRxiv}.
\bdoi{10.1101/2020.04.24.20078113}
\end{barticle}
\endbibitem

\bibitem{horby2020}
\begin{barticle}[author]
\bauthor{\bsnm{Horby},~\bfnm{Peter}\binits{P.}} \AND
  \bauthor{\bsnm{Landray},~\bfnm{Martin}\binits{M.}}
(\byear{2020}).
\btitle{Low-Cost Dexamethasone Reduces Death by up to One Third in Hospitalised
  Patients with Severe Respiratory Complications of COVID-19}.
\bjournal{RECOVERY Trial Press Release. Available at: https://www. ox. ac.
  uk/news/2020-06-16-low-cost-dexamethasone-reduces-death-one-thirdhospitalised-patients-severe.
  Accessed July}
\bvolume{5}
\bpages{2020}.
\end{barticle}
\endbibitem

\bibitem{klepac2020}
\begin{barticle}[author]
\bauthor{\bsnm{Klepac},~\bfnm{Petra}\binits{P.}},
  \bauthor{\bsnm{Kucharski},~\bfnm{Adam~J}\binits{A.~J.}},
  \bauthor{\bsnm{Conlan},~\bfnm{Andrew~JK}\binits{A.~J.}},
  \bauthor{\bsnm{Kissler},~\bfnm{Stephen}\binits{S.}},
  \bauthor{\bsnm{Tang},~\bfnm{Maria}\binits{M.}},
  \bauthor{\bsnm{Fry},~\bfnm{Hannah}\binits{H.}} \AND
  \bauthor{\bsnm{Gog},~\bfnm{Julia~R}\binits{J.~R.}}
(\byear{2020}).
\btitle{Contacts in Context: Large-Scale Setting-Specific Social Mixing
  Matrices from the BBC Pandemic project}.
\bjournal{medRxiv}.
\end{barticle}
\endbibitem

\bibitem{lageRKI}
\begin{bmisc}[author]
\bauthor{\bsnm{{Robert-Koch-Institut (RKI)}}}
(\byear{2020}).
\btitle{{Täglicher Lagebericht des RKI zur Coronavirus-Krankheit-2019
  (COVID-19), 10.08.2020}}.
\bnote{{https://www.rki.de/DE/Content/InfAZ/N/Neuartiges\_Coronavirus/Situationsberichte/2020-08-10-de.pdf?\_\_blob=publicationFile}}.
\end{bmisc}
\endbibitem

\bibitem{RKI2020}
\begin{bmisc}[author]
\bauthor{\bsnm{{Robert-Koch-Institut (RKI)}}}
(\byear{2020}).
\btitle{{SARS-CoV-2 Steckbrief zur Coronavirus-Krankheit-2019 (COVID-19) Stand:
  10.07.2020}}.
\bnote{{https://www.rki.de/DE/Content/InfAZ/N/Neuartiges\_Coronavirus/Steckbrief.html}}.
\end{bmisc}
\endbibitem

\end{thebibliography}

\newpage 
\appendix

\newpage 

\section{Appendix} \label{appendix}

\subsection*{Additional Results}

\begin{figure}[b]
\begin{center}
\textit{(i)}
\end{center}
\includegraphics[scale=0.75]{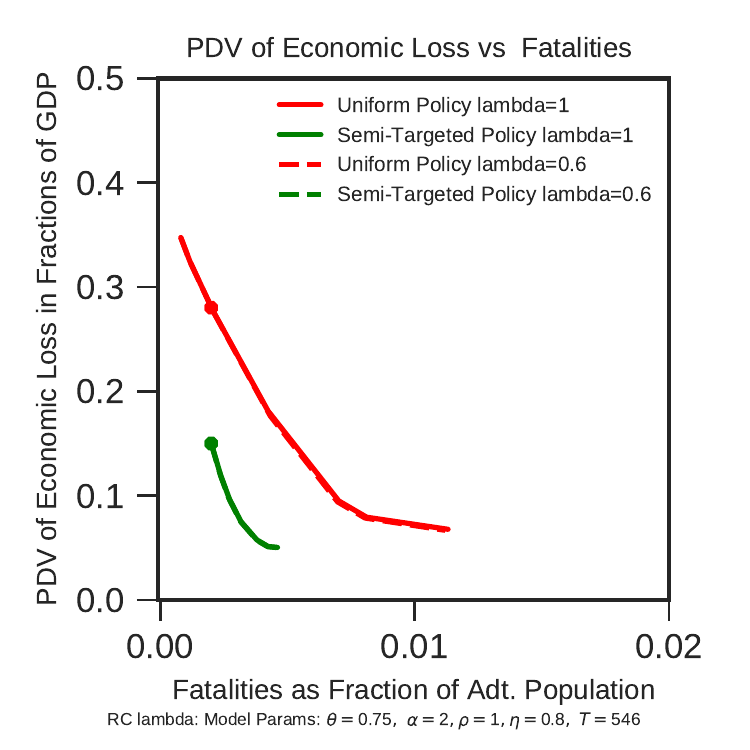} 
\includegraphics[scale=0.75]{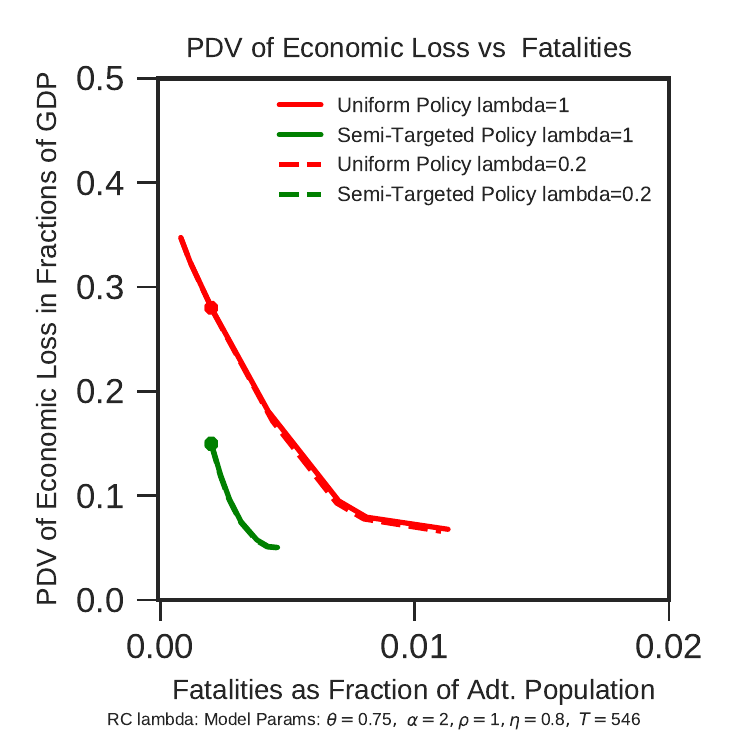} 
\begin{center}
\textit{(ii)}
\end{center}
\includegraphics[scale=0.5]{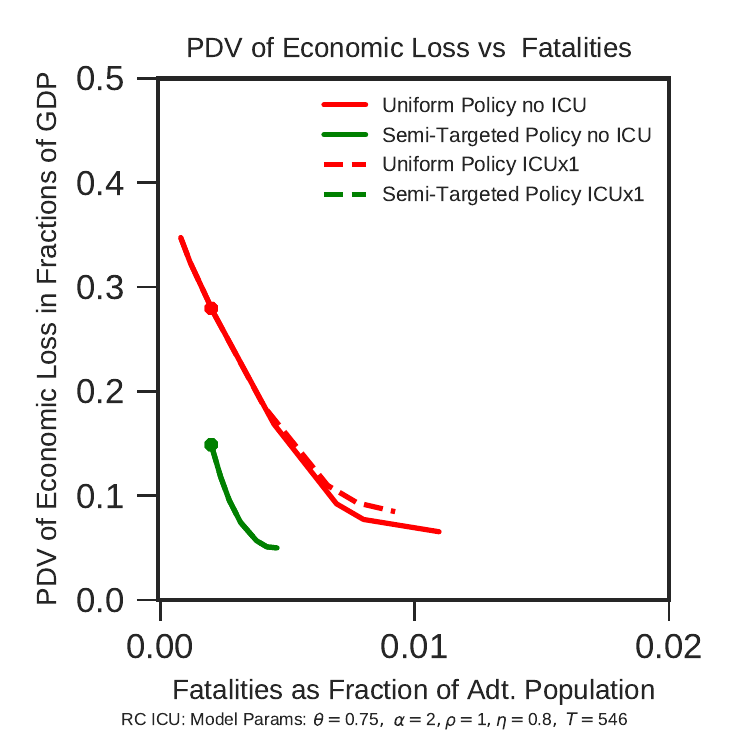} \includegraphics[scale=0.5]{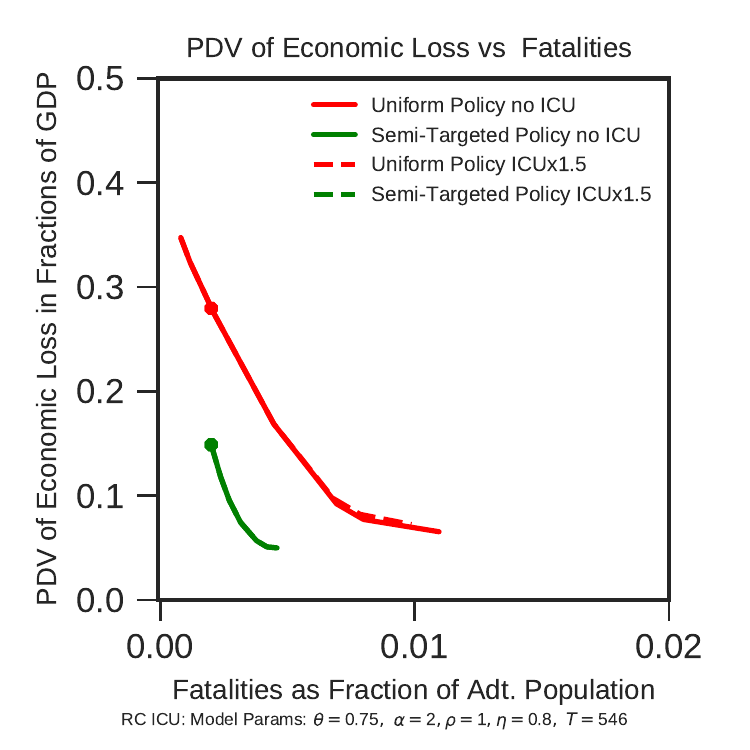} 
\includegraphics[scale=0.5]{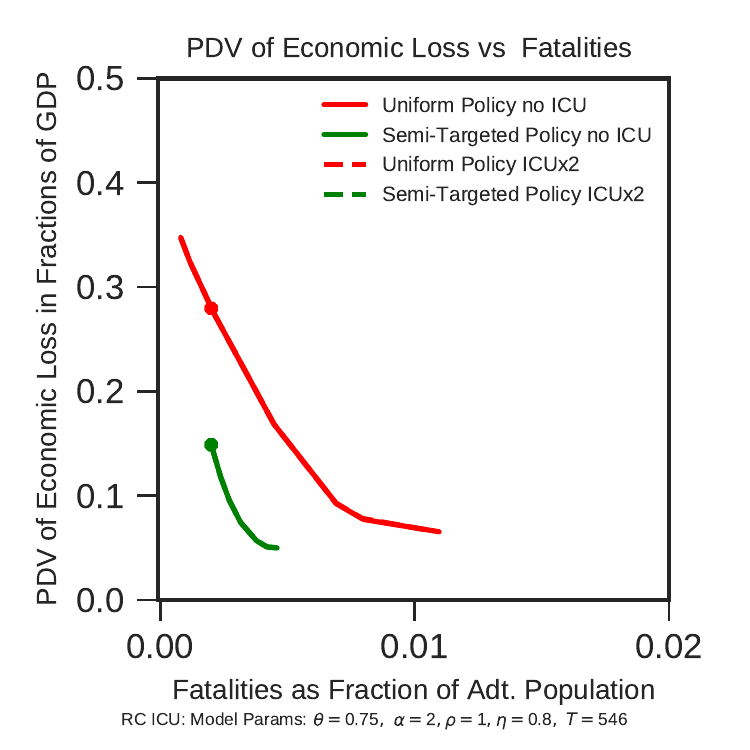} 
\caption{Panel ($i$): Variation in $\lambda=\lambda^\prime$ in Equation \ref{mort}, with $\lambda^\prime=0.6$ (left), $\lambda^\prime=0.2$ (right). The solid liner refers to the case with $\lambda=1$. The dashed lines refer to the policy frontier with the $\lambda=\lambda^\prime$. Panel ($ii$): Variation of the ICU constraint. The solid line indicates the optimal policy frontier without a binding ICU constraint. Dashed lines refer to binding ICU constraints with $\bar{H}(t)=0.02$ (left), $\bar{H}(t)=0.03$ (center), and $\bar{H}(t)=0.04$ (right).}
\label{lambdavar}
\end{figure}

Reducing the parameter $\lambda$ rotates the policy frontier to the left, bringing the menu of policy choices involving high mortality rates closer to the bliss point as illustrated in Figure \ref{lambdavar}, panel ($i$). Because targeted policies can reduce the mortality in the adult population effectively, reducing $\lambda$ has an effect in particular on the choice of uniform policies. 

Relaxing the hard ICU constraint allows lower mortality rates to be achieved at a given level of economic damage as can be concluded from Figure \ref{lambdavar}, panel ($ii$). Increasing the bound in the capacity constraint from $\bar{H}(t)=0.02$ to $\bar{H}(t)=0.04$ brings the policy frontier closer to the case without binding ICU constraints. This change can be observed for uniform shielding policies, whereas the impact on targeted policies is smaller and only observable for the case with a relatively tight constraint $\bar{H}(t)=0.02$.
 
\begin{figure}
\begin{center}
\textit{(i)}
\end{center}
 \includegraphics[scale=1]{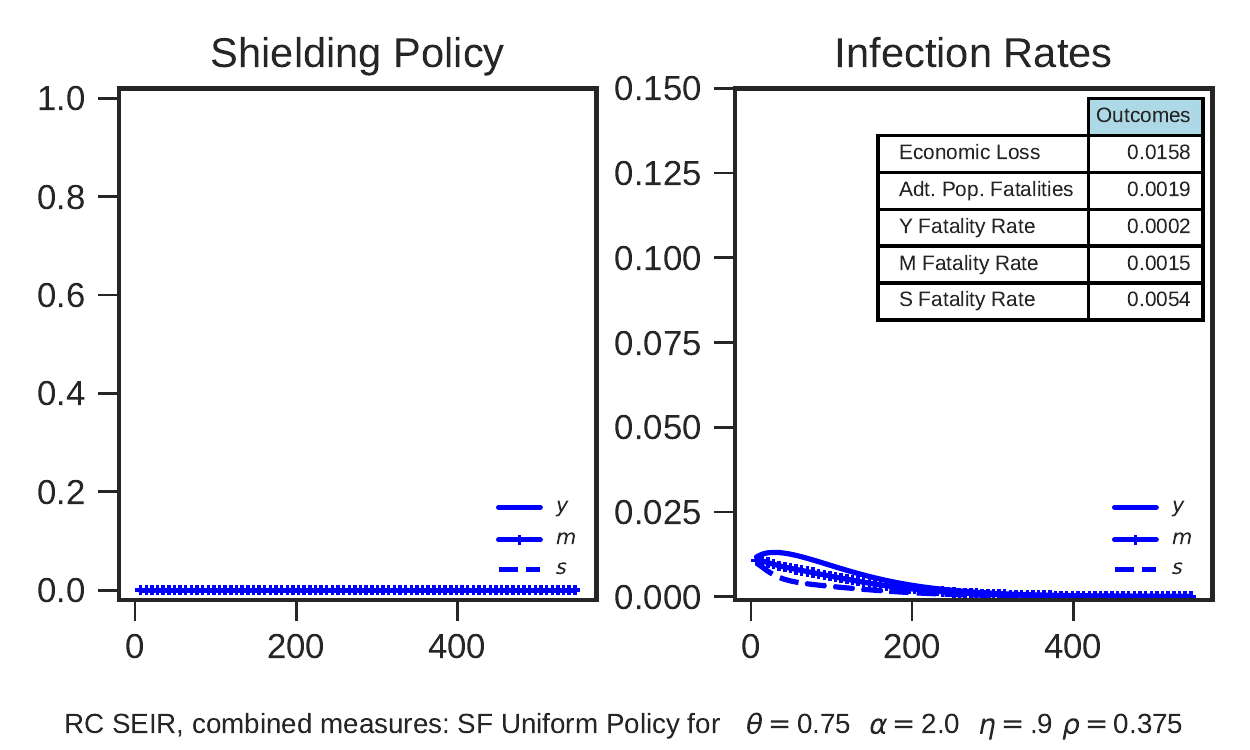}
\\
\begin{center}
\textit{(iii)}
\end{center}
 \includegraphics[scale=1]{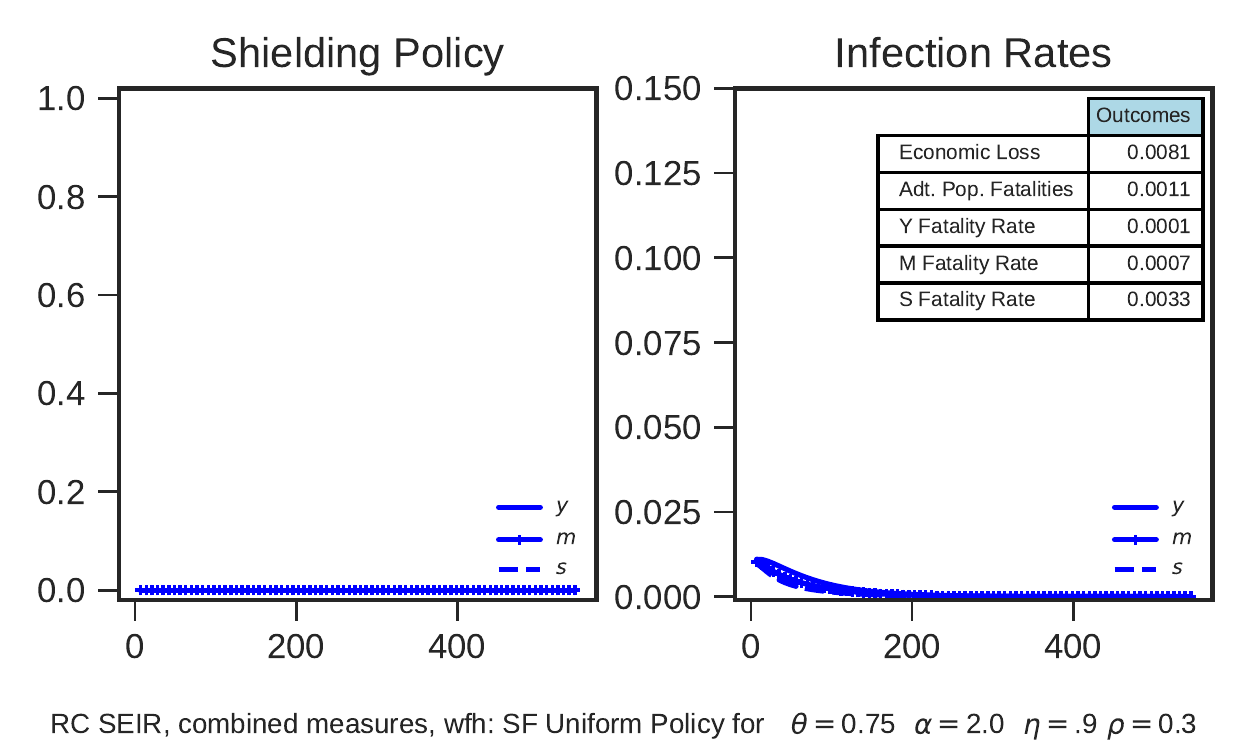}
\caption{Optimal uniform policy with safety focus. Combination of policy measures (comprehensive approach). Panel ($i$): Improved testing and isolation for infected ($\eta_I=0.7$) and exposed ($\eta_E=0.8$), reduced contact rates for interactions with the senior group ($\rho_{ys}=\rho_{ms}=0.2$). Panel ($iii$): Improved testing and isolation for infected ($\eta_I=0.7$) and exposed ($\eta_E=0.8$), reduced contact rates for interactions with the senior group ($\rho_{ys}=\rho_{ms}=0.2$), and improved conditions for working from home ($\pi_1=20\%, \pi_2=5\%, \xi=0.4$.)}
\label{unifcomb} 
\end{figure}

\begin{figure}
\begin{center}
\textit{(i)}
\end{center}
 \includegraphics[scale=0.8]{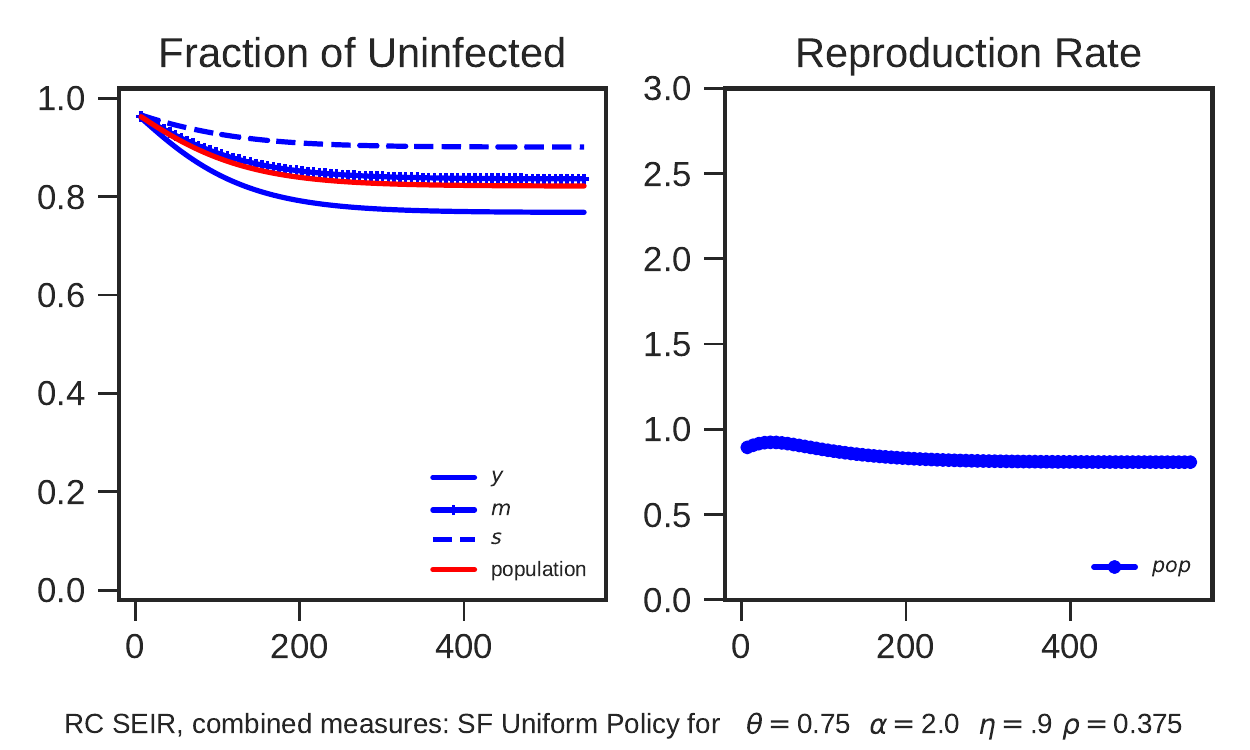}
\\
\begin{center}
\textit{(ii)}
\end{center}
 \includegraphics[scale=0.8]{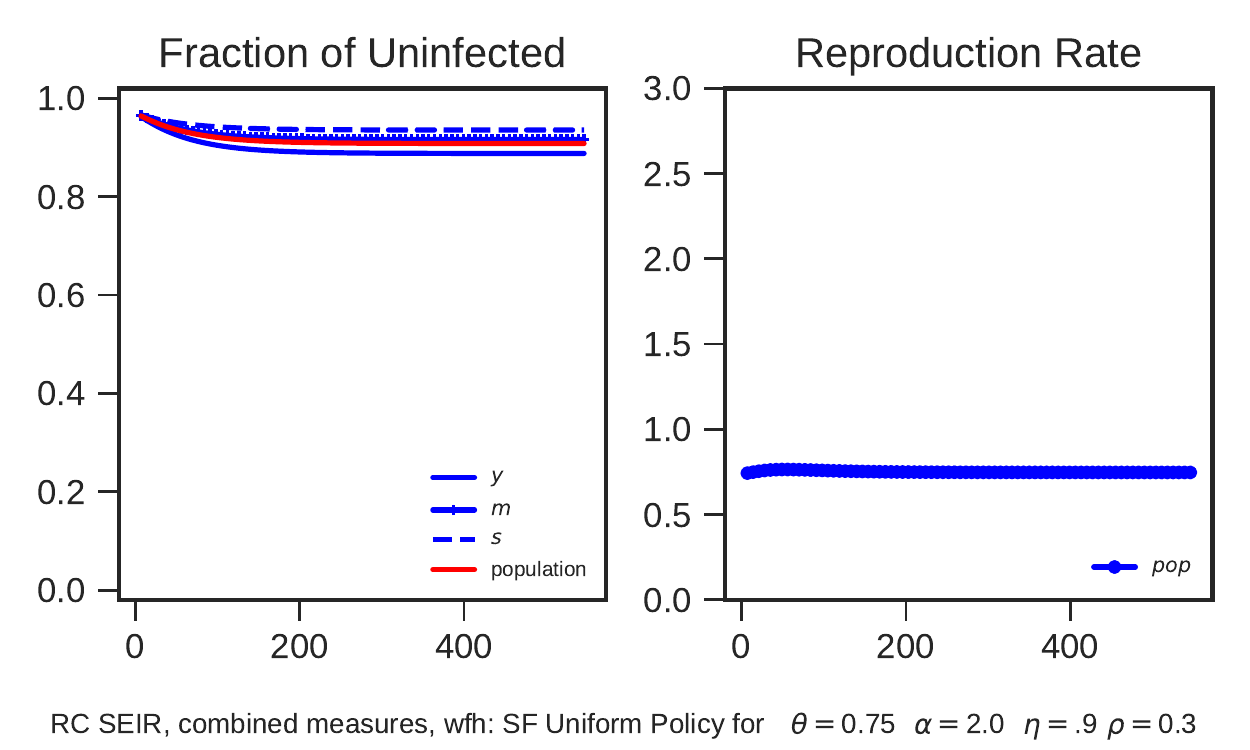}
\caption{Share of uninfected (left) and reproduction rate $R(t)$ (right) in the setting with combination of policy measures (comprehensive approach). Uniform shielding policies. Panel ($i$): Improved testing and isolation for infected ($\eta_I=0.7$) and exposed ($\eta_E=0.8$), reduced contact rates for interactions with the senior group ($\rho_{ys}=\rho_{ms}=0.2$). Panel ($ii$): Improved testing and isolation for infected ($\eta_I=0.7$) and exposed ($\eta_E=0.8$), reduced contact rates for interactions with the senior group ($\rho_{ys}=\rho_{ms}=0.2$), and improved conditions for working from home ($\pi_1=20\%, \pi_2=5\%, \xi=0.4$.)}
\label{unifcombepi} 
\end{figure}

\begin{figure}
\centering
\includegraphics[scale=0.75]{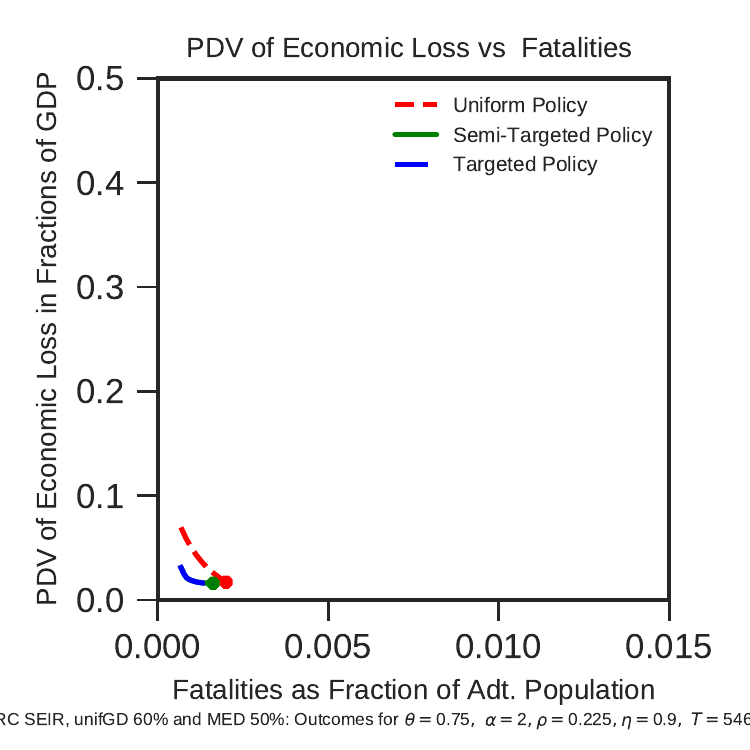} 
 \includegraphics[scale=0.75]{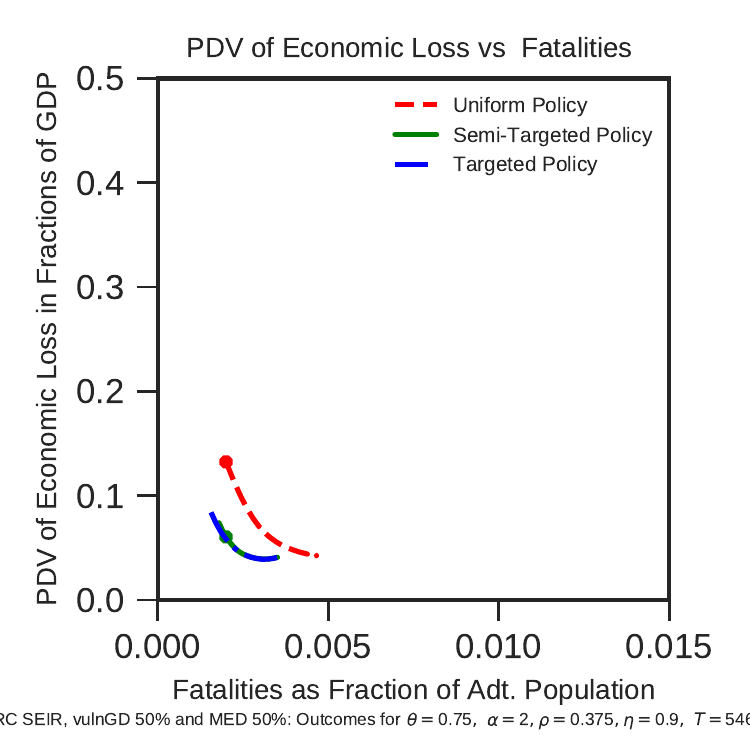}
\caption{Combination of group distancing and decreased mortality of the elderly by 50\% due to improved medical treatment. Efficient frontier with uniform distancing policy (left) and distancing targeted towards the vulnerable (right). }
\label{GDMED}
\end{figure}

\begin{figure}
\centering
\begin{center}
\textit{(i)}
\end{center}
\includegraphics[scale=1]{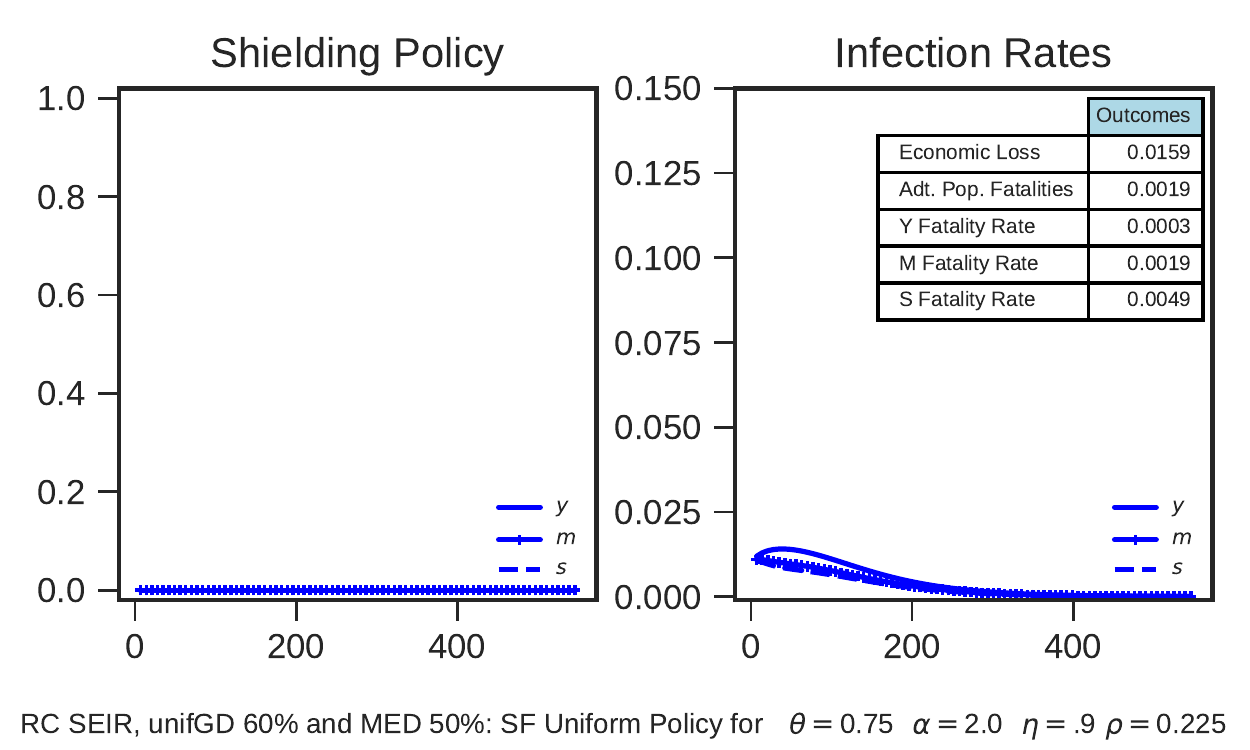} \\
\begin{center}
\textit{(ii)}
\end{center}
\includegraphics[scale=1]{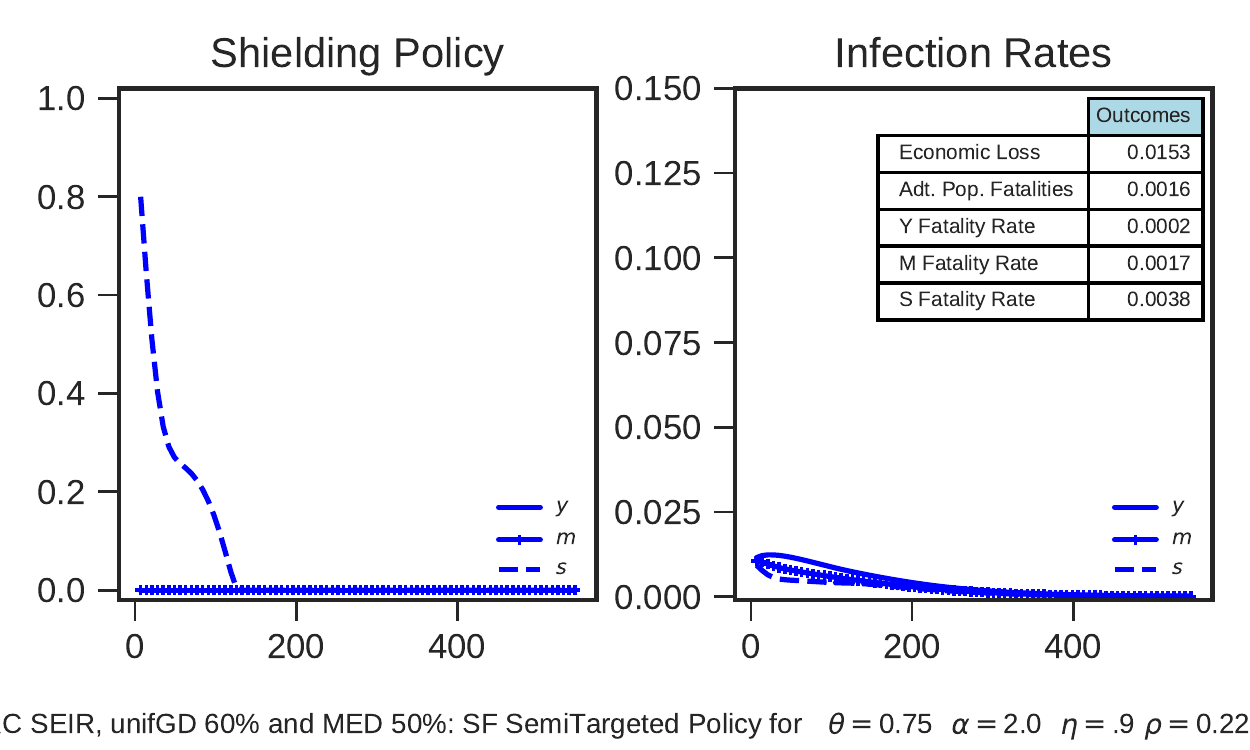}
\caption{Optimal safety-focused shielding policy with combination of a uniform group distancing policy (reduction of all elements in $\rho$ by 40\%) and improved medical treamtent (50\% lower mortality rate for the senior group). Panel ($i$): Optimal semi-targeted shielding policy. Panel ($ii$): Optimal semi-targeted shielding policy.}
\label{unifGDMED}
\end{figure}

\begin{figure}
\centering
\begin{center}
\textit{(i)}
\end{center}
\includegraphics[scale=1]{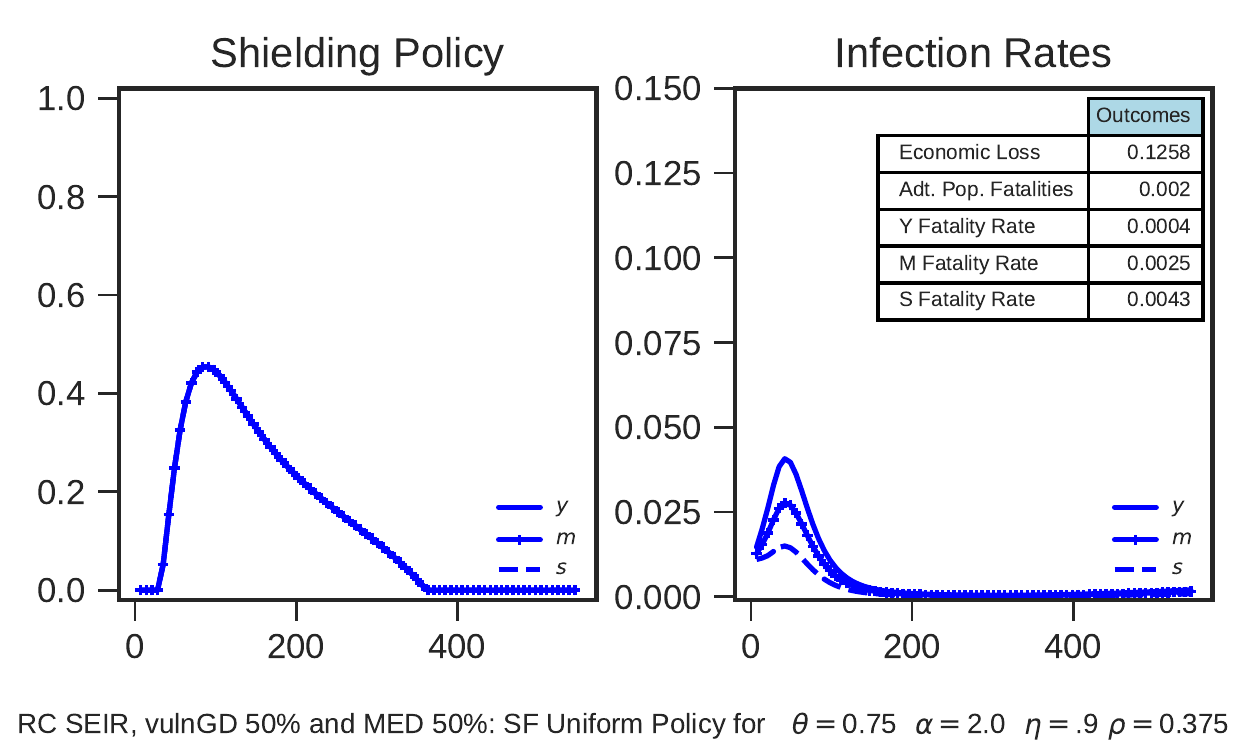} \\
\begin{center}
\textit{(ii)}
\end{center}
\includegraphics[scale=1]{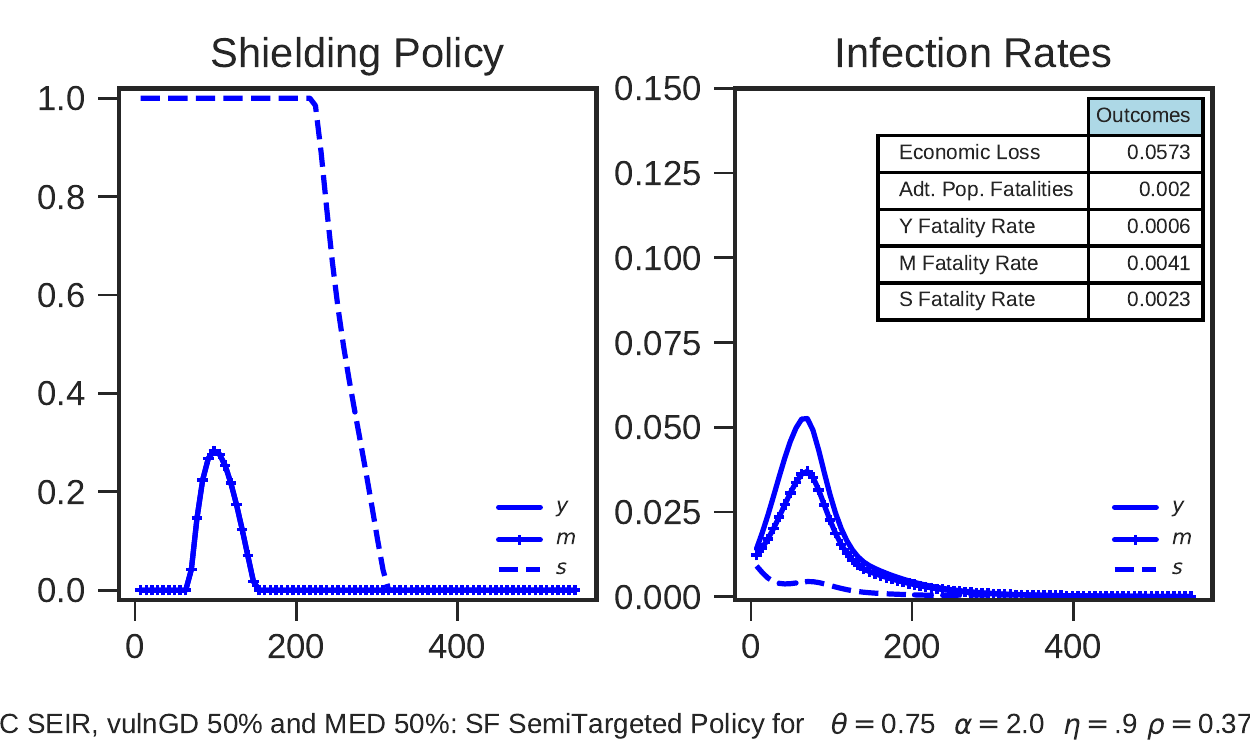}
\caption{Optimal safety-focused shielding policy with combination of group distancing towards the vulnerable (reduction of contact rates $\rho_{ys}, \rho_{ms}$ by 50\%) and improved medical treatment (50\% lower mortality rate for the senior group). Panel ($i$): Optimal uniform  policy. Panel ($ii$): Optimal semi-targeted  policy.}
\label{vulnGDMED}
\end{figure}

\end{document}